\documentclass[final,5p,times,twocolumn,authoryear]{elsarticle}

\usepackage{amssymb}
\usepackage{float}
\usepackage{lipsum}
\usepackage{ulem}

\usepackage{float}
\usepackage{caption}
\usepackage{supertabular}

\usepackage{orcidlink}

\usepackage{longtable}
\usepackage{adjustbox}
\usepackage{rotating}
\usepackage{tikz}
\usepackage{geometry}

\usetikzlibrary{shapes.geometric, arrows, positioning}
\usepackage{booktabs}
\usepackage{amssymb}
\usepackage{graphicx}
\usepackage{adjustbox}
\usepackage{tabularx}
\usepackage{subcaption}
\usepackage{natbib}
\usepackage{pdflscape}

\usepackage{amsmath}
\usepackage{hyperref}
\usepackage{subcaption} 
\journal{High Energy Astrophysics}

\begin{document}

\begin{frontmatter}


\title{Multi-Model Framework for Reconstructing Gamma-Ray Burst Light Curves}


\author[1]{A. Kaushal\fnref{fn1}}
\affiliation[1]{organization= {Department of Computer Science and Engineering, UIET-H, Panjab University},
                addressline={Punjab, 146001, India}}

\author[2]{A. Manchanda\fnref{fn1}}
\affiliation[2]{organization= {Centre for Astrophysics and Supercomputing, Swinburne University of Technology},
                city={Victoria},
                postcode={3122},
                country={Australia}}

\author[3,4,5,6]{M. G. Dainotti\corref{cor2}\fnref{fn1}}
\cortext[cor2]{Corresponding author, maria.dainotti@nao.ac.jp}
\affiliation[3]{organization={Division of Science, National Astronomical Observatory of Japan},
               addressline={2-21-1 Osawa, Mitaka},
               city={Tokyo},
               postcode={181-8588},
               %state={},
               country={Japan}}

\fntext[fn1]{Equal contribution}

\affiliation[4]{organization={The Graduate University for Advanced Studies (SOKENDAI)},
               addressline={Shonankokusaimura, Hayama, Miura District},
               city={Kanagawa},
               postcode={240-0115},
               %state={},
               country={Japan}}

\affiliation[5]{organization={Space Science Institute},
            addressline={4765 Walnut St Ste B},
            city={Boulder},
            postcode={80301},
            state={CO},
            country={USA}}            

\affiliation[6]{organization={Nevada Center for Astrophysics, University of Nevada},
            addressline={4505 Maryland Parkway},
            city={Las Vegas},
            postcode={89154},
            state={NV},
            country={USA}}

\author[8]{K. Gupta\fnref{fn1}}
\affiliation[8]{organization= {Department of Computer Science, PSIT College},
                city={Uttar Pradesh},
                postcode={209305},
                country={India}}

\author[9]{Z. Nogala}
\affiliation[9]{organization= {Department of Mathematics, University of Wroclaw},
                city={Wroclaw},
                postcode={50-384},
                country={Poland}}
                
\author[10]{A. Madhan}
\affiliation[10]{organization= {BE CSE, Visveswaraya Technological University},
                city={Karnataka},
                postcode={590018},
                country={India}}

\author[11]{S. Naqi}
\affiliation[11]{organization= {B. Tech, Information Technology, KNIT Sultanpur},
                city={Uttar Pradesh},
                postcode={228118},
                country={India}}

\author[12]{Ritik Kumar}
\affiliation[12]{organization= {Indian Institute of Science},
                city={Karnataka},
                postcode={560012},
                country={India}}  

\author[13]{V. Oad}
\affiliation[13]{organization= {Vishwakarma Government Engineering College},
                city={Ahmedabad},
                postcode={382424},
                country={India}}  
                
\author[14]{N. Indoriya}
\affiliation[14]{organization= {Department of Electrical Engineering and Computer Science, Indian Institute of Science Education and Research},
                city={Bhopal},
                postcode={462066},
                country={India}}  

\author[7]{Krishnanjan Sil \orcidlink{0009-0000-8423-4039}\fnref{fn1}}
\affiliation[7]{organization= {Department of Physics, Ramakrishna Mission Vivekananda Centenary College},
                city={Kolkata},
                postcode={700118},
                country={India}}        
% \author[15]{H. Gupta}
% \affiliation[15]{organization= {MIT WPU School of Computer Science \& Engineering},
%                 city={Pune},
%                 postcode={411038},
%                 country={India}} 
                            
% \author[16]{S. Roy}
% \affiliation[16]{organization= {National Institute of Technology Agartala},
%                 city={Tripura},
%                 postcode={799046},
%                 country={India}}  
           
\author[17]{D. H. Hartmann}
\affiliation[17]{organization= {Department of Physics and Astronomy, Clemson University},
                city={Clemson},
                postcode={SC 29634},
                country={USA}} 

\author[20]{M. Bogdan}
\affiliation[20]{
    organization= {Department of Mathematics, University of Wroclaw},
    city={Wroclaw},
    postcode={50-384},
    country={Poland}} 
    
\author[18,19]{A. Pollo}
\affiliation[18]{organization={Astronomical Observatory of Jagiellonian University in Kraków},
             addressline={Orla 171},
             city={Krakow},
             postcode={30-244},
             country={Poland}}
             
\affiliation[19]{organization={National Centre for Nuclear Research},
            city={Warsaw},             postcode={02-093},
           country={Poland}}

% \affiliation[16]{
%     organization= {Department of Statistics, Lund University},
%     city={Lund},
%     postcode={SE-221 00},
%     country={Sweden}} 

\author[21]{J.X. Prochaska}
\affiliation[21]{
    organization= {University of California},
    addressline={Santa Cruz, 1156 High Street},
    city={Santa Cruz},
    postcode={CA 95064},
    country={USA}}

\author[22]{N. Fraija}
\affiliation[22]{
    organization= {National Autonoma University of Mexico},
    addressline={Circuito Interior},
    city={Mexico City},
    postcode={04510},
    country={Mexico}
}

%\author[11]{N. Fraija}
%\affiliation[11]{organization={National Autonoma University of Mexico },
 %            addressline={Circuito Interior},
  %           city={Mexico City},
   %          postcode={04510}}

% \textcolor{yellow}{Our findings demonstrate that Isotonic Regression achieves the highest uncertainty reduction for all three parameters (36.3\% for $\log T_a$, 36.1\% for $\log F_a$, and
% 43.6\% for $\alpha$) outperforming all the other models. 
% The CNN-BiLSTM model shows consistent improvements across all GRB parameters with the lowest outlier rate for $\alpha$ (0.550\%), surpassing the performance of the LSTM model in \citealt{ manchanda2025gammarayburstlightcurve}. The DGP model offers reliable uncertainty reduction across all parameters and improves upon the single-layer GP baseline.}

\begin{abstract}

Mitigating data gaps in Gamma-ray bursts (GRBs) light curves (LCs) is crucial for cosmological research, enhancing the precision of parameters, assuming perfect satellite conditions for complete LC coverage with no gaps. This analysis improves the applicability of the two-dimensional Dainotti relation, which connects the rest-frame end time of the plateau emission ($T_a$) and its luminosity ($L_a$), derived from the fluxes ($F_a$). The study expands on a previous 521 GRB sample by incorporating seven models: Deep Gaussian Process (DGP), Temporal Convolutional Network (TCN), Hybrid CNN with Bidirectional Long Short-Term Memory (CNN-BiLSTM), Bayesian Neural Network (BNN), Polynomial Curve Fitting, Isotonic Regression, and Quartic Smoothing Spline (QSS). Results indicate that QSS significantly reduces uncertainty across parameters—43.5\% for $\log T_a$, 43.2\% for $\log F_a$, and 48.3\% for $\alpha$, outperforming the other models where $\alpha$ denotes the slope post-plateau based on Willingale's 2007 functional form. The Polynomial Curve Fitting model demonstrates moderate uncertainty reduction across parameters, while CNN-BiLSTM has the lowest outlier rate for $\alpha$ at 0.77\%. These models broaden the application of machine-learning techniques in GRB LC analysis, enhancing uncertainty estimation and parameter recovery, and complement traditional methods like the Attention U-Net and Multilayer Perceptron (MLP). These advancements highlight the potential of GRBs as cosmological probes, supporting their role in theoretical model discrimination via LC parameters, serving as standard candles, and facilitating GRB redshift predictions through advanced machine-learning approaches.

\end{abstract}



\end{frontmatter}

%\tableofcontents



\section{Introduction}
\label{sec:intro}

GRBs are among the most luminous transients observed, reaching redshift $z=9.4$ \citep{Cucchiara2011}, making them valuable probes of the early universe. Crucial details about Population III stars can also be obtained from a detailed study of GRBs.

GRBs have traditionally been divided into two types based on the duration of their gamma-ray prompt emission, as measured by $T_{90}$. This measure shows how long it took to capture 90\% of all background-deducted counts, starting after the first 5\% of counts \citep{mazets1981catalog, kouveliotou1993classification}. The mergers of compact objects are usually linked to short GRBs (SGRBs), which are characterised by $T_{90}\leq 2$s \citep{duncan1992,Narayan1992,usov1992,thompson1994MNRAS.270..480T,levan2008intrinsic,metzger2011MNRAS.413.2031M,Bucciantini2012MNRAS.419.1537B,perna2016ApJ...821L..18P}. On the other hand, the collapse of large stars is associated with long GRBs (LGRBs), which have $T_{90} \ge 2$s \citep{woosley1993ApJ...405..273W,paczynski1998ApJ...494L..45P,macfayden1999ApJ...524..262M,bloom2002AJ....123.1111B,hjorth2003Natur.423..847H,Woosley2006ARA&A,woosley2006ApJ...637..914W,kumar2008Sci...321..376K,hjorth2012grb..book..169H,bucciantini2008,cano2017AdAst2017E...5C,lyman2017MNRAS.467.1795L,perna2018ApJ...859...48P,aloy2021MNRAS.500.4365A,Ahumada2021NatAs...5..917A}.

The high-energy prompt phase has been interpreted by internal shell collision or magnetic reconnection \citep{Vestrand2005Natur,Blake2005Natur,Beskin2010ApJ,2012MNRAS.421.1874G,2014Sci...343...38V} and the long-lasting multi-wavelength afterglow phase as the interaction of the shells with the circumburst medium \citep{costa1997,vanParadijs1997,Piro1998,Gehrels2009ARA&A}. Instruments on board the Neil Gehrels Swift Observatory (Swift, \citealt{Gehrels2004ApJ...611.1005G}), BAT, XRT, and UVOT, have been crucial for rapidly identifying GRBs and enabling follow-up across X-ray to optical bands \citep{barthelmy2005burst,burrows2005swift,Roming2005}. Furthermore, novel features of GRB LCs have been discovered by Swift's quick multi-wavelength afterglow follow-up \citep{Tagliaferri2005,Nousek2006,Troja2007}.

Most X-ray LCs exhibit a plateau after the prompt phase \citep{Zhang2006,OBrien2006,Nousek2006,Sakamoto2007,Liang2007,willingale2007testing,Dainotti2008,dainotti2010a,Dainotti2016,dainotti2017a, dereli2024unraveling}, which can be modeled using a broken power-law (BPL) \citep{Zhang2006, Zhang2007ApJ...655L..25Z,Racusin2009}, a smoothly BPL, or the phenomenological model proposed by \citealt{willingale2007testing}(W07). 
Section ~\ref{section:sampling} details the W07 and the critical parameters. The plateau phase is particularly notable and is frequently explained using the framework of the magnetar model \citep{Zhang2001, 2014MNRAS.443.1779R, Rea2015, Stratta2018}, where precise measurements of $T_a$ (time at the end of plateau) are necessary to confirm the validity of this model. Moreover, a significant anti-correlation between rest-frame time $T^{*}_{X,a}$ and the corresponding X-ray luminosity $L_X$, known as the Dainotti 2D relation, has been identified \citep{Dainotti2008, dainotti2010a, dainotti2011a,dainotti2013determination, dainotti2015, dainotti2017a, Tang2019ApJS..245....1T, Wang:2021hcx, Zhao2019ApJ...883...97Z, Liang2007, Li2018b}. 

This 2D L-T relation is further developed into the 3D Dainotti relation \citep{Dainotti2016,dainotti2017a,dainotti2020a, 2022ApJS..261...25D}, incorporating prompt luminosity $L_{X, peak}$, and aiding cosmological parameter constraints \citep{dainotti2023b, dainotti2022g, dainotti2022b, 2022MNRAS.512..439C, 2022MNRAS.510.2928C}. The reduction of the uncertainties on the plateau parameter by 47.5\% could match the cosmological precision of $\Omega_M$ from SNe Ia \citep{dainotti2020x} within 8 years \citep{dainotti2022g, Betoule}, compared to the 16 years required at current observation rates \citep{dainotti2022g}, highlighting the prospects of a practical LC reconstruction (LCR) approach.
This enhancement can be performed assuming that the data augmentation would come from an ideal situation in which the satellite in question would be a perfect one with no gaps.

A critical barrier to using GRBs for population studies or cosmology is the presence of observational gaps in LCs, due to instrumental constraints or follow-up delays. These gaps hinder reliable model testing, such as the standard fireball model \citep{panaitescu2000analytic, piran1999gamma} or evaluating closure relations \citep{willingale2007testing,evans2009methods,racusin2009jet,kumar2010external, srinivasaragavan2020investigation, dainotti2021closure, ryan2020gamma, tak2019closure}. LCR methods have emerged as a powerful solution to this problem.

Previous works, such as \citealt{dainotti2023stochastic} and \citealt{sourav2023predicting}, have introduced probabilistic and deep learning techniques, such as Gaussian Processes (GP) \citep{GP2003gaussian} and Long Short-Term Memory (LSTM; \citealt{hochreiter1997long}), to reconstruct the temporal gaps in the LCs. \citealt{manchanda2025gammarayburstlightcurve} expanded on these approaches by evaluating ten different models for LCR. Building upon these advancements, we extend the LCR framework by introducing seven more models: Deep Gaussian Process (DGP), Temporal Convolutional Network (TCN), Hybrid model of Convolutional Neural Network with Long Short-Term Memory (CNN-BiLSTM), Bayesian Neural Network (BNN), Polynomial Curve Fitting, QSS (Quartic Smoothing Spline), and Isotonic Regression.

This paper is organized as follows: Section \S\ref{section:methods} offers an in-depth description of the dataset used and the different models applied to reconstruct GRB LCs.
%The paper comprises the following sections: sec. \S\ref{section:methods} discusses the utilized dataset and reconstruction models, entailing the W07, Bi-MAMBA, MLP, Fourier Transform and GP-RF methods. 
The uncertainty, performance, and outliers in Sec. \S\ref{section:results}. Sec. \S\ref{section:conclusion} provides the synopsis and conclusions on the observed efficacy of each model. The appendices \S\ref{sec:appendixA} \& \S\ref{sec:appendixB} discuss attempts to build another promising model and solutions to resolve a systematic shift problem discovered during analysis of a simulated GRB dataset, respectively.

\section{Methodology}\label{section:methods}

This section outlines the methodology adopted for the reconstruction of GRB LCs. We first motivate the selection of the models, then describe the data set and preprocessing steps, followed by a detailed explanation of the theoretical framework and implementation of each model.

\subsection{Motivation for the models}

A brief description as well as the motivation for selecting these models are provided below.

\subsubsection{DGP}
Deep Gaussian Processes (DGPs) represent another powerful approach and we expect that they might perform better than the GP or the GP-Random Forest (GP-RF) hybrid model for reconstructing the GRB LCs. Although single GP models are good for handling basic patterns and are faster to train, they may miss the deeper, complex structures in the LCs. DGPs work in layers, learning more abstract and detailed features step by step, and they also give a clear idea of how confident the model is in its predictions. 

\subsubsection{TCN}
Temporal Convolutional Networks (TCNs) are used for GRB LCR because they are proficient in identifying intricate temporal trends and long-range dependencies in time-series data \citep{Bai2018}. GRB LCs often feature sudden bursts and fluctuations across various time scales. TCNs are well-suited to handle this kind of data because they avoid issues like the weights of the model either increasing exponentially or going toward zero, namely exploding or vanishing gradients, which can occur in traditional models like Recurrent Neural Networks (RNNs). They can learn patterns in GRB data more efficiently, making them particularly useful for predicting the evolution of LCs over time.

\subsubsection{CNN-BiLSTM}
CNN-BiLSTM models integrate the advantages of convolutional neural networks and long short-term memory networks. We use this hybrid CNN-BiLSTM model with the expectation that it would perform better than the standard Bi-LSTM in reconstructing GRB LCs. The CNN component first detects and emphasizes important local features in the data, such as rapid bursts, while the LSTM network then models how these features evolve. This separation of feature extraction and sequence modeling can make CNN-BiLSTM more focused and reliable, especially in noisy LCs. In contrast, Bi-LSTM processes the entire sequence bidirectionally without explicitly identifying which input, such as flux and time parts are most relevant, making it computationally heavier and less effective at filtering noise.

\subsubsection{BNN}
Bayesian Neural Networks (BNNs) are chosen for GRB LCR because they offer a way to estimate uncertainty in predictions. Since GRB data can be noisy, sparse, and unpredictable, it's crucial to understand the confidence level of the model's predictions. BNNs allow for uncertainty estimation by incorporating probabilistic distributions over the model's parameters \citep{Blundell2015}. This helps in situations where data is limited or uncertain, providing a more reliable model for LCR by quantifying both the mean and variance of predictions.

\subsubsection{Polynomial Curve Fitting}
Polynomial Curve Fitting is used to fit a polynomial function to the data and is particularly effective when the LC does not exhibit rapid fluctuations or irregular bursts. Polynomial fitting can capture the general trend of the LC, making it a simple yet useful approach when the data is relatively well-behaved or to serve as a baseline model for more complex methods.

\subsubsection{Isotonic Regression}
Isotonic regression is a non-parametric method that can fit a monotonic curve to the data, which makes it well-suited for situations where the relationship between time and intensity is not linear but still follows a consistent trend \citep{Barlow1972}. This method is flexible and can handle noisy data better than simple linear regression, making it useful for reconstructing the LCs in GRBs with flares and breaks.

\subsubsection{Quartic Smoothing Spline}

Even though contemporary machine learning models have strong predictive powers for time-series analysis, issues with model interpretability, the requirement for large training datasets, and the potential for overfitting sparse, noisy observations can make it difficult to apply these models to scientific data, like GRB LCs \citep{Mehta_2019}. From the class of non-parametric statistical frameworks, the QSS offers a strong substitute in this situation \citep{hastie2009elements}. The QSS stands out for its mathematical transparency and structural simplicity, as it offers explicit and principled control over the bias-variance trade-off by solving a constrained optimization problem where the model's smoothness is defined by a constraint on the sum of squared discontinuity jumps in its \textbf{fifth derivative}—a higher-order criterion well-suited for complex dynamics \citep{1993csfw.book.....D}. In contrast to many purely data-driven models, it naturally produces a continuously differentiable ($C^3$) reconstruction and imposes a strong and often physically desirable smoothness constraint \citep{deboor2001practical}. Consequently, the QSS is a robust and interpretable framework that provides a crucial benchmark for reconstructing complex temporal dynamics, particularly when the primary goal is to derive concrete insights from noisy and irregular observational data.

\subsection{Dataset and the Willingale model}\label{section:sampling}

This study utilizes the same dataset as described in \citealt{dainotti2023stochastic, manchanda2025gammarayburstlightcurve}, comprising 521 GRBs. The data originates from the Swift BAT-XRT repository \citep{Evans2009,Evans2007} and were originally compiled from: 455 GRBs observed by Swift between 2005 and 2019 \citep{2020ApJ...903...18S,dainotti2020a, giovanna2024inferring}, 4 GRBs detected by Fermi-LAT \citep{dainotti2024optical} and 62 additional Swift GRBs from 2019 to 2023 \citep{narendra2024grb}. The final sample includes 521 unique GRBs, of which 230 have known redshift. The sample was selected to include GRBs with a well-defined plateau phase in their X-ray afterglow LCs and a sufficient number of data points (typically more than five) to allow meaningful reconstruction. This 521 GRBs dataset represents approximately 32.33\% of all the GRBs detected by Swift from 2005 to 2023.
All LCs exhibit temporal gaps, with a minimum gap size of 0.03 in the log-time scale adopted for the reconstruction task.

%\begin{table*}[htbp] 
 %   \centering
  %  \begin{tabular}{lcccccr}
   % \hline
   %    GRB NAME&	$z$	&	$T_{90}$	&	$\frac{T_{90}}{(1+z)}$	&	LONG/SHORT	&	Type	&	Reference	\\
   %    \hline
   %    GRB060510B&	4.90&	262.95&	44.57&	L	&	XRR&	[1] 	\\
   %    GRB060526&	3.22&	275.20&	65.21&	L	&	XRR	&	[2]	\\
   %    GRB060604&	2.14&	96.00&	30.62&	L	& XRR	&	[1] 	\\
   %    GRB060605&	3.73&	539.12&	113.91&	L	&	XRR&	[2]	\\
    %   GRB060607A&	3.07&	102.20&	25.11&	SEE&	SEE&	[1]	\\
     %  GRB060614&	0.13&	109.104&	96.95&	SEE& SEE/XRR&	[1], [3] 	\\
    %   GRB060707&	3.14&	66.65&	15.08&	L	& XRR	&	[2]	\\
      % GRB060708&	1.92&	9.96&	3.41&	L	&	XRR&  [1]		\\
      % GRB060714&	2.71&	116.06&	31.28&	L	&	XRR	&	[2]	\\
      % \hline
    %\end{tabular}
    %\caption{The dataset comprising all the 521 GRBs, detailing redshift ($z$), duration of burst in the observer frame ($T_{90}$), rest-frame duration ($\frac{T_{90}}{(1+z)}$), classification into long vs short bursts, GRB type, and relevant references. The entire table is available online.}
  %  \label{tab:sample}
%\end{table*}

The W07 function has been used to model LCs of GRBs, defined in Eq. \ref{eqn1} and introduced by \citealt{willingale2007testing}:

\begin{equation}
f(t) = \left \{
\begin{array}{ll}
\displaystyle{F_i \exp{\left ( \alpha_i \left( 1 - \frac{t}{T_i} \right) \right )} \exp{\left (
- \frac{t_i}{t} \right )}} {\hspace{1cm}\rm for} \ \ t < T_i, \\
\\
\displaystyle{F_i \left ( \frac{t}{T_i} \right )^{-\alpha_i}
\exp{\left ( - \frac{t_i}{t} \right )}} {\hspace{1cm}\rm for} \ \ t \ge T_i. \\
\end{array}
\right .
\label{eqn1}
\end{equation}

Here, the symbols $T_i$ and $F_i$ refer to the time and flux, respectively. This corresponds to the end of the plateau phase, denoted with $a$; $\alpha_i$ is the represented temporal index parameter associated with $T_i$. The parameter $t_i$ corresponds to the onset of the rising phase, and the afterglow emission is represented by $t_a$.

% Here, the symbols $T_i$ and $F_i$ refer to the time and flux, respectively.  corresponding to either the conclusion of the prompt emission phase ($p$) or the plateau phase ($a$). The temporal index parameter associated with $T_i$ is represented by $\alpha_i$. The parameter $t_i$ signifies the onset of the rise phase, while $t_a$ corresponds to the afterglow emission phase.

The LCs analyzed in this study have been processed to remove the prompt emission segment due to its significant variability and the challenges in modeling it effectively. 

The 545 GRBs, of which 521 were used and classified into four distinct classes as mentioned in \citealt{dainotti2023a, manchanda2025gammarayburstlightcurve, dainotti2023stochastic}, are:
\begin{itemize}
 \item \textbf{Good GRBs:} GRBs aligning closely with the W07 model and represent 55.3\% of the dataset.
 \item \textbf{Flares/Bumps:} GRBs showing flares and bumps throughout the afterglow phase account for 24.06\% of the dataset.
 \item \textbf{Break:} GRBs with a single break towards the end of the LC, making up 13.14\% of the sample.
 \item \textbf{Flares/Bumps + Double Break:} GRBs that display a combination of flares/bumps and a double break constitute 7.5\% of the dataset.
\end{itemize}

\begin{figure*}
\begin{center}
\includegraphics[width=.45\textwidth]{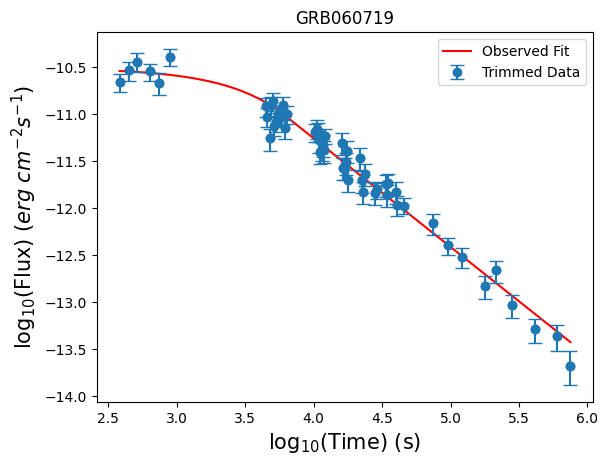}
\includegraphics[width=.45\textwidth]{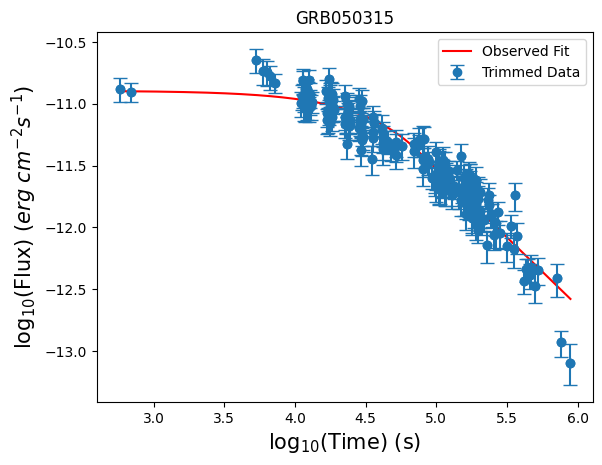}
\includegraphics[width=.45\textwidth]{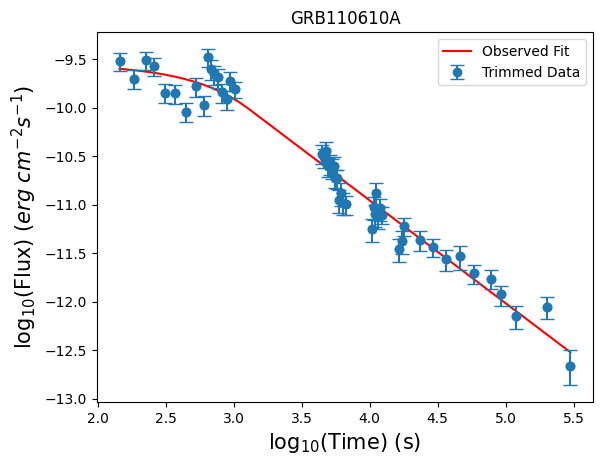}
\includegraphics[width=.45\textwidth]{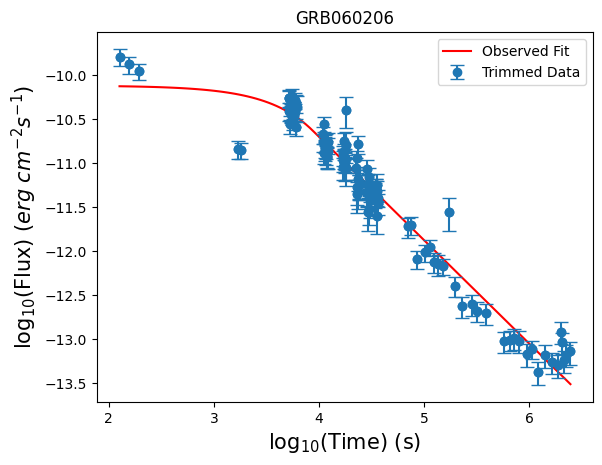}  
\end{center}
    \caption{The GRB LCs can be classified into four types based on the afterglow feature: i) Good GRBs (upper left); ii) GRB LCs that have a break in the terminal end (upper right); iii) Flares or Bumps in the afterglow (bottom left); iv) Flares or Bumps that have Double Break at the end of the LC (bottom right).References: [1] \citealt{2008ApJS..175..179S}; [2] \citealt{2018ApJ...866...97B}; [3] \citealt{2022ApJ...938...41D}.}
    \label{fig:categories}
\end{figure*}

The parameters for each GRB LC have been derived from the work of \citealt{2020ApJ...903...18S}. Fig. \ref{fig:categories} illustrates the LCs for the four identified classes and their corresponding W07 model fits. The blue points refer to the observed data that has been trimmed of the prompt emission.

%%\begin{figure*}[] 
%\begin{center}
%\includegraphics[width=.45\textwidth]{figures/GRB050822.png}
%\includegraphics[width=.42\textwidth]{figures/residual_GRB050822.png}  
%\end{center}
 %   \caption{The left image shows the LC of GRB050822 starting from the plateau emission, and the best-fit W07 model is displayed in red. The right plot shows the log(flux) residual histogram, and the best fit Gaussian distribution is displayed in black.}
  %  \label{fig:residual}
%\end{figure*}

\subsection{Dataset Preprocessing}
\label{sec:data-process}
We follow a similar preprocessing and training pipeline as described in \citealt{dainotti2023stochastic, manchanda2025gammarayburstlightcurve}, which ensures consistency across all models and facilitates a fair comparison. We used a random subset of 16 GRBs, four from each GRB category, to tune our model by minimizing the Mean Squared Error (MSE) loss. This tuning was conducted using optuna framework \citep{2019optuna} and the parameters obtained were used for the reconstruction of the entire 521 GRBs dataset. After obtaining the parameters, models, which required automatic hyperparameter tuning, are trained independently on individual GRB LCs, using the log$_{10}$-transformed time and flux values. The data is scaled to [0,1] using min-max normalization, described in Eq. \ref{eq: minmax-scaling}:

\begin{equation}
\label{eq: minmax-scaling}
    X_{i} = \frac{X_{i} - X_{min}}{X_{max} - X_{min}};  i \in D,
\end{equation}
where $X_{min}$ and $ X_{max}$ are the minimum and the maximum value in the training dataset $D$.

For QSS \& DGP Z-score normalization was performed, described as:
\begin{equation}
\label{eq: standard-scaling}
    X_{i} = \frac{X_{i} - X_{mean}}{X_{std}};  i \in D,
\end{equation}
where $X_{mean}$ and $X_{std}$ are the mean and standard deviation of the training dataset $D$.

After training the model and reconstructing the temporal gaps, we estimate the aleatoric uncertainty in the reconstructed flux values in a similar way as \citealt{dainotti2023stochastic, manchanda2025gammarayburstlightcurve}. DGP and QSS models do not require Monte Carlo (MC) simulations for uncertainty quantification, while for the BNN and TCN models, 100 MC simulations were done, while for the rest of the models, 1000 MC simulations were made. This captures the inherent data variability of the LCs. The 95\% confidence interval (CI) is computed using the 2.5th and 97.5th percentiles of the distribution of the reconstructed values at each time point, providing a reliable estimate of the aleatoric uncertainty. In CNN Bi-LSTM, Polynomial Curve Fitting, TCN, and Isotonic Regression models, uncertainty are computed from the observational errors of input data. Deep GP and QSS models' uncertainty is calculated from the residuals.

% The performance of the models is evaluated using the 5-fold Cross-Validation (CV) technique. For each GRB, we perform the 5-fold CV separately and average over the complete dataset of 521 GRBs to obtain the train and test mean squared error (MSE). 

\subsection{The Machine-Learning Approach}
This section provides a brief overview of various ML-based models for LC reconstruction.

\subsubsection{Deep Gaussian Model}
\begin{figure}[h]
  \centering
  \includegraphics[width=0.5\textwidth]{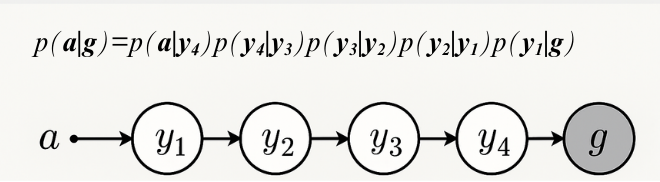} 
  \caption{Deep GP as a Markov chain}
  \label{fig:deepgp}
\end{figure}

This method layers the GP \citep{rasmussen2006gp} regression technique, enabling deep learning for probabilistic outputs tailored to regression tasks. A Deep Gaussian Process (DGP) \citep{damianou2013deep} extends the traditional GP framework by stacking multiple GP layers, where each layer models latent functions whose outputs serve as inputs to the next. Unlike a single-layer GP, which assumes a joint Gaussian distribution between inputs and outputs, a DGP introduces intermediate layers of latent variables—unobserved, inferred quantities that capture complex underlying structure—breaking the joint Gaussianity. This hierarchical setup enables the model to capture increasingly complex patterns in LC data by learning a distribution over the latent space at each layer. As shown in Fig.~\ref{fig:deepgp}, we can view the DGP as a form of Markov Chain, where each layer conditionally depends only on its immediate predecessor, thus simplifying the propagation of uncertainty through the network.

In our model, we use two DGP layers, each consisting of a scalable single-layer approximate GP. The model is implemented using \texttt{GPyTorch}, a Gaussian Process library built on top of PyTorch. The input is standardized before training to ensure stable optimization. For smooth and effective transitions, we use the Radial Basis Function (RBF) kernel with a length scale of 2.0. The RBF kernel, also known as the squared exponential kernel, measures similarity between inputs based on their Euclidean distance and assumes that points closer in the input space are more strongly correlated. Within a single-layer GP, we employ a GPyTorch-enabled Cholesky variational strategy to learn the optimal inducing point locations. Inducing points are a small set of representative inputs that help approximate the full GP, making it faster and more scalable. The Cholesky strategy improves stability and efficiency by factorizing the covariance matrix, allowing the model to handle larger datasets without sacrificing performance. We use 35 inducing points in the first layer and 25 for the second layer, selected through manual hyperparameter tuning on different GRB types. We tested several combinations on all types of GRBs and selected the configuration that minimized reconstruction error while maintaining computational efficiency. All evaluations were done under consistent training settings and preprocessing steps to ensure a generalization comparison. Each layer outputs a latent distribution whose mean is passed as input to the next layer, modeling a two-layered function:
\[
y = g(h(x)).
\]
This hierarchical formulation allows the model to capture complex, non-linear behaviors that might be missed by a shallow GP model.

To model the observed log flux values, we use a Gaussian likelihood:
\[
p(\mathbf{y} \mid \mathbf{f}) = \mathcal{N}(\mathbf{y} \mid \mathbf{f}, \sigma_n^2 \mathbf{I}),
\]
where $\mathbf{y}$ denotes the observed outputs, $\mathbf{f}$ represents the latent function outputs, $\sigma_n^2$ is the learnable noise variance, and $\mathbf{I}$ is the identity matrix, assuming Gaussian noise. The noise is initialized at 0.01 and constrained within the interval $[10^{-4}, 0.05]$ to prevent the model from overfitting or underfitting the noise levels across different GRB types.

% Training is performed over 500 iterations with an initial learning rate of 0.005. To stabilize optimization and prevent oscillations or overshooting, we apply a StepLR learning rate scheduler that reduces the learning rate at fixed intervals. For reliable performance estimation, we calculate the 5-fold CV MSE.

Training is performed over 500 iterations with an initial learning rate of 0.005. To stabilize optimization and prevent oscillations or overshooting, we apply a StepLR learning rate scheduler that reduces the learning rate at fixed intervals. 

This model configuration performs well across most GRBs. Notably, adding a third GP layer improves performance for 34 GRBs with a large number of data points, as deeper latent representations can be learned.

Compared to standard GPs, DGPs can model non-stationary, non-Gaussian, and multi-modal behavior more effectively. While GPs are limited in expressiveness, DGPs offer greater flexibility at the cost of increased computational complexity.

\subsubsection{Temporal Convolutional Neural Network}

\begin{figure}[h]
    \centering
    \includegraphics[width=0.5\textwidth]{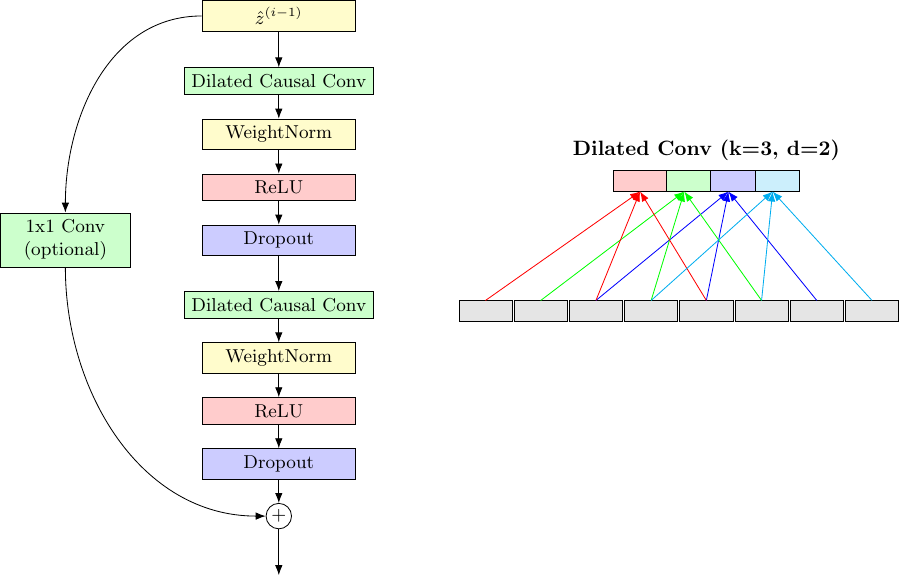} 
    \caption{a) Residual block of a Temporal Convolutional Network. (b) Expanded view of a dilated convolution layer}
    \label{fig:pdf-figure}
\end{figure}

The Temporal Convolutional Network (TCN) \citep{Bai2018AnEE} is a convolutional neural network architecture designed for sequence modeling tasks. TCN uses causal convolutions, where the output at time step t is computed exclusively from elements at time t and earlier, preventing information leakage from future time steps and enhancing convolutions through dilation, where filters (learnable weights that detect patterns) are applied with regularly spaced gaps across the input sequence. Residual connections \citep{he2016deep} involve adding the input of a layer directly to its output, skipping one or more intermediate layers as shown in Fig.~\ref{fig:pdf-figure}(a). The expanded view of the dilation convolution layer is illustrated in Fig.~\ref{fig:pdf-figure}(b). For a 1D input sequence \( \mathbf{x} \in \mathbb{R}^n \) and a filter \( f : \{0, \dots, k-1\} \rightarrow \mathbb{R} \), the dilated convolution \( F \) at position \( s \) is given by:

\begin{equation}
   F(s) = (\mathbf{x} *_d f)(s) = \sum_{i=0}^{k-1} f(i) \cdot \mathbf{x}_{s - d \cdot i}. 
\end{equation}

This technique allows the model to learn identity mappings, which helps prevent the vanishing gradient problem and facilitates more stable gradient flow during backpropagation. The TCN uses residual connections to improve training stability and ensure efficient convergence.

Unlike RNNs, which process sequences step by step, TCNs operate over the entire sequence in parallel. This parallelism improves computational efficiency and reduces training time. In terms of resource efficiency, TCNs generally require less memory during training compared to gated RNNs such as LSTMs \citep{Hochreiter1997LongSM} and Gate Recurrent Unit, GRUs, \citep{Cho2014LearningPR}. This is because TCNs share convolutional filters across time steps and do not need to store internal states or gate activations. 

In this study, the TCN was implemented using the \texttt{keras-tcn} library in combination with Keras' Sequential. The network was created with 64 filters, a kernel size of 5, a dropout rate of 0.1, a single stack of residual blocks, and dilation factors set to $(1, 2, 4, 8, 16, 32)$. The TCN output was passed to a Dense layer with a linear activation function of shape $(1, 1, 1)$ where the first, second and third dimensions represent batch size, the time stamps, and the input, respectively.
Hyperparameter optimization was performed through grid search on a randomly selected subset of GRB data, using the following ranges:
\begin{itemize}
    \item \text{nb\_filters\_range = [16, 32, 64, 128]}
    \item \text{kernel\_size\_range = [1, 3, 5]}
    \item \text{dropout\_range = [0.1, 0.5, 0.7]}
\end{itemize}

The ideal set of values for all GRBs was determined by combining the values that performed the best hyperparameters on this subset. Using the Adam optimizer with a batch size of 8 and MSE as the loss function, the model was trained across 100 epochs. Both input features and target values were normalized using MinMaxScaler to ensure consistent scaling. 
% To evaluate performance, 5-fold CV was employed.

\subsubsection{Convolutional Neural Network - Bidirectional Long Short Term Memory}

Convolutional Neural Networks (CNNs) \citep{lecun1998gradient, zhang2019cnn} and Bidirectional Long Short-Term Memory (Bi-LSTM) networks have individually demonstrated strong performance in extracting spatial and temporal features, respectively, making their hybrid combination particularly powerful for sequential pattern learning. CNN-BiLSTM models leverage this synergy by integrating convolutional feature extraction with temporal sequence modeling, enabling superior performance in tasks involving complex and noisy time-series data \citep{wang2017time, yao2017deepsense}.

In the CNN-BiLSTM architecture, the input LC is initially passed through one or more convolutional layers that act as localized feature detectors. These layers apply multiple filters to capture short-range dependencies and local patterns such as flares, peaks, or breaks in GRB LCs. Filters are like pattern detectors that slide over the input data and look for patterns in a local region of the input. This local processing is effective in denoising the signal and emphasizing key morphological structures relevant for downstream reconstruction \citep{kiranyaz20211dcnn}. A typical convolution operation in a 1D CNN can be defined as:
\begin{equation}
    y_i = \sum_{j=0}^{k-1} w_j \cdot x_{i+j},
\end{equation}
where \(w_j\) are the filter weights, \(x\) is the input sequence, \(k\) is the kernel size, and \(i\) indexes the position in the output feature map.

The compressed feature maps from the CNN are then fed into a Bi-LSTM block that processes the sequence in both forward and backward directions. This dual perspective allows the model to capture long-range dependencies while incorporating context from both past and future time steps \citep{graves2005bidirectional}. For GRB LCs, where information from both earlier and later intervals can improve reconstruction accuracy, especially in the presence of complex events like multiple flares, this bidirectional processing is critical.

The Bi-LSTM component consists of three stacked layers, each with 100 hidden units. Within each LSTM cell, an input gate (\(i_t\)), forget gate (\(f_t\)), and output gate (\(o_t\)) regulate information flow:
\begin{align}
  i_t &= \sigma\bigl(W_i[x_t,\,h_{t-1}]+b_i\bigr),\\
  f_t &= \sigma\bigl(W_f[x_t,\,h_{t-1}]+b_f\bigr),\\
  o_t &= \sigma\bigl(W_o[x_t,\,h_{t-1}]+b_o\bigr).
\end{align}
These gates control what information is written to, retained in, and released from the cell state, enabling stable learning of long-term dependencies.

Each Bi-LSTM layer returns a full sequence of hidden states in both directions, which are concatenated at each time step:
\begin{equation}
  h_t^{\text{Bi-LSTM}} = \bigl[h_t^{\rightarrow};\;h_t^{\leftarrow}\bigr].
\end{equation}

In our implementation, we use:
\begin{itemize}
  \item A 1D convolutional layer with 64 filters, kernel size \(k=3\), ReLU activation, He-normal initialization, and \(L_2\)-regularization.
  \item A max-pooling layer with pool size 1 to preserve sequence length.
  \item Three stacked Bi-LSTM layers (100 units each, "100 units" refers to the number of hidden units (or neurons) within that specific LSTM layer.), utilizing ReLU activation for the first layer and Swish for the subsequent two. =The ReLU activation introduces non-linearity while being computationally efficient, enabling the model to capture temporal patterns without suffering from vanishing gradient issues. The ReLU function outputs zero for negative inputs and returns the input value for positive ones. In contrast, the Swish activation, defined as $f(x) = x \cdot sigmoid(x)$, often performs better in deeper networks because it is smoother and permits small negative activations instead of truncating them completely. The \textit{sigmoid} component squashes its output values to the range (0, 1), thereby modulating the gradient flow more smoothly during training. 
   \item A final dense layer with a single unit to reconstruct the flux value.
\end{itemize}

We use the Adam optimizer with a learning rate of 0.0001 for training. This hyperparameter was determined during the hyperparameter tuning phase as mentioned in section \ref{sec:data-process}.

% and model generalization was evaluated via 5-fold CV with model re-initialization per fold to avoid information leakage.

Fig.~\ref{fig:cnn_bilstm_workflow} illustrates the end-to-end workflow of our CNN-BiLSTM architecture.

\begin{figure}[htbp]
  \centering
  \resizebox{\linewidth}{!}{%
    \begin{tikzpicture}[
        node distance=0.5cm and 0.5cm,
        block/.style={
          rectangle, rounded corners, draw=black, thick,
          minimum width=2.0cm, minimum height=1.5cm, align=center, fill=gray!10
        },
        arrow/.style={-stealth, thick}
      ]
      \node[block] (in)    {Input Sequence\\$(T \times F)$};
      \node[block, right=of in]  (conv)  {Conv1D\\64 filters, \(k=3\)};
      \node[block, right=of conv] (pool)  {MaxPool1D\\size=1};
      \node[block, right=of pool] (bilstm) {3\(\times\) Bi-LSTM\\100 units each};
      \node[block, right=of bilstm](out)   {Dense Output\\1 unit};
      \draw[arrow] (in)   -- (conv);
      \draw[arrow] (conv) -- (pool);
      \draw[arrow] (pool) -- (bilstm);
      \draw[arrow] (bilstm) -- (out);
    \end{tikzpicture}%
  }
  \caption{Workflow of the CNN-BiLSTM model for GRB light-curve reconstruction.}
  \label{fig:cnn_bilstm_workflow}
\end{figure}

\subsubsection{Bayesian Neural Network}

Bayesian neural networks (BNNs) are made up of a neural network and a stochastic model. The latter introduces either a stochastic activation or a stochastic weight. It is possible to simulate multiple models of parameters $\theta$ with an associated probability distribution $p(\theta)$. Comparing their predictions shows uncertainty: agreement indicates low uncertainty; disagreement indicates high uncertainty. Training a BNN involves computing a posterior distribution over the weights given data, denoted as $p(\theta \mid D)$, where $\theta$ represents the weights and $D$ is the training dataset. Exact inference is intractable, so variational inference is used. By minimizing the Kullback–Leibler divergence, a variable distribution $q(\theta)$ chosen from a family of distributions approximates the posterior $p(\theta \mid D)$:

\begin{equation}
    \text{KL}\bigl(q(\theta) \parallel p(\theta \mid D)\bigr).
\end{equation}

\begin{figure}
    \centering
    \includegraphics[width=0.8\linewidth]{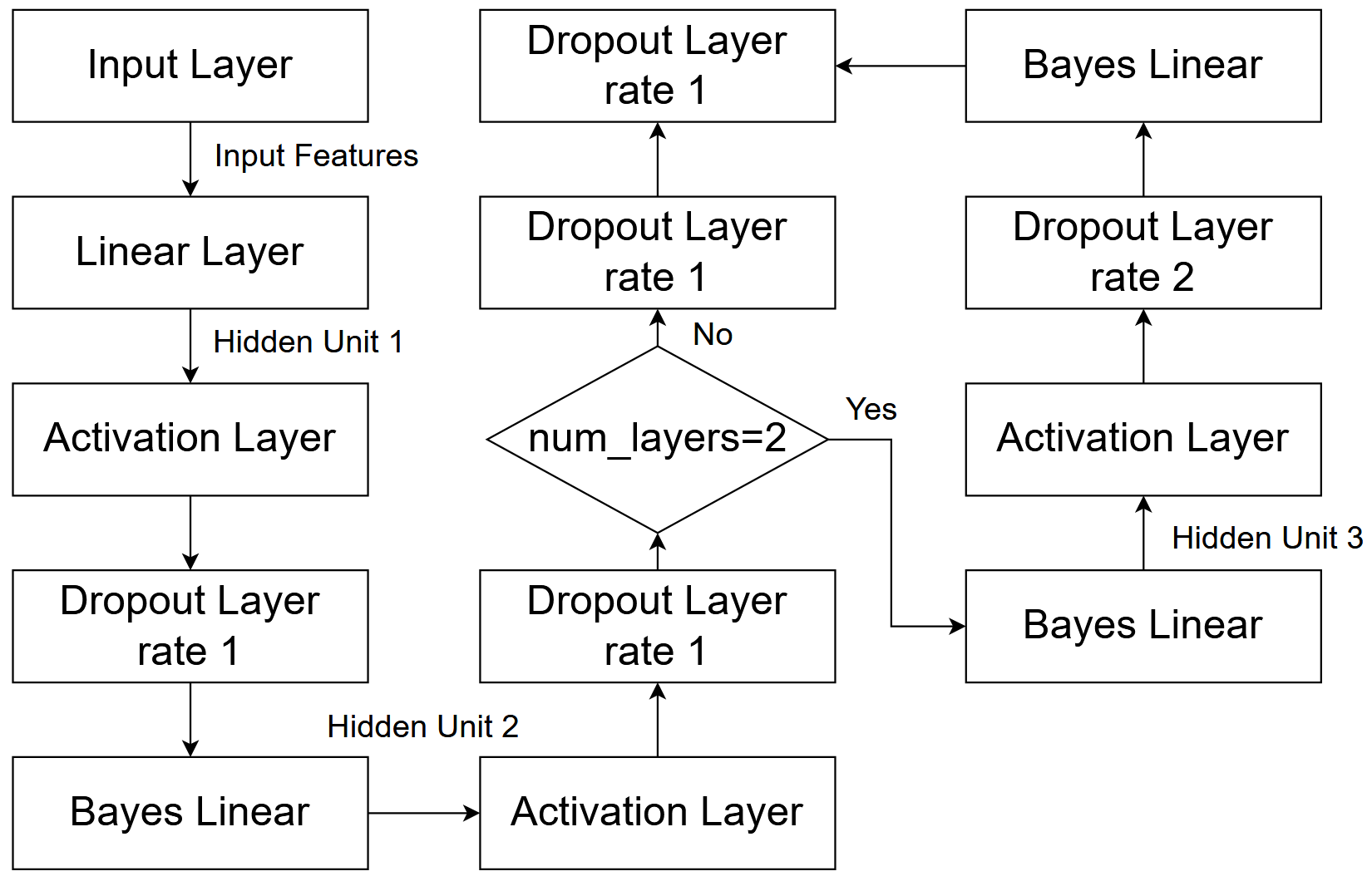}
    \caption{BNN architecture for GRB light-curve reconstruction}
    \label{fig:BNN_architecture}
\end{figure}

This converts the inference problem into an optimization one, often using reparameterization tricks to enable backpropagation through stochastic variables. Dropout-based approximations, such as Monte Carlo Dropout, implement a form of variational inference where dropout is applied at test time, sampling from an implicit posterior over the weights. These samples generate predictive distributions from which uncertainty can be estimated. A detailed architecture is illustrated in Figure \ref{fig:BNN_architecture}.
%BNNs yield two types of uncertainty:
%begin{itemize}
 % \item \textbf{Epistemic uncertainty}, arising from limited data and reducible with more observations. This is captured via the posterior over weights.
  %\item \textbf{Aleatoric uncertainty}, arising from inherent noise in the data and irreducible. This can be modeled by learning a distribution over outputs.
%\end{itemize}

Prediction involves marginalizing over the posterior:

\begin{equation}
    p(y \mid x, D) = \int p(y \mid x, \theta)\, p(\theta \mid D)\, d\theta,
\end{equation}

for an input vector $x$ and predicted output $y$.

Since this integral is intractable, it is approximated via Monte Carlo sampling for $T$ samples:

\begin{equation}
    p(y \mid x, D) \approx \frac{1}{T} \sum_{t=1}^{T} p(y \mid x, \theta_t), \quad \theta_t \sim q(\theta).
\end{equation}

This ensemble of predictions allows the computation of predictive mean and variance.

The model is implemented using PyTorch and \texttt{torchbnn}, structured with hidden layers (\texttt{num\_layers}). The first layer is a linear layer for mapping the input dimension to the subsequent hidden layers. Each layer afterwards is a \texttt{BNNHiddenLayer}, which wraps a \texttt{BNNLinear} layer with trainable Gaussian-distributed weights and biases. The hidden layers are defined by an activation function (Leaky ReLU, Tanh, or Swish) and include dropout for regularization. The output consists of a single Bayesian linear layer with two output features, which are subsequently interpreted as the predictive mean and the log-variance. Hyperparameters such as the number of layers, hidden units, dropout rate, and activation function are optimized using Optuna, see Table \ref{tab:optuna}. The values of \texttt{PRIOR\_MU} and \texttt{PRIOR\_SIGMA} are set to 0.0 and 0.1, respectively. The training process involves forward propagation with stochastic weight sampling, minimizing a composite loss function consisting of:
\begin{itemize}
  \item The negative log-likelihood of the Gaussian output.
  \item A KL-divergence term that regularizes the learned weights by penalizing deviations from a standard normal prior.
  \item Weighted MSE loss for aligning the predicted value ($\hat{y}$) with true value ($y$) according to the weights ($w_i$) assigned for each predicted class ($i$).
  \item Additionally, a log-variance penalty term is added to prevent the predicted variances from becoming excessively large or unstable during training.
\end{itemize}

\begin{equation}
\mathcal{L'}(\theta) =
\underbrace{- \left[ \log p(D \mid \theta) \right]}_{\text{Negative Log-Likelihood}}
+ \underbrace{\mathrm{KL}\left(q(\theta) \,\|\, p(\theta)\right)}_{\text{KL Divergence}} + \underbrace{\frac{\sum_{i = 0}^{n}{w_i  (y-\hat{y})^{2}}}{\sum_{i = 0}^{n} w_i}}_{\text{Weighted MSE Loss}}
\end{equation}

\begin{equation}
\mathcal{L}(\theta) = \mathcal{L'}(\theta) + \underbrace{\lambda_{\text{log-var}}\sum_{i=1}^{N}{(\log{\sigma^2}})^2}_{\text{Log-Variance Penalty}}.
\end{equation}

The KL-divergence term is scaled by a fixed multiplicative factor \( \lambda_{\text{KL}} = 10^{-3} \) to balance model fit and uncertainty.
 After hyperparameter optimization, the model is retrained on the full LC of every GRB separately. For prediction, multiple stochastic forward passes are conducted to generate a distribution of outputs, from which prediction intervals and aleatoric uncertainty estimates are derived.

\begin{table}[h]
\centering
\begin{tabular}{|l|c|}
\hline
\textbf{Hyperparameter} & \textbf{Optuna Values/Range} \\
\hline
Number of layers & (1, 2, 3) \\
Hidden Unit 1    & (32, 64, 128) \\
Hidden Unit 2    & (32, 64, 128) \\
Hidden Unit 3    & (16, 32, 64) \\
Dropout Rate     & (0.1, 0.2, 0.3,  0.4) \\
\hline
\end{tabular}
\caption{Hyperparameter ranges used in Optuna optimization}
\label{tab:optuna}
\end{table}

\subsection{The statistical approach}
This section provides a brief overview of various statistical-based models for LC reconstruction.

\subsubsection{Polynomial Curve Fitting}
In previous sections, we analyzed with various sophisticated approaches. However, now, we aim to simplify the curve-fitting process. Our approach involves fitting an n-degree polynomial to the LC, starting with an initial fine-tuning phase. Similar polynomial regression techniques have been employed in previous studies, as noted by \citep{10.5120/ijca2024923400}. An n-degree polynomial can be defined as:

\begin{equation}
    \label{eq:general-curvefit}
    y = c_1 + c_2 x + c_3x^2 ... c_nx^{n-1}.
\end{equation}

We deliberately use a general polynomial form without imposing any specific curve function, allowing the model to capture the underlying trend in GRBs without bias toward a predefined shape. Importantly, the choice of a general curve does not contradict or attempt to draw definitive conclusions about the physical interpretations or trends established in GRB-related literature.

The objective is to select the degree $n$ such that the model achieves an optimal balance between flexibility and generalization. To do this, we apply \textit{MinMaxScaling} given by the Eq. \ref{eq: minmax-scaling} and then randomly sample four instances from each GRB class and determine the polynomial degree that minimizes MSE loss while maintaining model simplicity. In our analysis, a second-degree polynomial yielded the best performance. Apart from choosing the polynomial degree, we also fine-tune the initial parameters $c1$, $c2$, and $c3$, which serve as the starting point for optimization. The resulting polynomial equation takes the form:

\begin{equation}
    \label{eq:finetuned-curvefit}
    y = -0.35 + 0.80 x - 0.026x^2.
\end{equation}

This polynomial is then fitted to the data using the \textit{$scipy.optimize.curve\_fit$} function, which estimates the optimal coefficients $\{c1,c2,c3\}$ for each GRB.

\begin{figure}[h!]
    \begin{center}
        \includegraphics[width=.45\textwidth]{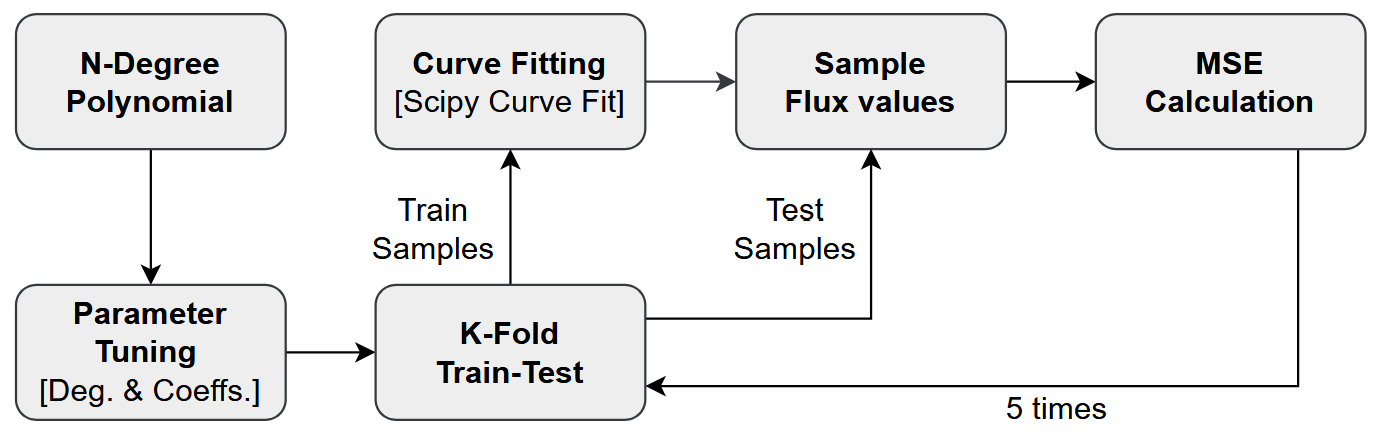}
    \caption{Workflow of the Polynomial Curve Fit model for GRB light-curve reconstruction.}
    \label{curvefit:workflow}
    \end{center}
\end{figure}

%We evaluate the model’s performance using 5-fold CV based on the MSE loss. 
Fig. \ref{curvefit:workflow} presents an overview of the entire process.

\subsubsection{Isotonic regression}

\begin{figure}[h!]
    \begin{center}
        \includegraphics[width=.45\textwidth]{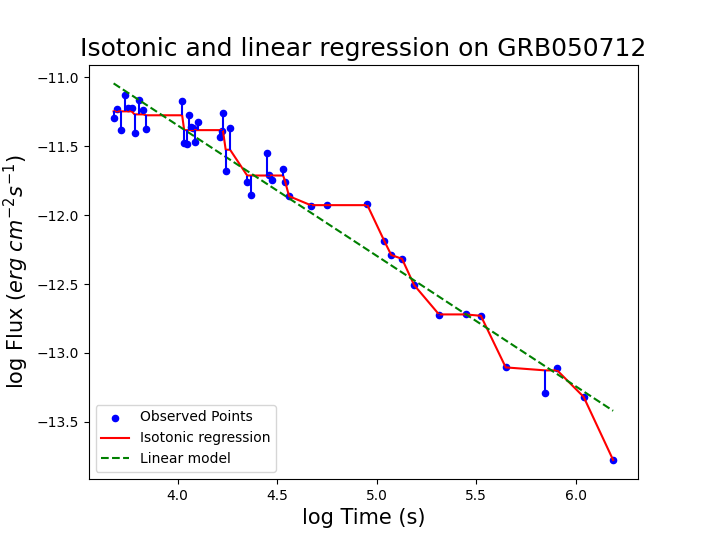}
    \caption{Weighted least-square method minimizes the residuals, which are marked with blue lines}
    \label{fig:Iso}
    \end{center}
\end{figure}

This approach uses a statistical, non-parametric method to estimate monotonic (either decreasing or increasing) one-dimensional data. Unlike classical regression, which assumes a linear functional form, the non-parametric approach does not presume the structure. Therefore, this method is more flexible than standard regression and fits the data more accurately, as it is displayed in Fig. \ref{fig:Iso}.

Given that the LCs generally show a decreasing trend over time, in our implementation, a non-increasing isotonic regression model is applied \citep{Best1990}.

Let $(x_1, y_1), \ldots, (x_n, y_n)$ be observations such that\newline  $x_1 < x_2 < \ldots < x_n$, and for any $i > j$, it holds that $y_j \geq y_i$. With such an assumption, the isotonic regression seeks a non-increasing function $f: X \rightarrow \mathbb{R}$ that minimizes the weighted least squares criterion:
\begin{equation}
    \min_{f}\sum^n_{j=1}w_j(y_j - f(y_j))^2,
    \label{eq: iso1}
\end{equation}

where $w_j \geq 0$ is the added weight to the observation.

Typically, weights are set to 1, but in our implementation, we assign a weight of 0.1 for the outliers to reduce their influence on the fitted curve.

To find the most optimal function $f$, the pool adjacent violators (PAV) algorithm is used. This procedure is implemented in the Python $\texttt{sklearn}$ package in $\texttt{IsotonicRegression}$ in the fitting method \citep{deLeeuw2009}.
Let us present this algorithm briefly.\newline Let $\hat{y}_1, \hat{y}_2, \ldots, \hat{y}_n$ denote the fitted values obtained from isotonic regression corresponding to $x_1, \ldots x_n$, respectively. Consequently, the function defined by the Eq. \ref{eq: iso1} is expressed as:

\begin{equation}
f(x_i) = \hat{y}(x_i) =\hat{y}_i. 
\end{equation}

In our case, we perform non-increasing isotonic regression, which imposes the condition:

\begin{equation}
\hat{y}_i \geq \hat{y}_{i+1} \quad \text{for all } i = 1, \ldots, n-1.
\end{equation}

The algorithm proceeds as follows:
\begin{enumerate}
    \item \textbf{Initialization:}  
    Set the initial fitted values equal to the observed values: 
    \begin{equation}
    \hat{y}_i = y_i \quad \text{for all } i = 1, \ldots, n.
    \end{equation}
    
    \item \textbf{Check for monotonicity:}  
    For each pair $(\hat{y}_i, \hat{y}_{i+1})$, verify whether the non-increasing condition holds:
    \begin{equation}
    \hat{y}_{i+1} \leq \hat{y}_i.
    \end{equation}
    
    \item \textbf{Merge in case of violation:}  
    If the condition is violated (i.e., $\hat{y}_{i+1} > \hat{y}_i$), merge the two values using a weighted average:

    \begin{equation}
    \hat{y}_{\text{merged}} = \frac{w_i \cdot \hat{y}_i + w_{i+1} \cdot \hat{y}_{i+1}}{w_i + w_{i+1}}.
    \end{equation}
    
    Replace both $\hat{y}_i$ and $\hat{y}_{i+1}$ with $\hat{y}_{\text{merged}}$, therefore $\hat{y}_i= \hat{y}_{i+1} =\hat{y}_{\text{merged}}$ and treat them as a single block with combined weight $w_i + w_{i+1}$ (i.e., validate the next steps of the algorithm with $\hat{y}_{\text{merged}}$ and the updated weight.) 

    \item \textbf{Iterative verification of merged values:}  
    After merging, verify whether the updated values violate the non-increasing condition with the preceding values. If a violation occurs, continue merging recursively to the left.

    \item \textbf{Repeat until no violations remain}. 
\end{enumerate}

Therefore, the sequence $\hat{y}_1, \ldots, \hat{y}_n$ is the fitted values resulting from isotonic regression, such that they satisfy the monotonicity $\hat{y}_1 \geq \hat{y}_2 \geq \ldots \geq \hat{y}_n$.

After training the model, the procedure of predicting a new value $x$ is given by:

\begin{equation}
    \hat{y}(x) = \begin{cases} \hat{y}_1 & \text{if }x \leq x_1, \\ 
\hat{y}_i + \frac{x - x_i}{x_{i+1} - x_i}(\hat{y}_i - \hat{y}_{i+1}) & \text{if } x_i\leq  x \leq x_{i+1}, \\
\hat{y}_n & \text{if }x \geq x_n,
\end{cases}
\end{equation}

where $x_1, x_2, \ldots, x_n$ are the data points used to fit the isotonic regression, satisfying the strictly increasing order \newline $x_1 < x_2 < \ldots < x_n$.

\subsubsection{Quartic Smoothing Spline Model}

A non-parametric framework called the Quartic Smoothing Spline (QSS) was created to recover the intrinsic temporal structure of GRB LCs from observational data that is noisy and irregularly sampled. A piecewise fourth-degree polynomial basis, which offers a potent blend of high-order smoothness and local flexibility, defines the model's architecture. Instead of minimizing a single penalized functional, the model finds the optimal representation of the LC, $S(x)$, by solving a constrained optimization problem. It minimizes the sum of squared residuals to the data, subject to a smoothness constraint based on the spline's derivatives \citep{1993csfw.book.....D}:
\begin{equation}
  \min_{S} \sum_{i=1}^{N}\bigl[y_i - S(x_i)\bigr]^2 \quad \text{subject to} \quad \sum_{j} \left( \Delta^{(5)}S(\tau_j) \right)^2 \le s,
\end{equation}
where $(x_i,y_i)$ represent the observed log-time and log-flux data pairs. The novelty of this framework lies in its definition of smoothness: for a quartic spline ($k=4$), the constraint is placed on the sum of the squared discontinuity jumps of the \textbf{fifth derivative} ($\Delta^{(5)}S$) at each interior knot $\tau_j$. This provides a powerful modeling dynamic where the spline's local flexibility is regularized by a high-order smoothness condition, a trade-off well-suited for the complex morphology of GRBs.

% \subsubsection*{Model Architecture and Fitting Procedure}
The construction and implementation of the QSS model are governed by the following structural choices and procedures:
\begin{itemize}
    \item \textbf{Basis Representation:} The fundamental building block of the model is a basis of degree-four ($k=4$) B-spline functions \citep{deboor2001practical}. The final spline is a linear combination of these basis functions:
    \begin{equation}
      S(x) \;=\; \sum_{j=1}^{n} c_j\,B_j(x),
    \end{equation}
    where $\{B_j(x)\}$ are the quartic B-spline basis functions and the coefficients $\{c_j\}$ are determined via the constrained optimization. This architectural choice endows the model with its characteristic properties.

    \item \textbf{Inherent Structural Properties:} The degree-four basis offers two significant features. The first is enhanced local flexibility, which allows the model to accurately represent complex morphological features such as sharp, asymmetric peaks and flat plateaus. Because the resulting curve is three times continuously differentiable ($C^3$), the second is higher-order smoothness, which ensures that the rate of change of curvature is continuous and can represent more physically plausible transitions in the GRB emission process.

    \item \textbf{Smoothing and Regularization:} The smoothing factor, $s$, serves as the upper bound for the smoothness constraint in the optimization problem. With $N$ representing the number of data points, we set $s = N$. For astrophysical time-series data, this choice offers an empirical balance between bias and variance, making it a robust method for regularization \citep{hastie2009elements}.
    
    \item \textbf{Implementation and Evaluation:} The model is implemented using the \texttt{scipy.interpolate.UnivariateSpline} routine. The solver places knots at the data locations and efficiently computes the coefficient vector $\mathbf{c}$. The final, continuous representation $S(x)$ is then evaluated on a dense grid to produce the high-resolution reconstructed LC.
\end{itemize}

% \subsubsection*{Noise Modeling and Uncertainty Estimation}
To provide a complete probabilistic reconstruction, we model the uncertainty inherent in the observations. The residuals between the fitted spline and the data, $r_i = y_i - S(x_i)$, are calculated to estimate a residual standard deviation, $\sigma_{\rm resid}$:
\[
  \sigma_{\rm resid} \;=\; \sqrt{\frac{\sum_{i=1}^{N}r_i^2}{N}}.
\]
To generate realistic synthetic data points, we first characterize the original observational log-flux errors by fitting both Normal and Laplace distributions and selecting the one with the highest likelihood. Noise is then sampled from this best-fit distribution to be added to the mean spline prediction. For uncertainty visualization, 95\% confidence bands are constructed around the mean spline fit as $S(x) \;\pm\; 1.96\,\sigma_{\rm resid}$, which are then transformed from log-space back to linear flux units for physical interpretation \citep{eilers1996flexible}. The model's generalization performance is validated using 5-fold cross-validation. The entire workflow is summarized in Fig. \ref{fig:quartic_spline_flowchart}.

\begin{figure}[htbp]
\centering
\begin{tikzpicture}[
    node distance=.5cm and .5cm,
    block/.style={
      rectangle, rounded corners, draw=black, thick,
      minimum width=.5cm, minimum height=.5cm, align=center, fill=gray!10
    },
    arrow/.style={-stealth, thick}
  ]
  % Defining nodes
  \node[block] (load) {Load Time \& Flux Data};
  \node[block, below=of load] (transform) {Log–Log Transform};
  \node[block, below=of transform] (normalize) {Normalize to Zero Mean \\ \& Unit Variance};
  \node[block, below=of normalize] (fit) {Fit Quartic Spline \\ (\texttt{k=4}, \texttt{s=N})};
  \node[block, below=of fit] (evaluate) {Evaluate on Dense Grid};
  \node[block, below=of evaluate] (residuals) {Compute Residuals \& \\ $\sigma_{\rm resid}$};
  \node[block, below=of residuals] (noise) {Sample \& Add Noise};
  \node[block, below=of noise] (ci) {Construct 95\% CI};
  \node[block, below=of ci] (output) {Output Reconstructed Curve};

  % Drawing arrows
  \draw[arrow] (load) -- (transform);
  \draw[arrow] (transform) -- (normalize);
  \draw[arrow] (normalize) -- (fit);
  \draw[arrow] (fit) -- (evaluate);
  \draw[arrow] (evaluate) -- (residuals);
  \draw[arrow] (residuals) -- (noise);
  \draw[arrow] (noise) -- (ci);
  \draw[arrow] (ci) -- (output);
\end{tikzpicture}
\caption{Workflow of the quartic smoothing spline model for GRB light‐curve reconstruction, from raw data loading through spline fitting to uncertainty quantification.}
\label{fig:quartic_spline_flowchart}
\end{figure}

\subsection{Parameter Uncertainty and Evaluation Metrics}

In training our reconstruction model, we do not impose any analytical form on the GRB afterglow, such as the W07 profile or a broken power law. The machine learning framework is therefore inherently model-independent, allowing it to learn the temporal structure directly from the data rather than being constrained by a predetermined shape. Nevertheless, to determine whether the reconstruction leads to improved physical parameter estimates, a quantitative benchmark is required.

For this purpose, we use the W07 as a reference and take its fitted parameters as the baseline for comparison. Following \citealt{2020ApJ...903...18S}, the uncertainty associated with each parameter  $p = (\log T_{a}, \log F_{a}, \alpha)$, are defined as:

 \begin{equation}
 \Delta p = \frac{\max(p) - \min(p)}{2}.
  \label{p}
 \end{equation}

It is clear that with data augmentation, since the uncertainties, $\delta$ scales as $1/\sqrt{N}$ where N is the number of data points, in principle one can increase this number, and thus reduce the uncertainty. However, we have been very careful in data augmentation, because we aim to preserve the features of the initial LCs so that the increase in the number of data points should be proportional to the existing data points. Indeed, if we increase the number of data points without accounting for this recipe, systematic biases will result. Thus, we have worked out a strategy that balances data augmentation, preserves the original model we assume (in this case, the \citep{willingale2007testing} function) when the LCs are reconstructed, and still reduces the final uncertainties in the given model's parameters. For details about the method used, see Appendix B. 
This configuration may represent current data only if we assume the satellite we are considering does not undergo any orbital gap. So, data augmentation will solve the issue of gaps if perfect seeing will be available.
One can also consider the same situation if you allow future satellites that, when combined, provide full data coverage. This analysis was performed using the W07 model, a benchmark. Indeed, the same procedure can be repeated for any other model, as we have already done in \citep{dainotti2023stochastic}, in which a broken power-law was tested. The additional model testing or simulations of future satellites are beyond the scope of the current paper.

This definition captures the observational limitations inherent in GRB afterglow measurements, where incomplete LC coverage introduces non-negligible statistical uncertainty. In the absence of fully sampled LCs, this metric provides a consistent basis for evaluating whether our reconstruction effectively reduces parameter uncertainties relative to those obtained from the W07 model.

To quantify this improvement, we compute the error fraction (EF) for each parameter before and after reconstruction \citealt{dainotti2023stochastic}:

\begin{equation}
  EF_{\log_{10}(T_{a})}=\left|\frac{\Delta{\log_{10}(T_{a})}}{\log_{10}(T_{a})}\right|,
  \label{eqn5}
\end{equation}

\begin{equation}
  EF_{\log_{10}(F_{a})}=\left|\frac{\Delta{\log_{10}(F_{a})}}{\log_{10}(F_{a})}\right|, 
  \label{eqn6}
\end{equation}

\begin{equation}
 EF_{\alpha_{a}}=\left|\frac{\Delta{\alpha_{a}}}{\alpha_{a}}\right|.    
 \label{eqn7}
\end{equation}

We compute the reduction in the percentage of the error fractions in order to evaluate the enhancement in fit post-reconstruction.

\begin{equation}
 \%_{DEC}= \frac{\left|EF^{\rm{after}}_{X}\right|-\left|EF^{\rm{before}}_{X}\right|}{\left|EF^{\rm{before}}_{X}\right|}\times 100   
 \label{eqn8}.
\end{equation}

\section{Results}\label{section:results}

%Decreasing the uncertainty associated with the model parameters is an objective of LCR. To evaluate this, the error fractions, symbolized by $EF$, are calculated for every model parameter in both the primary datasets and after reconstructing them using the model. Eq. \ref{eqn5}, \ref{eqn6}, and \ref{eqn7} reflect the error fractions for the three parameters as stated in \citealt{dainotti2023stochastic}:

\begin{figure*}[htbp]
\begin{center}

    \includegraphics[width=.24\textwidth, height=.17\textwidth]{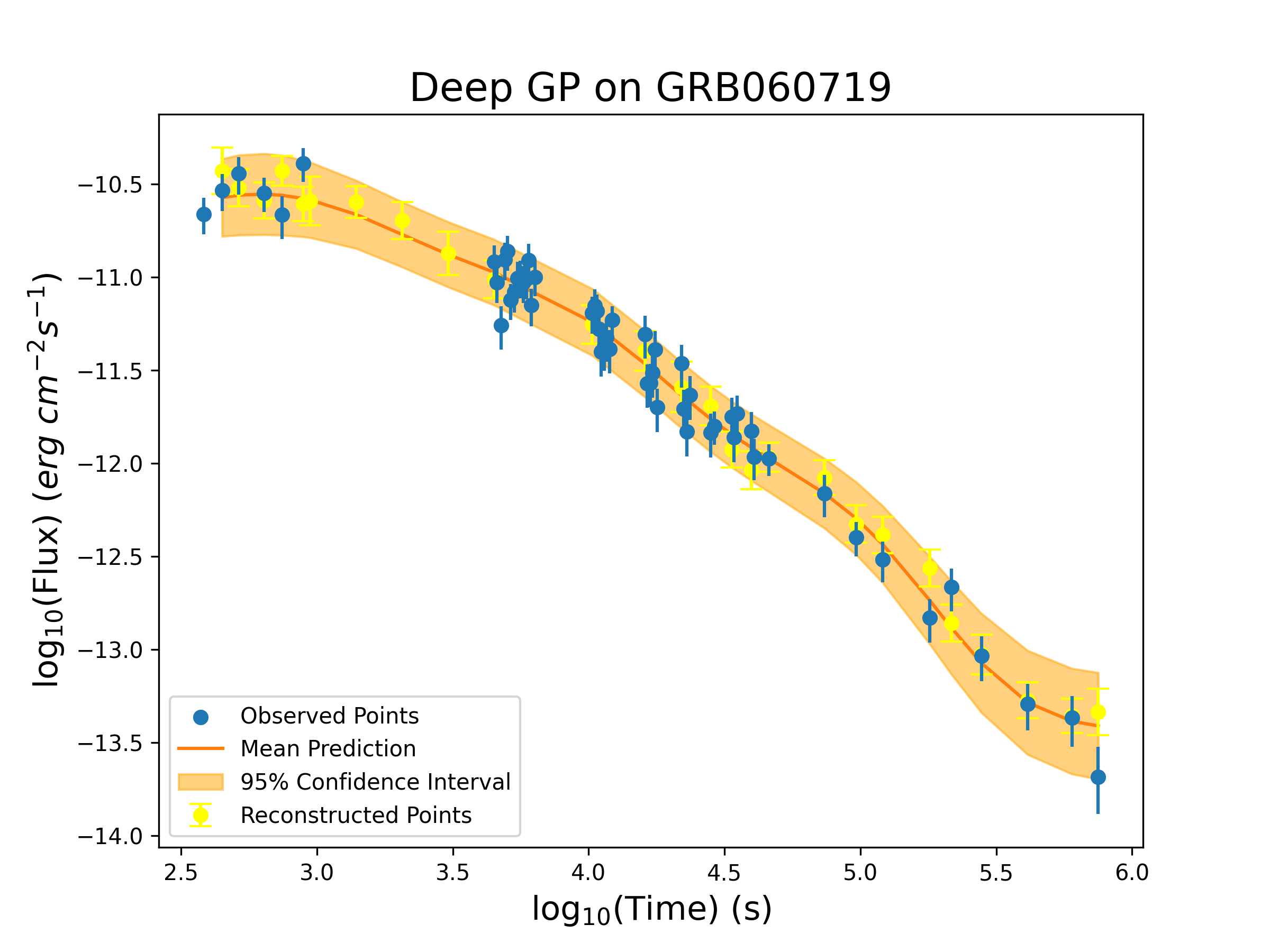}
    \includegraphics[width=.24\textwidth, height=.17\textwidth]{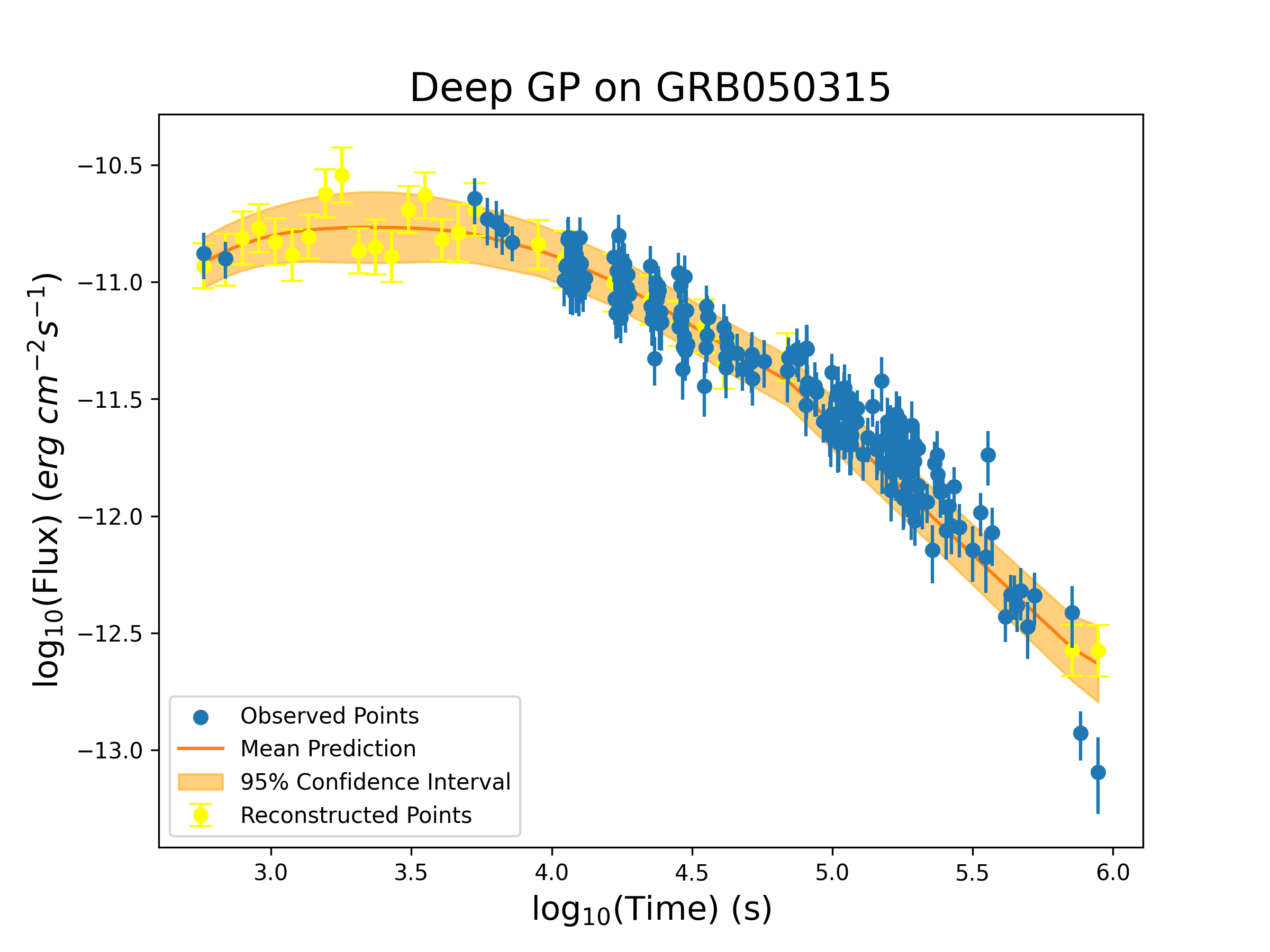}
    \includegraphics[width=.24\textwidth, height=.17\textwidth]{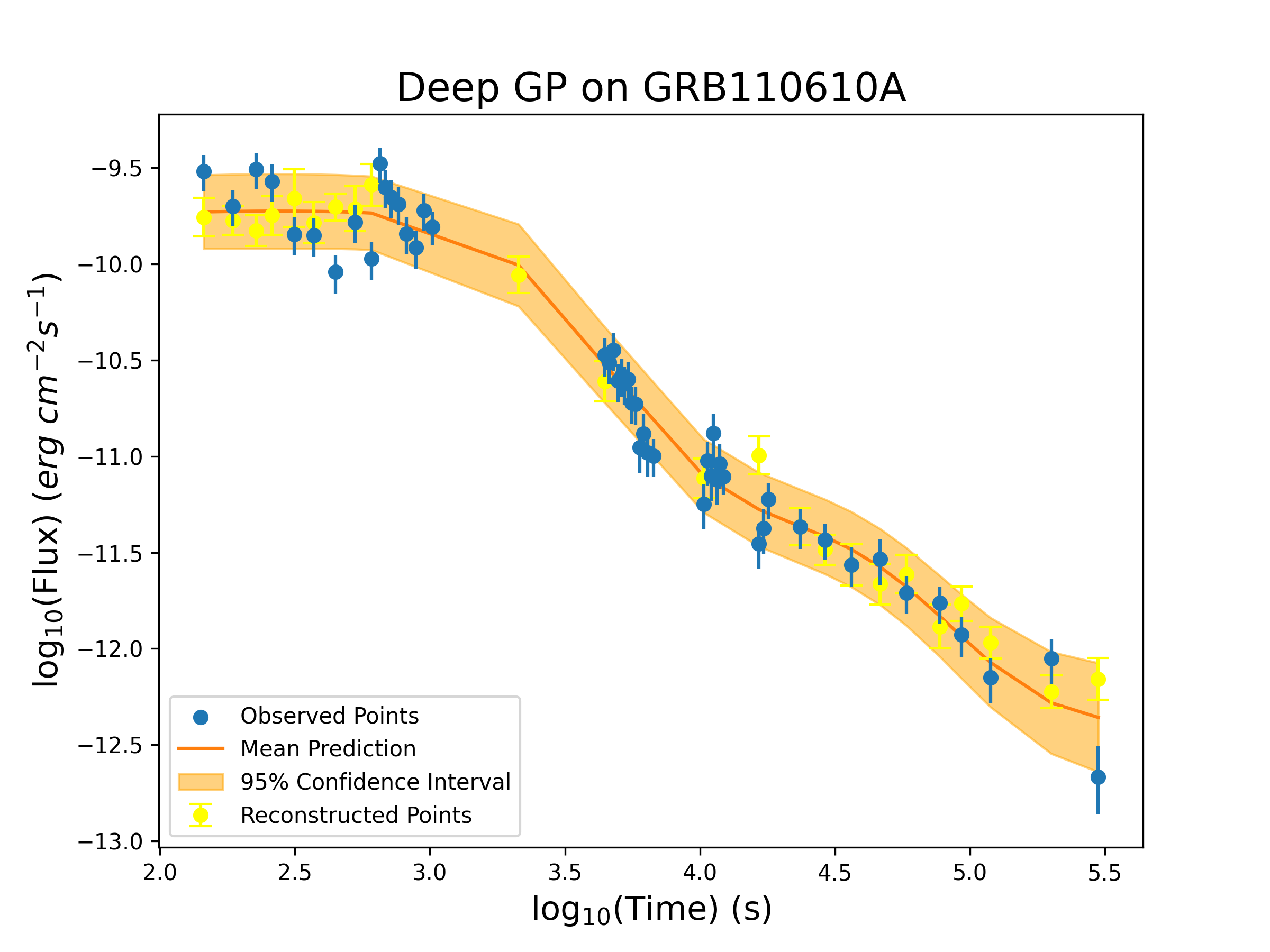}
    \includegraphics[width=.24\textwidth, height=.17\textwidth]{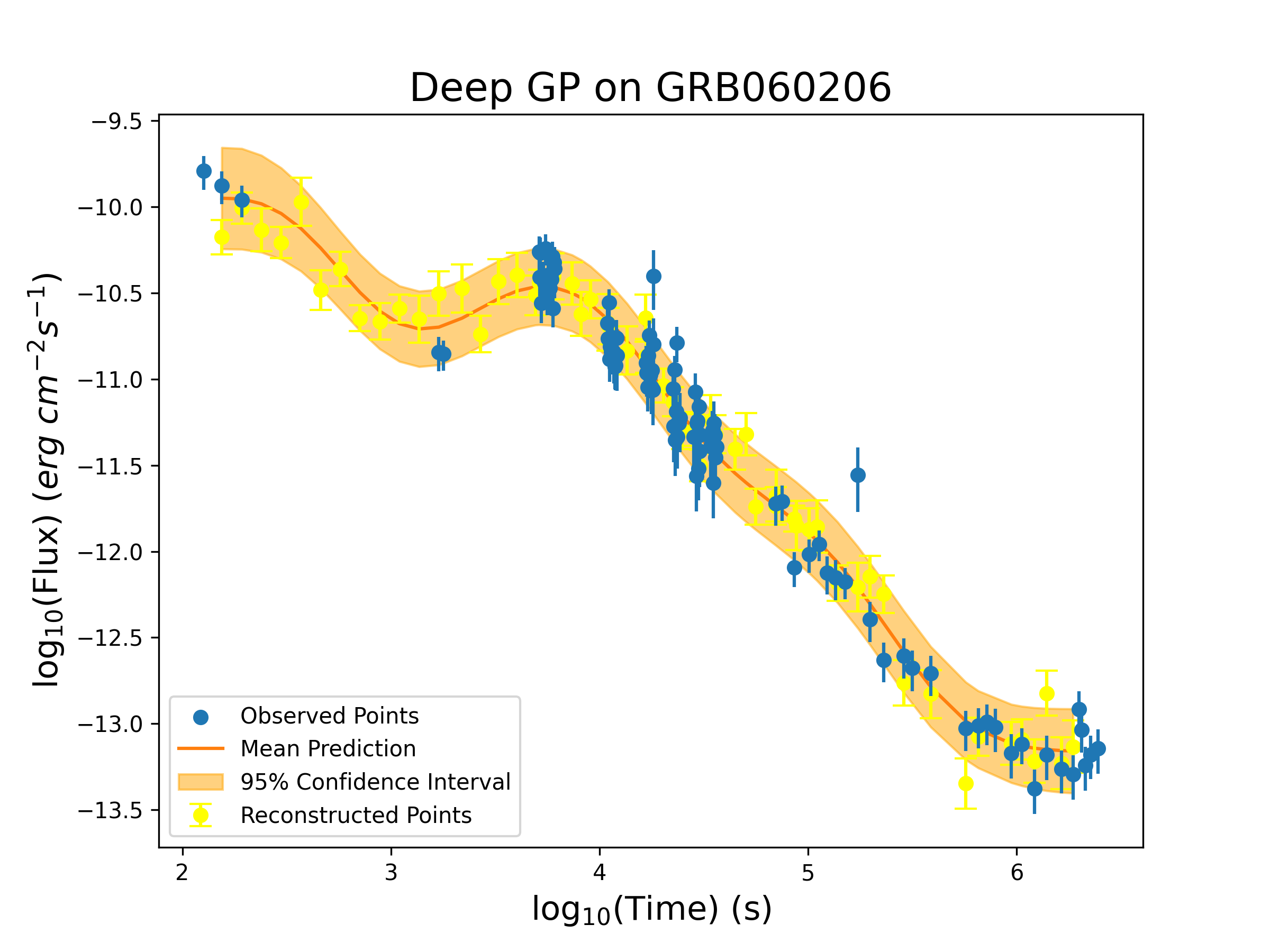}  

    \includegraphics[width=.24\textwidth, height=.17\textwidth]{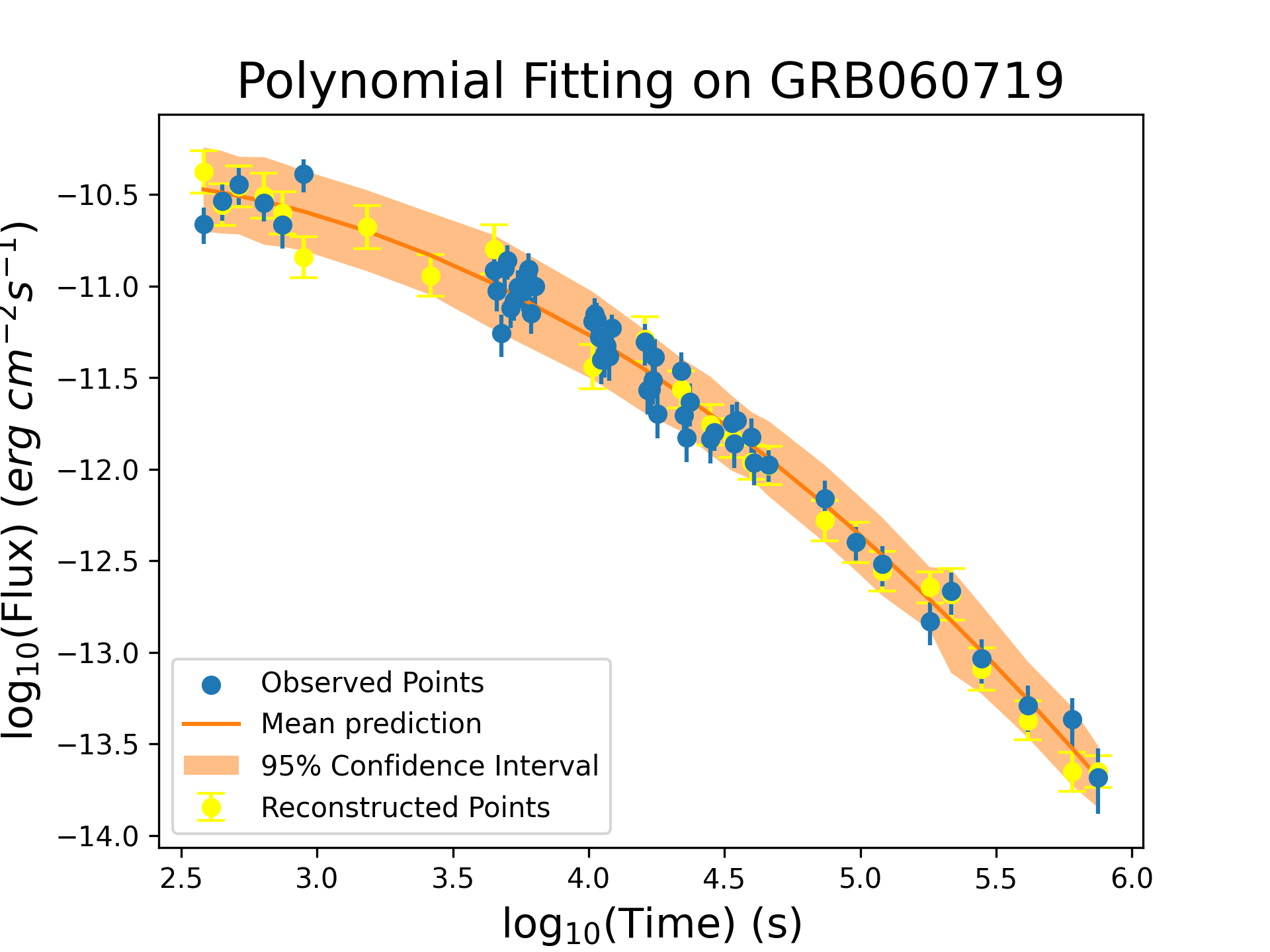}
    \includegraphics[width=.24\textwidth, height=.17\textwidth]{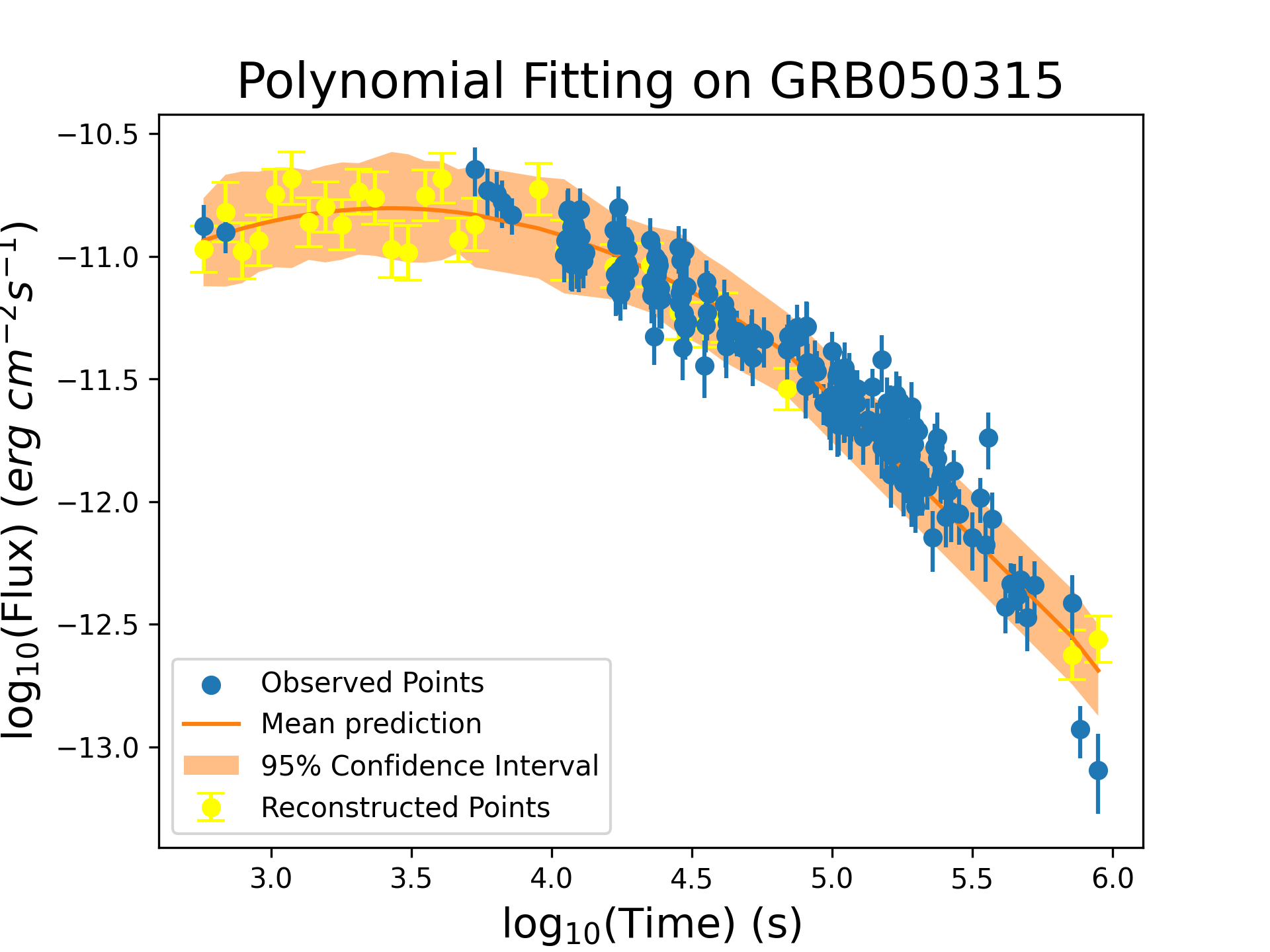}
    \includegraphics[width=.24\textwidth, height=.17\textwidth]{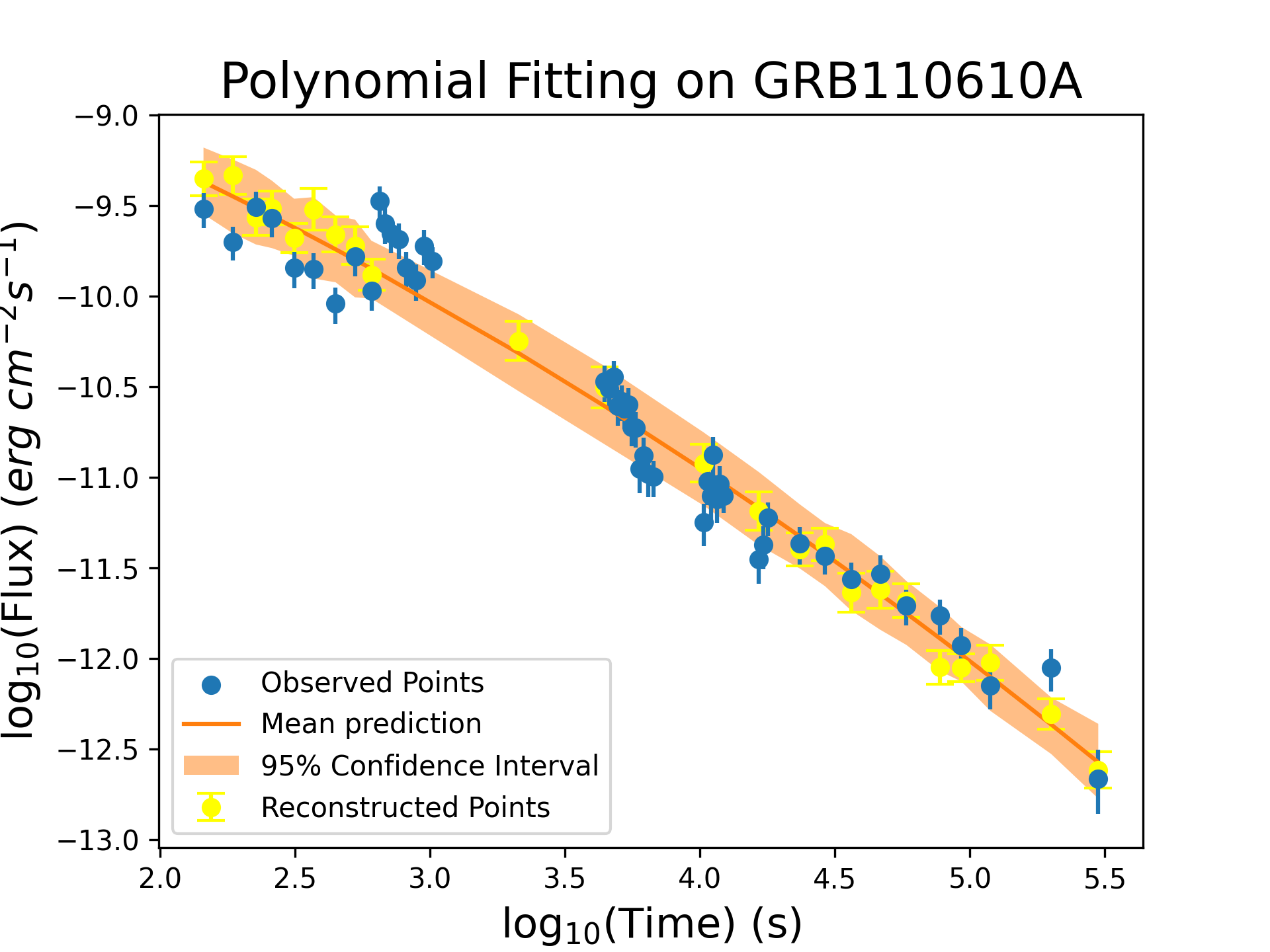}
    \includegraphics[width=.24\textwidth, height=.17\textwidth]{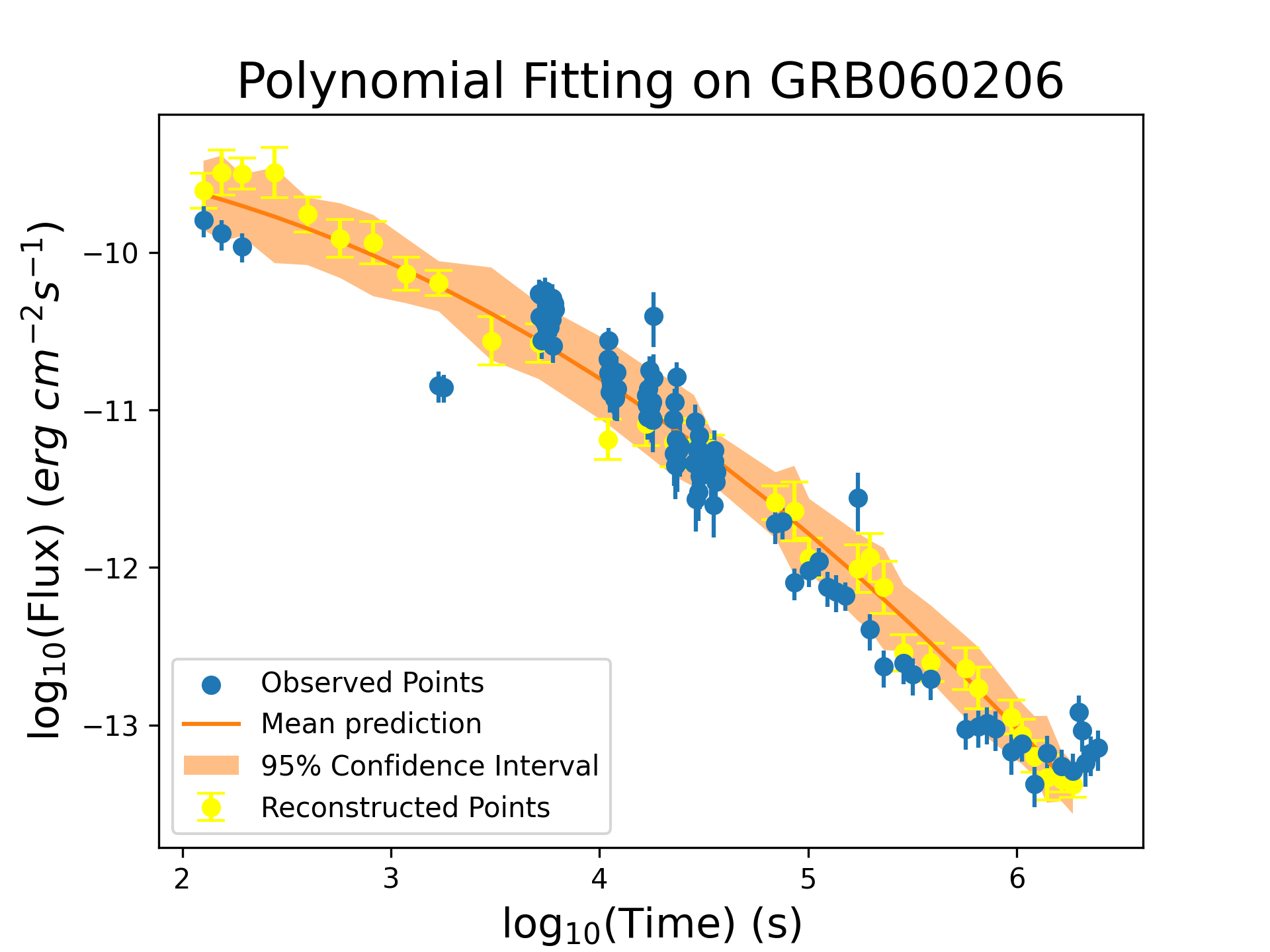}  
    
    \includegraphics[width=.24\textwidth, height=.17\textwidth]{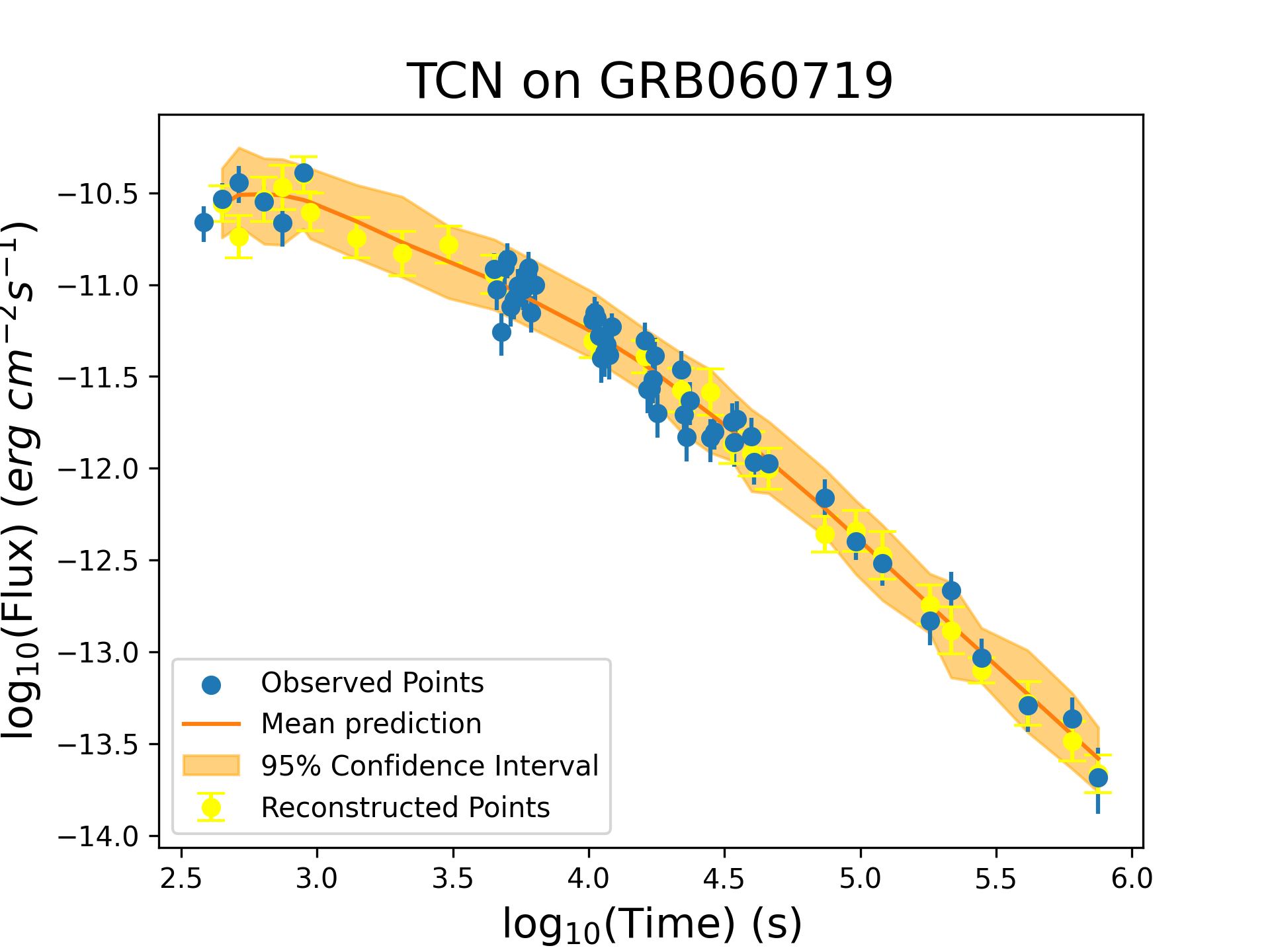}
    \includegraphics[width=.24\textwidth, height=.17\textwidth]{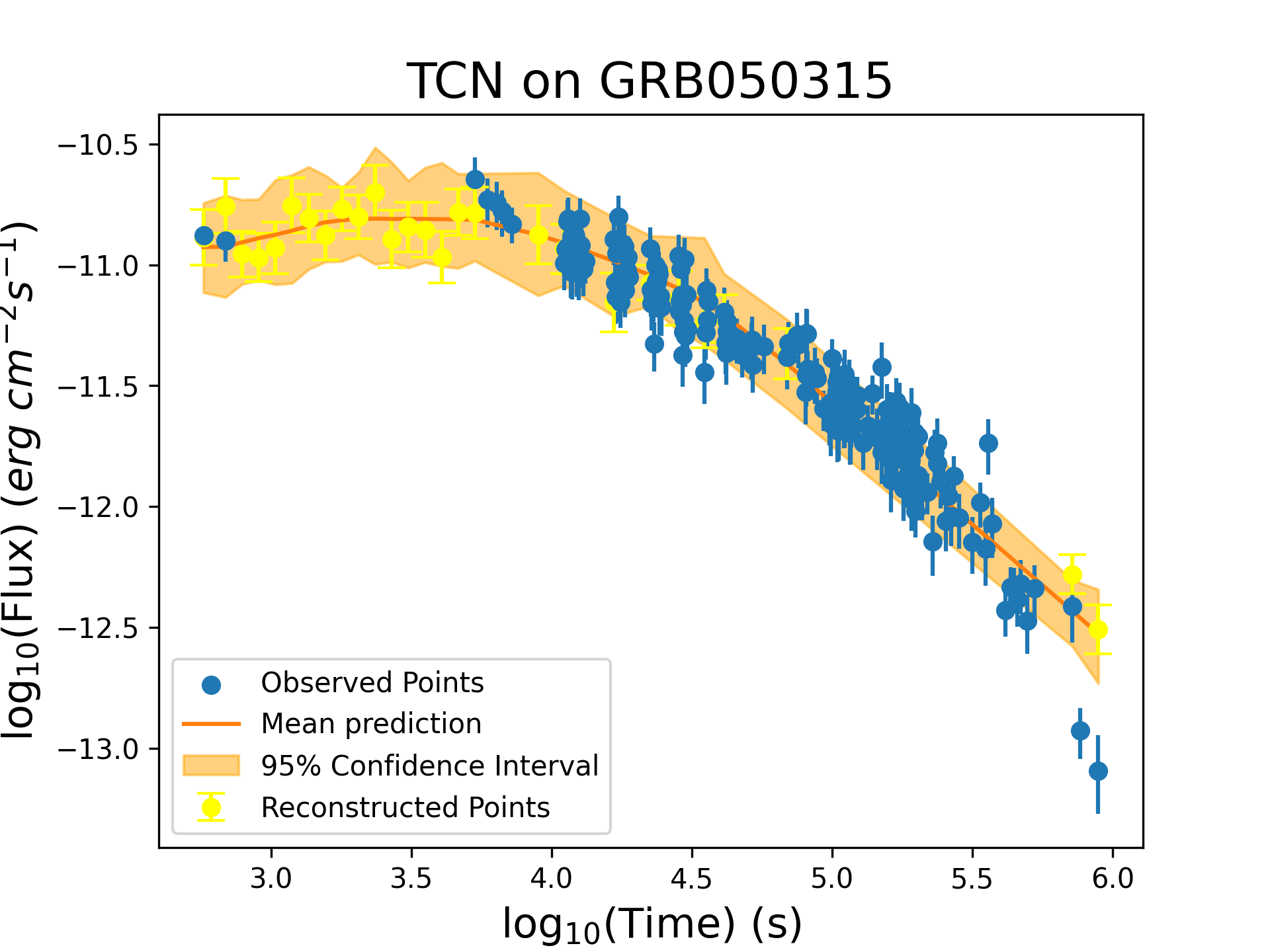}
    \includegraphics[width=.24\textwidth, height=.17\textwidth]{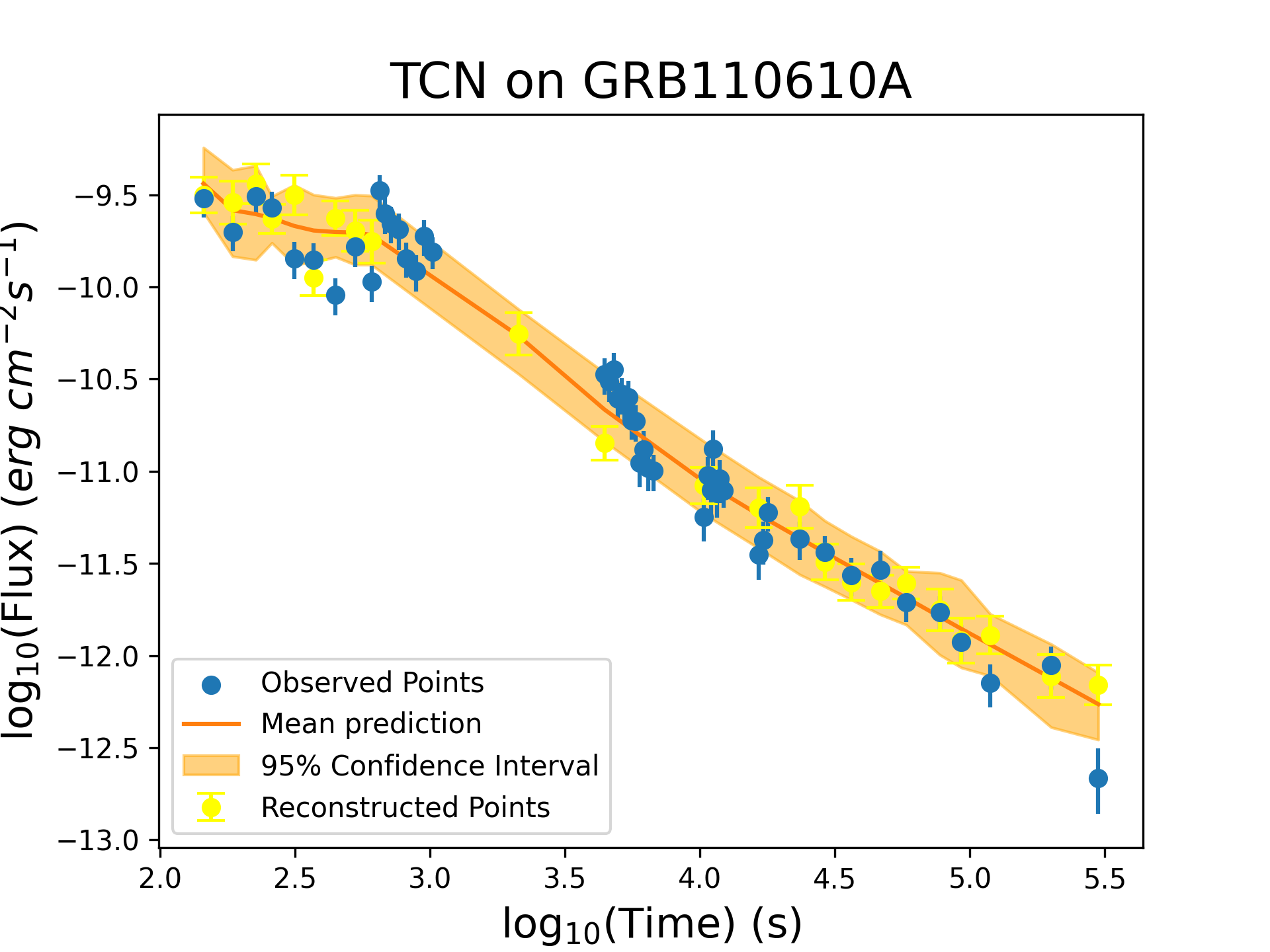}
    \includegraphics[width=.24\textwidth, height=.17\textwidth]{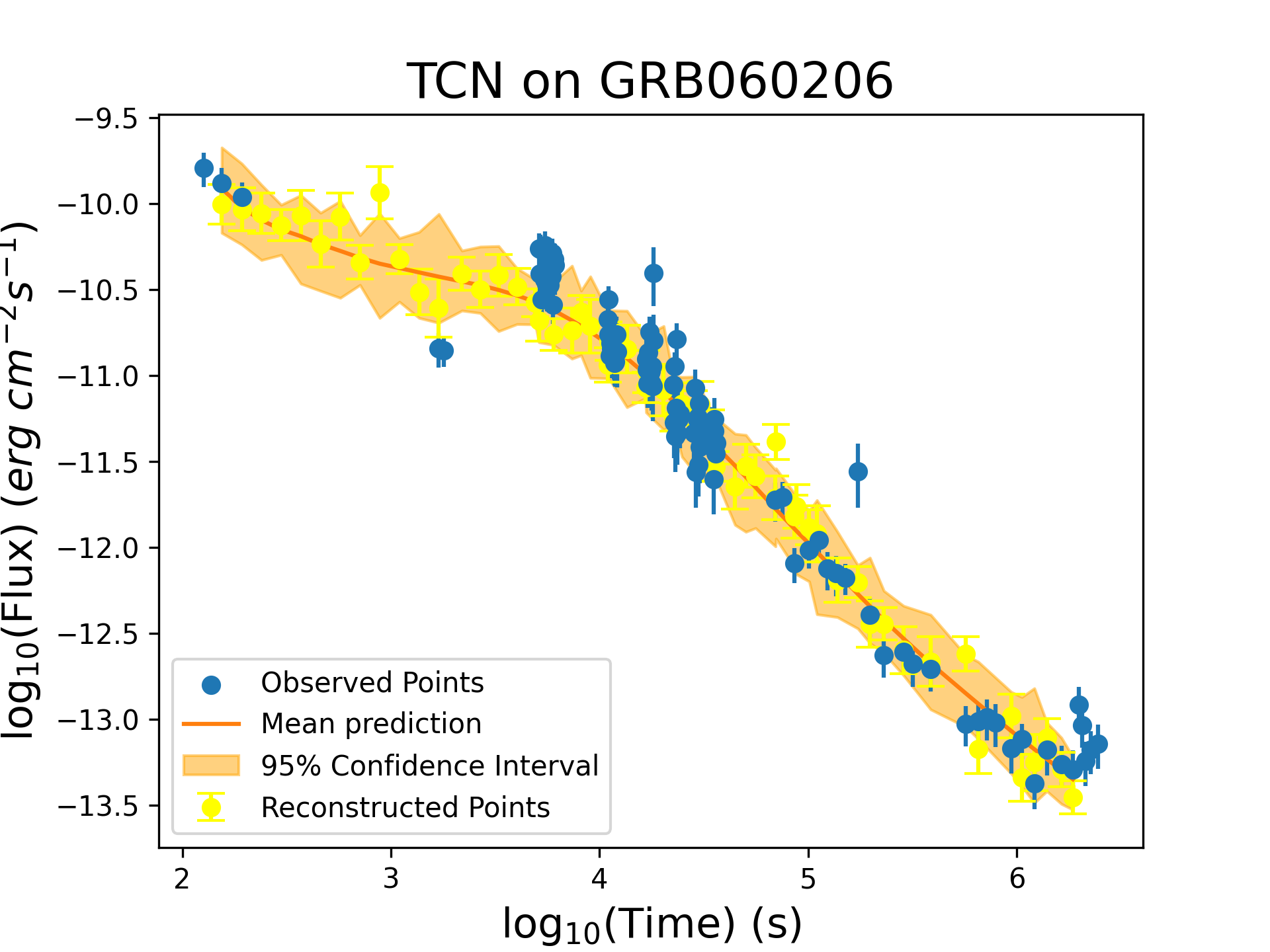}

    \includegraphics[width=.24\textwidth, height=.17\textwidth]{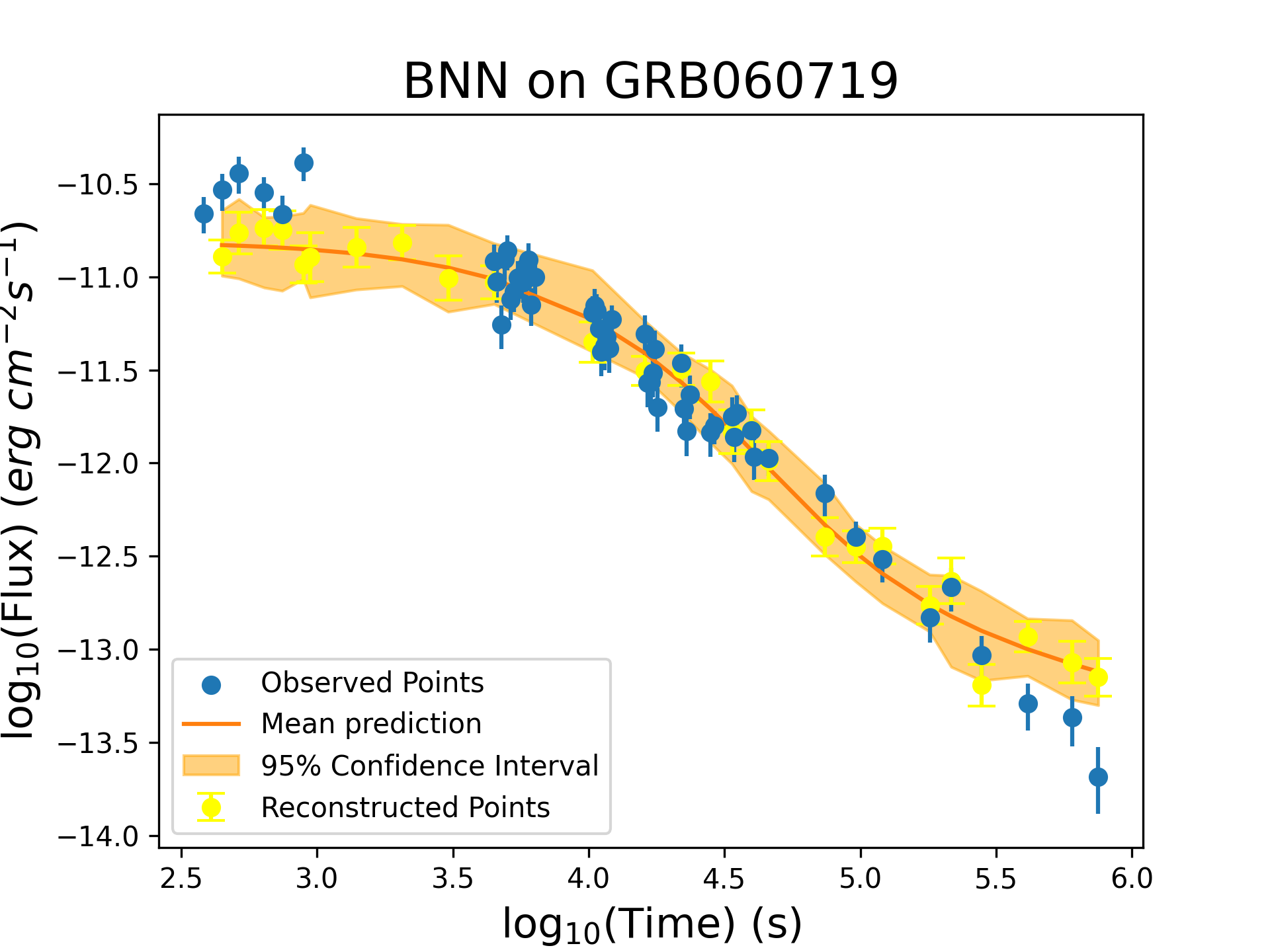}
    \includegraphics[width=.24\textwidth, height=.17\textwidth]{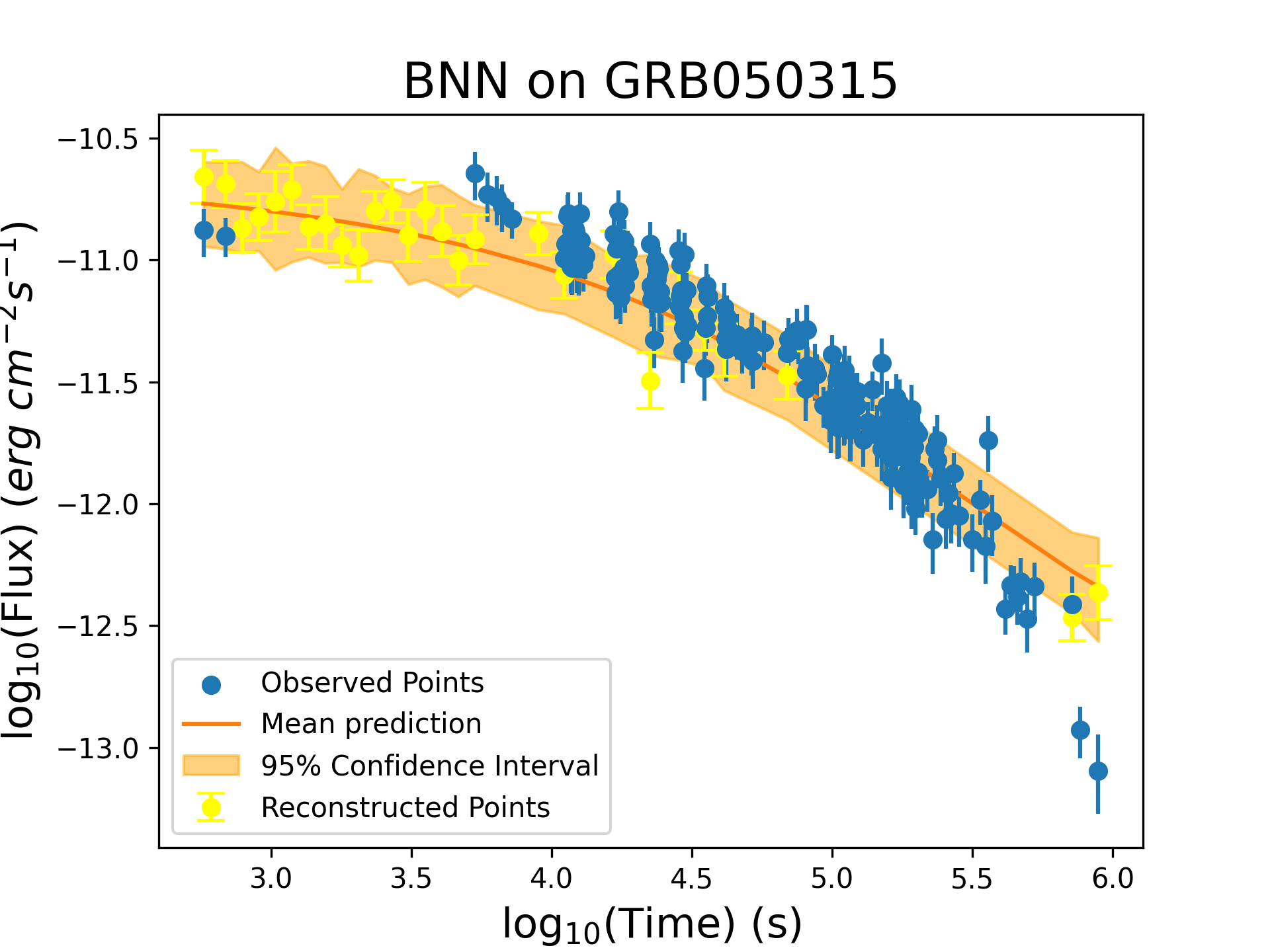}
    \includegraphics[width=.24\textwidth, height=.17\textwidth]{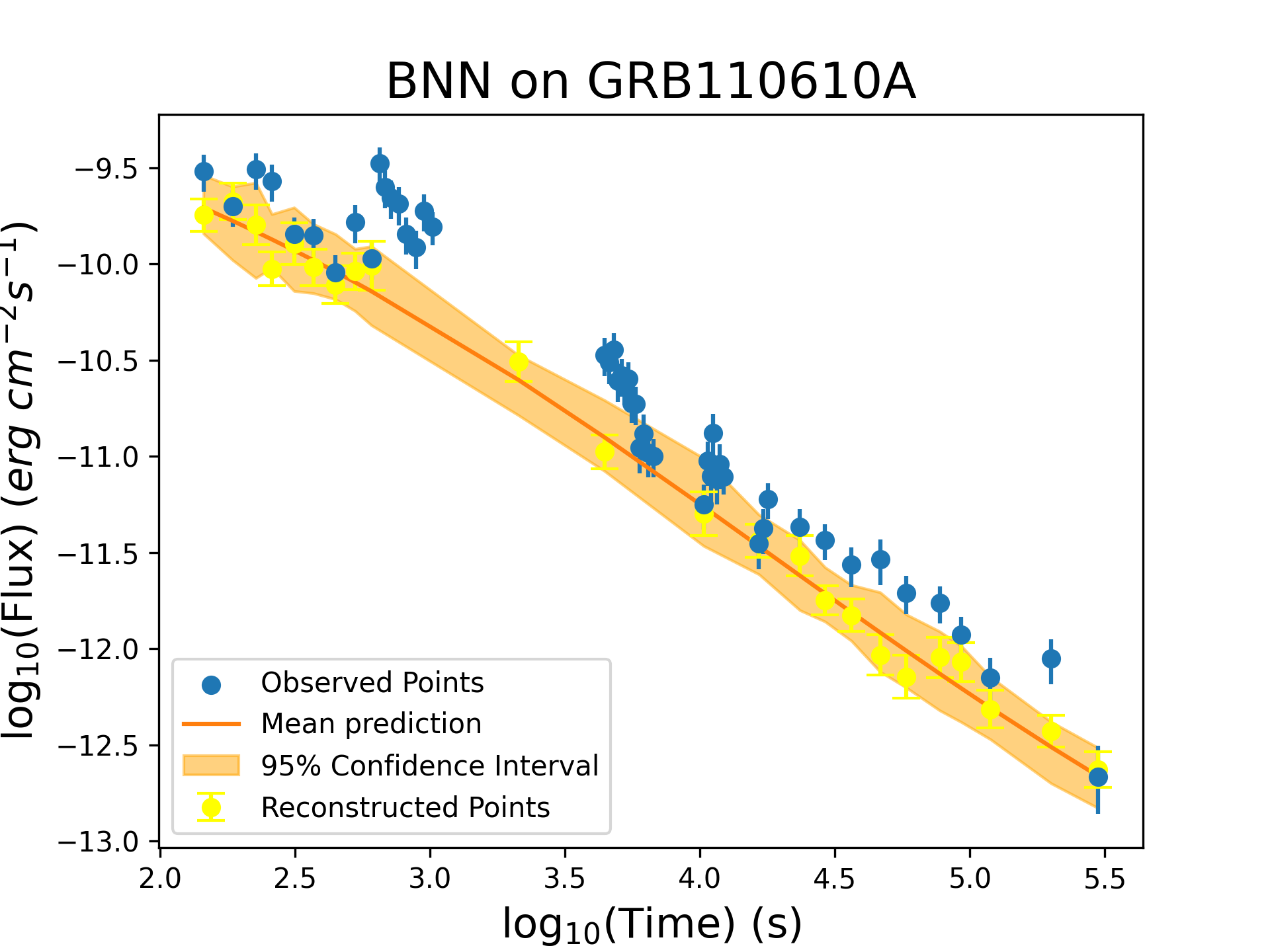}
    \includegraphics[width=.24\textwidth, height=.17\textwidth]{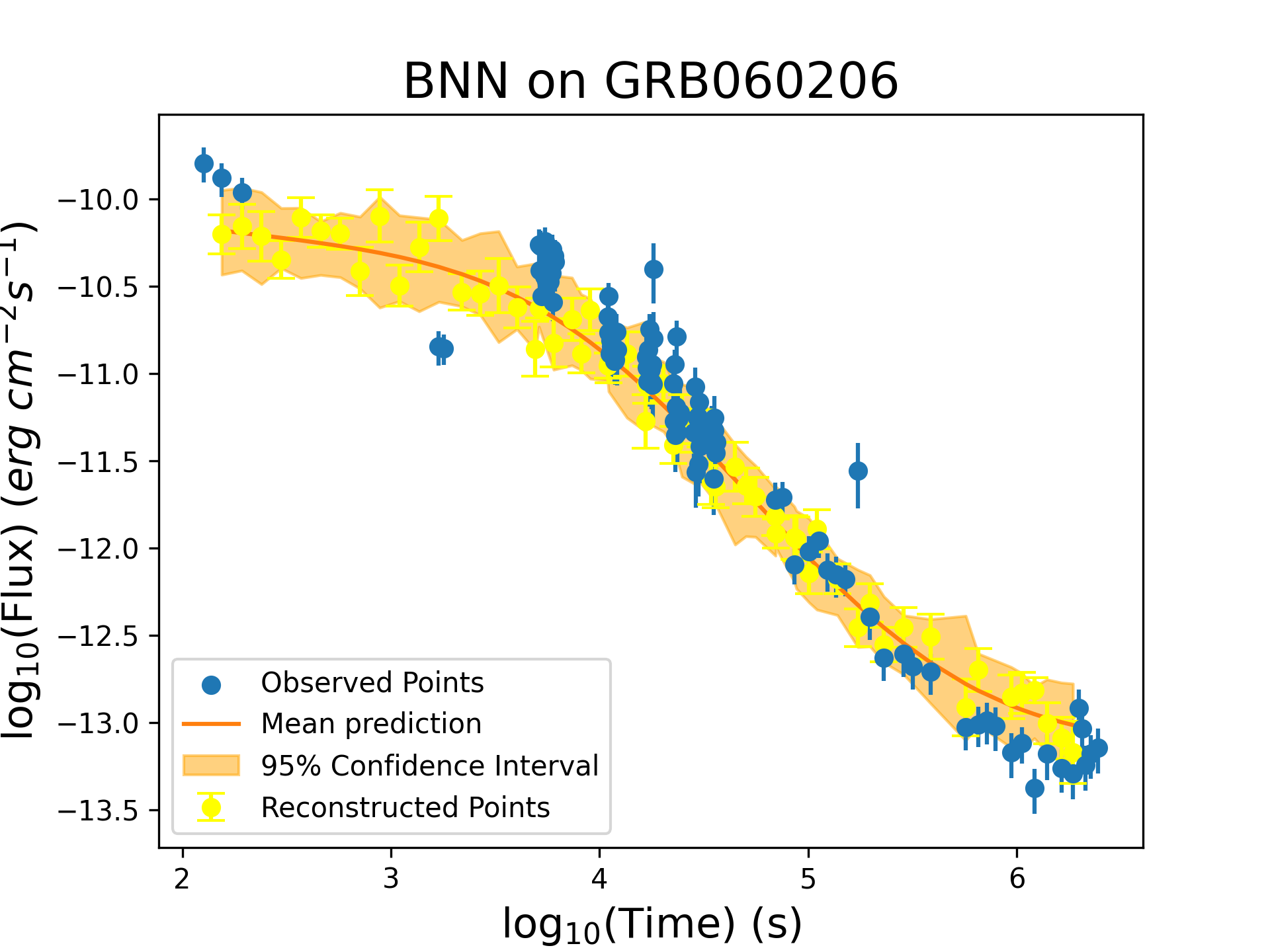}  

    \includegraphics[width=.24\textwidth, height=.17\textwidth]{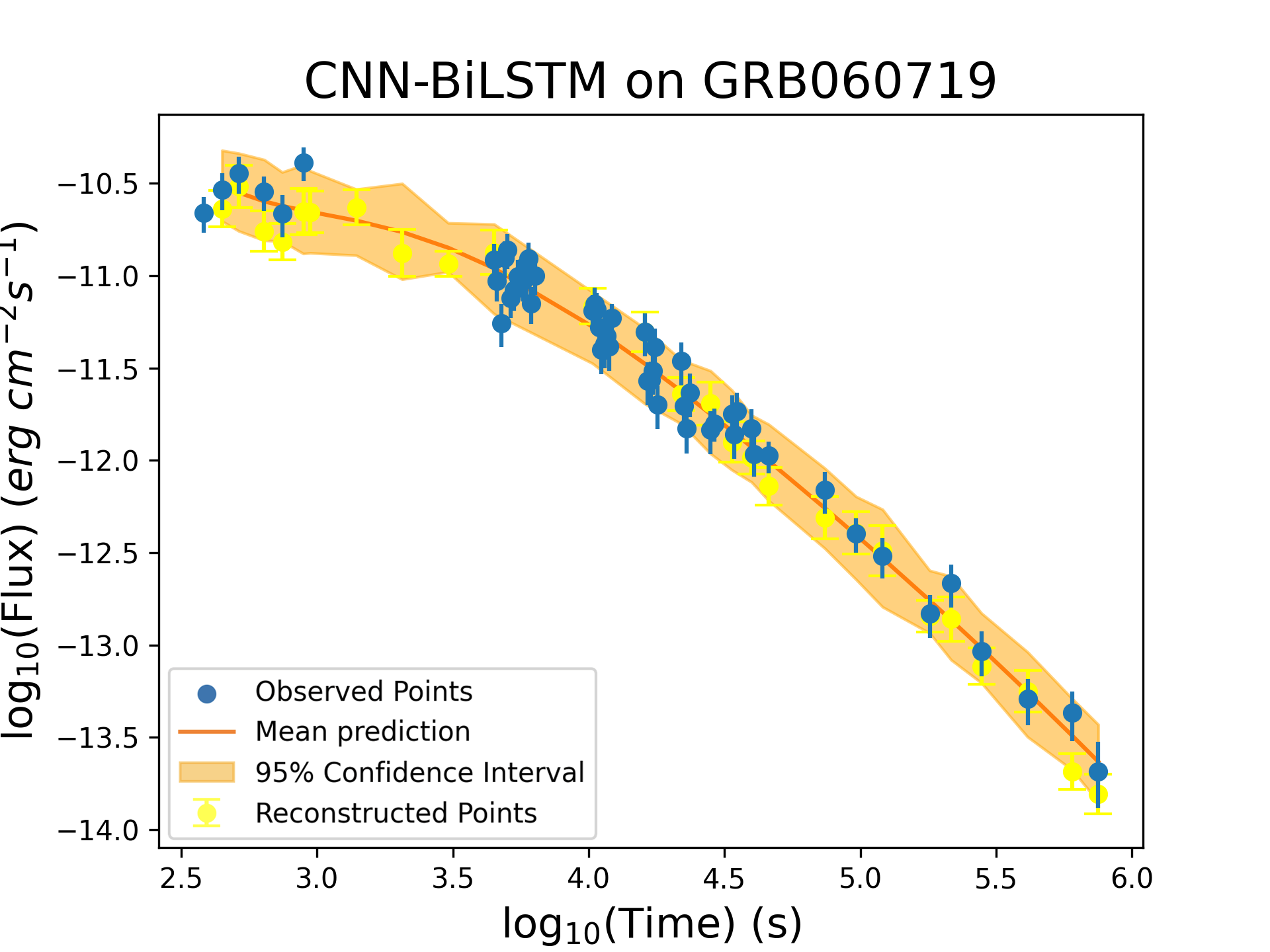}
    \includegraphics[width=.24\textwidth, height=.17\textwidth]{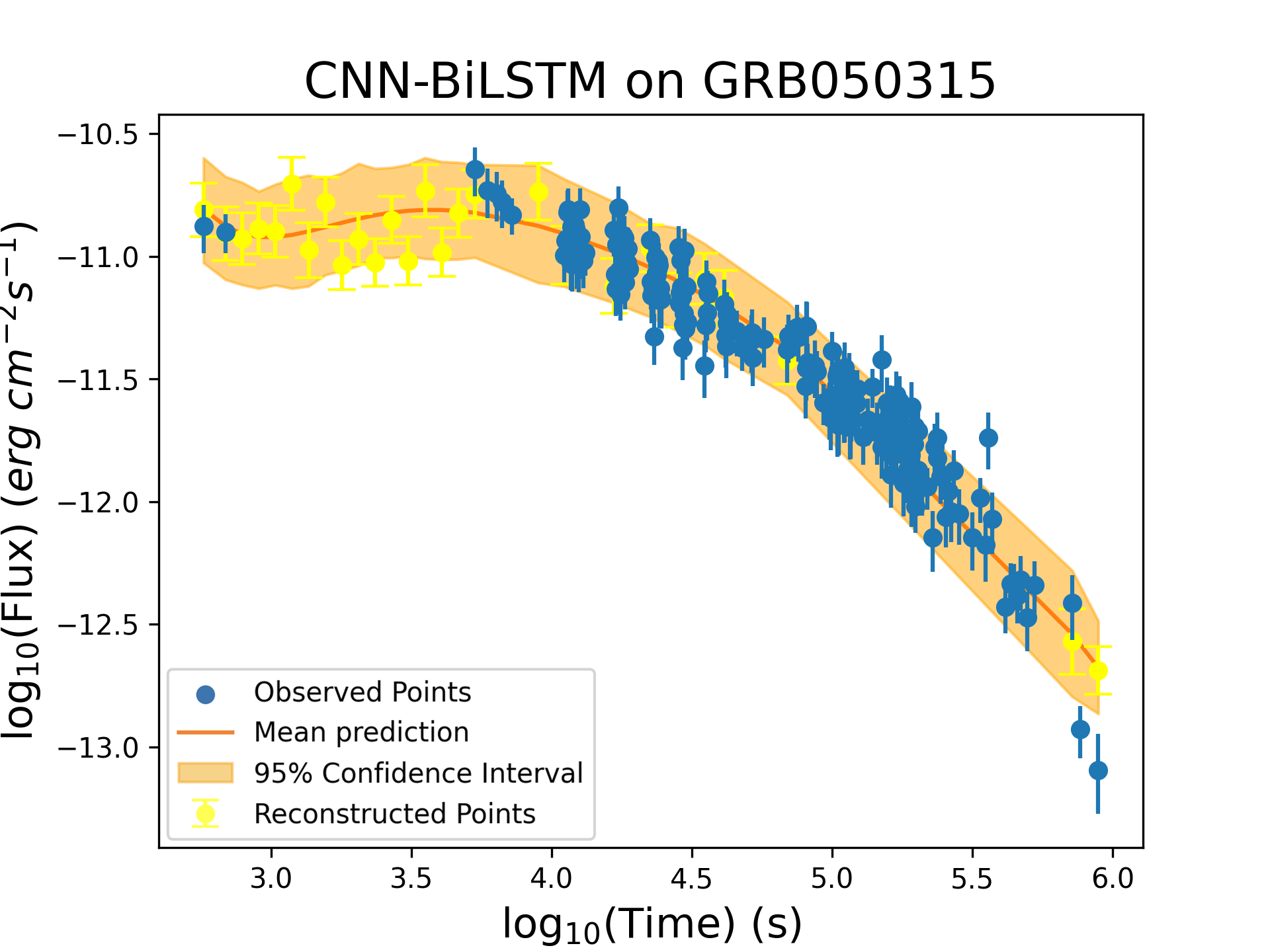}
    \includegraphics[width=.24\textwidth, height=.17\textwidth]{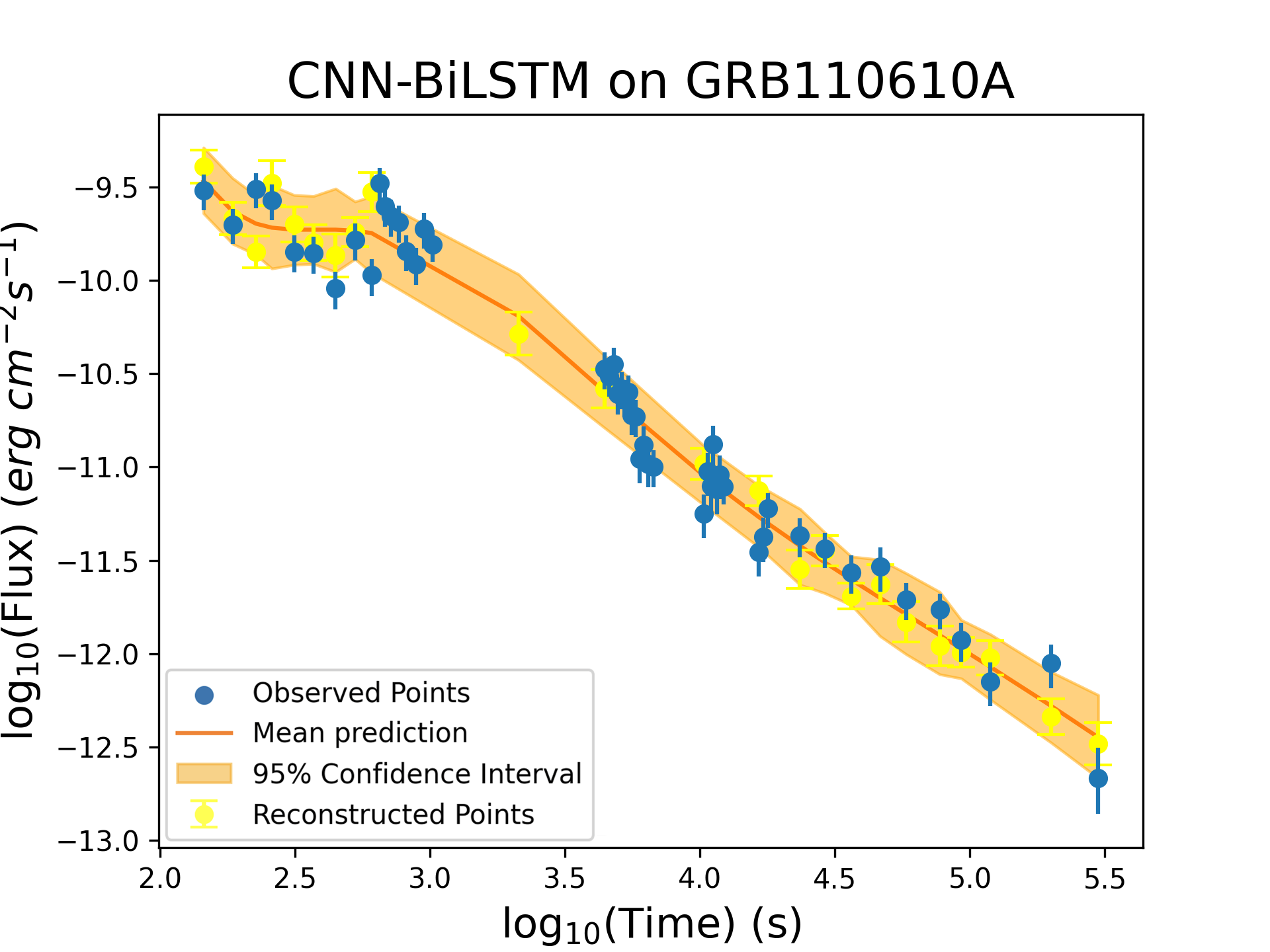}
    \includegraphics[width=.24\textwidth, height=.17\textwidth]{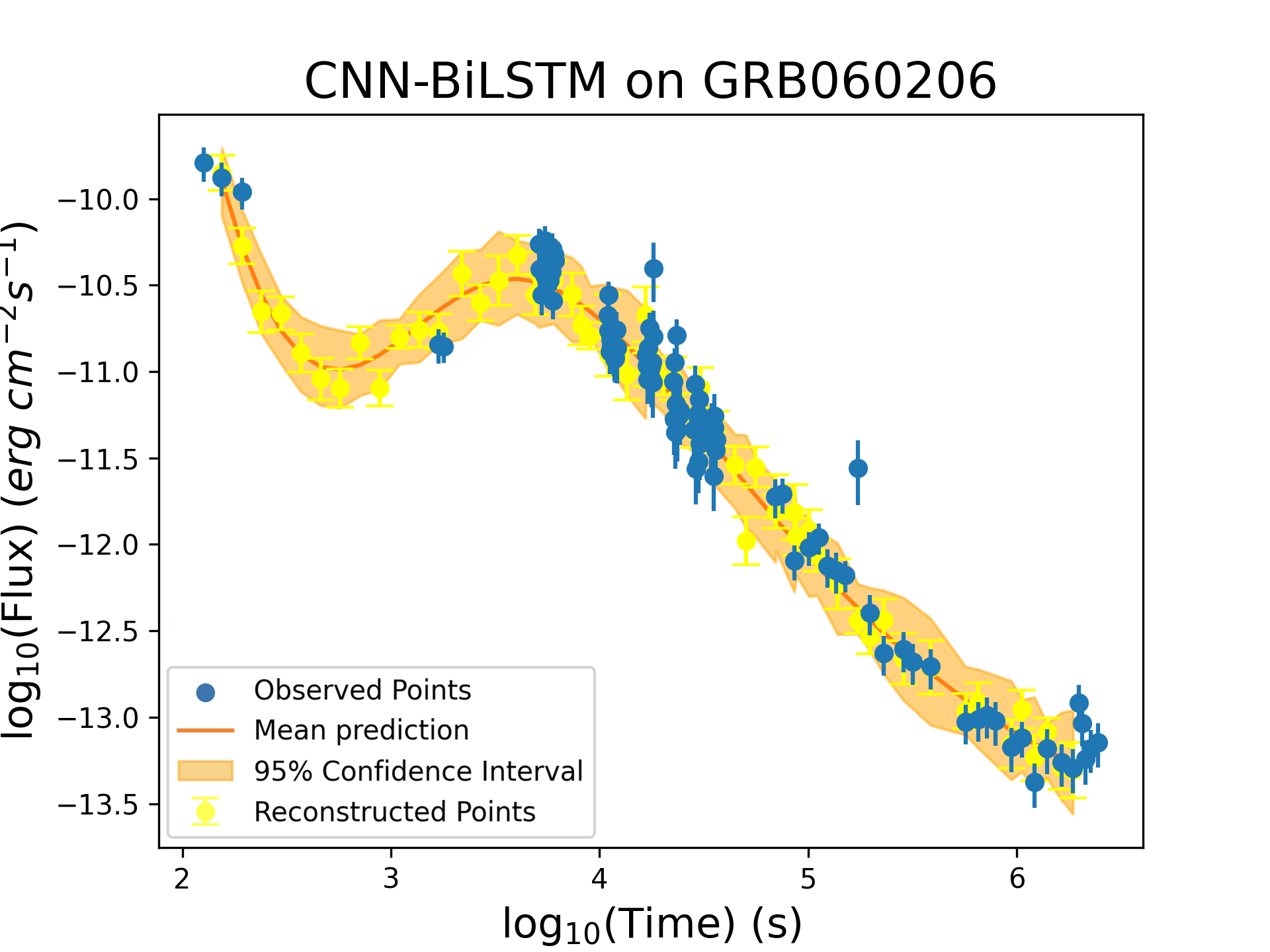}

    \includegraphics[width=.24\textwidth, height=.17\textwidth]{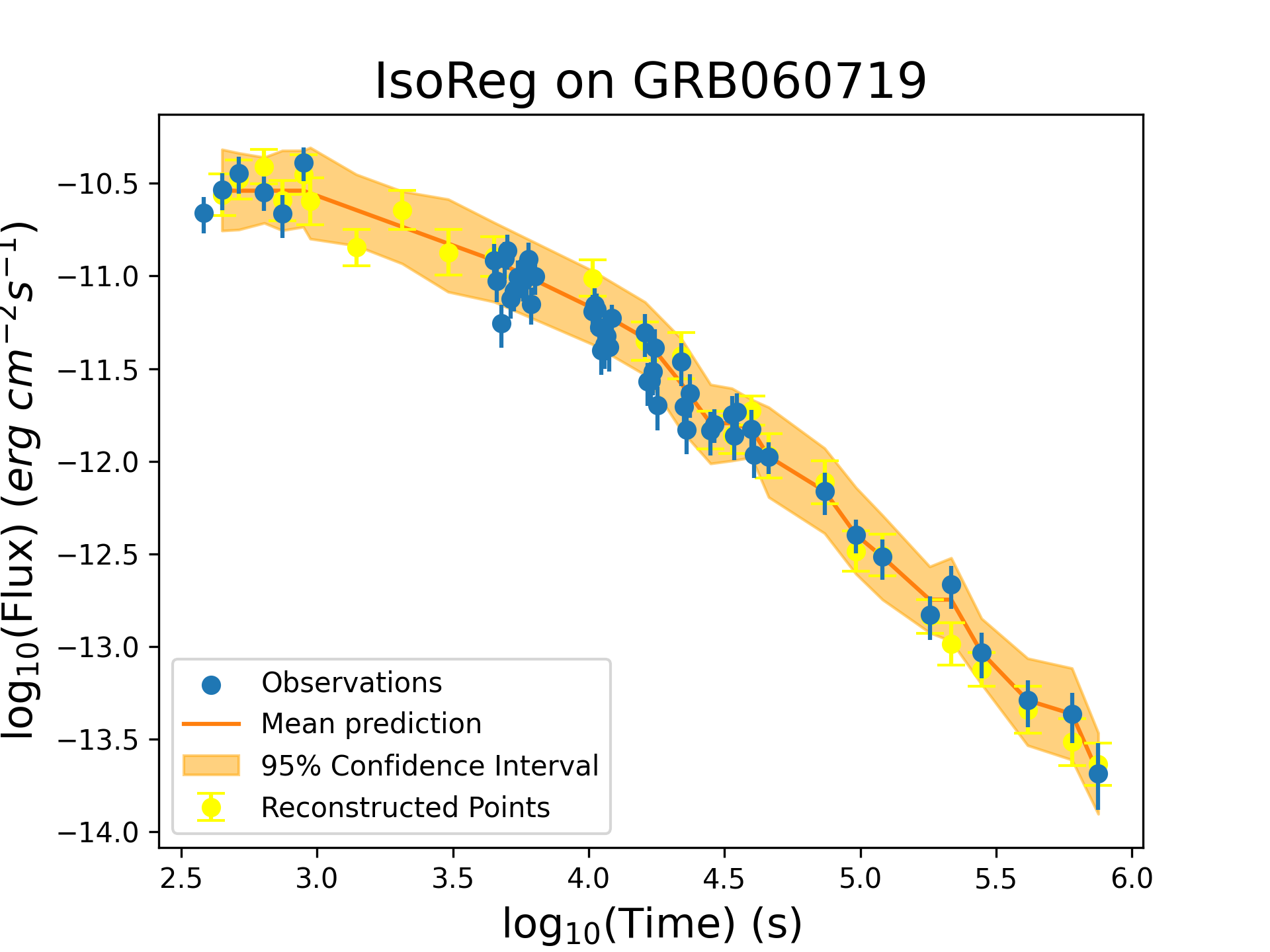}
    \includegraphics[width=.24\textwidth, height=.17\textwidth]{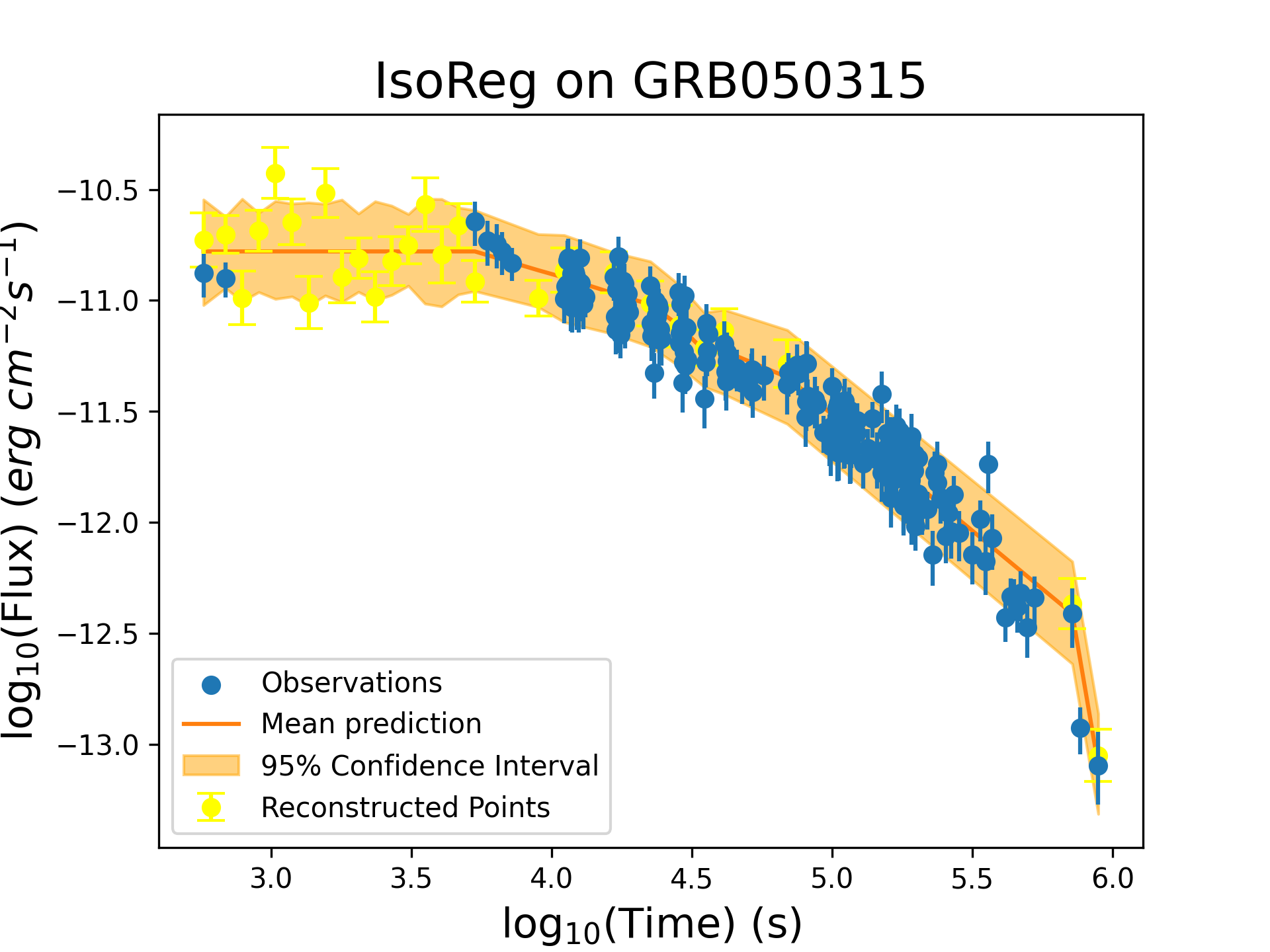}
    \includegraphics[width=.24\textwidth, height=.17\textwidth]{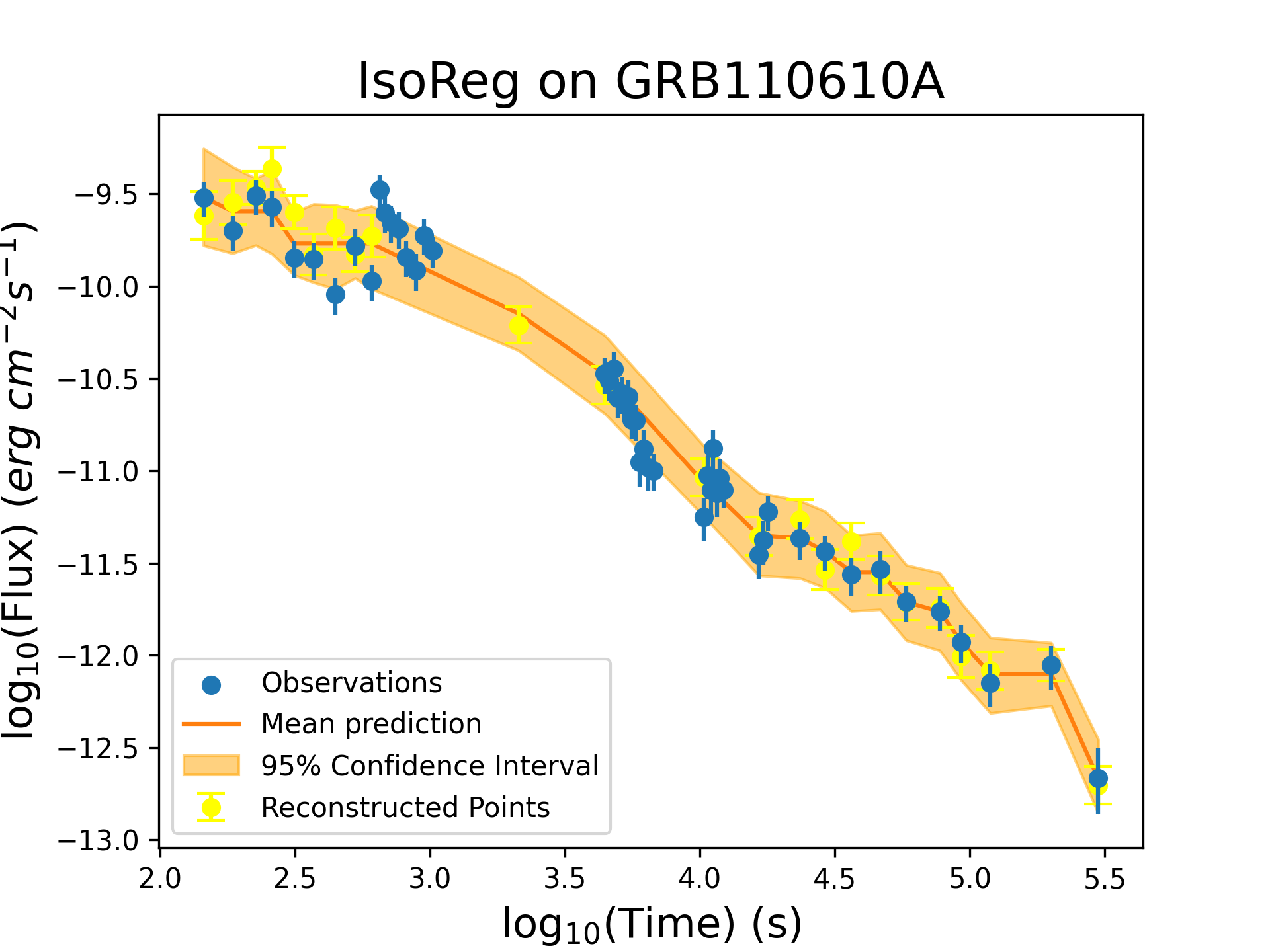}
    \includegraphics[width=.24\textwidth, height=.17\textwidth]{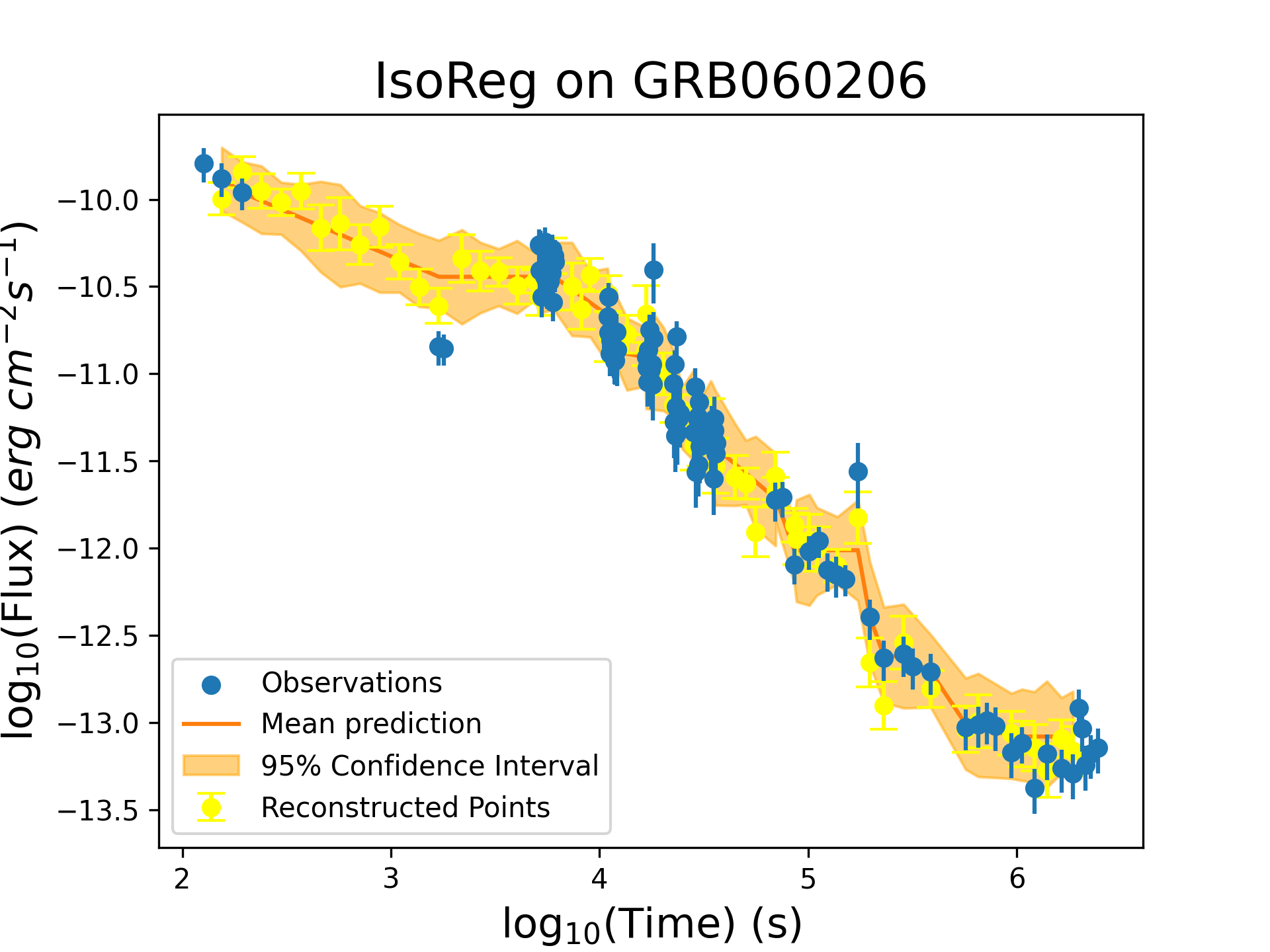}

    \includegraphics[width=.24\textwidth, height=.17\textwidth]{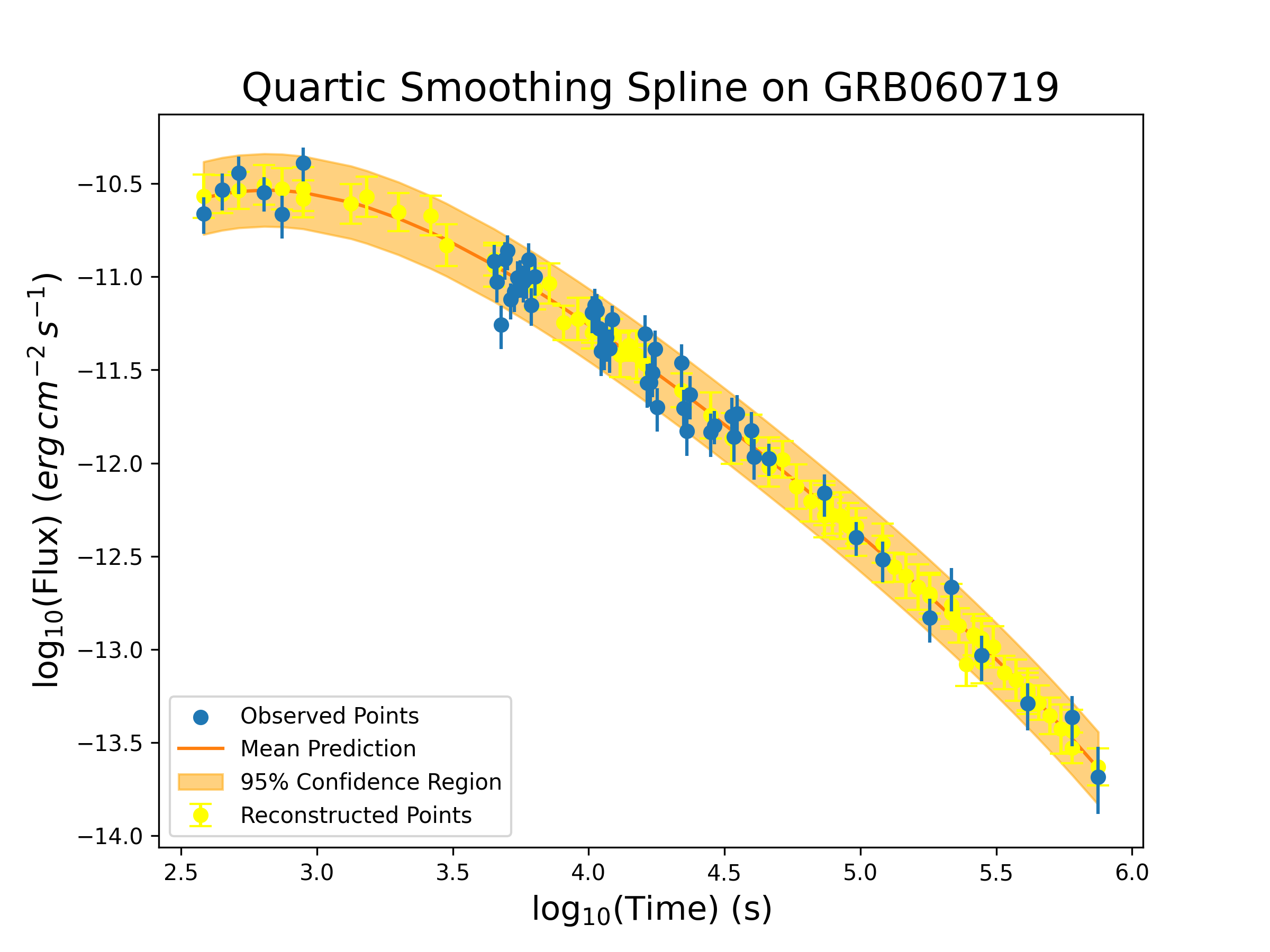}
    \includegraphics[width=.24\textwidth, height=.17\textwidth]{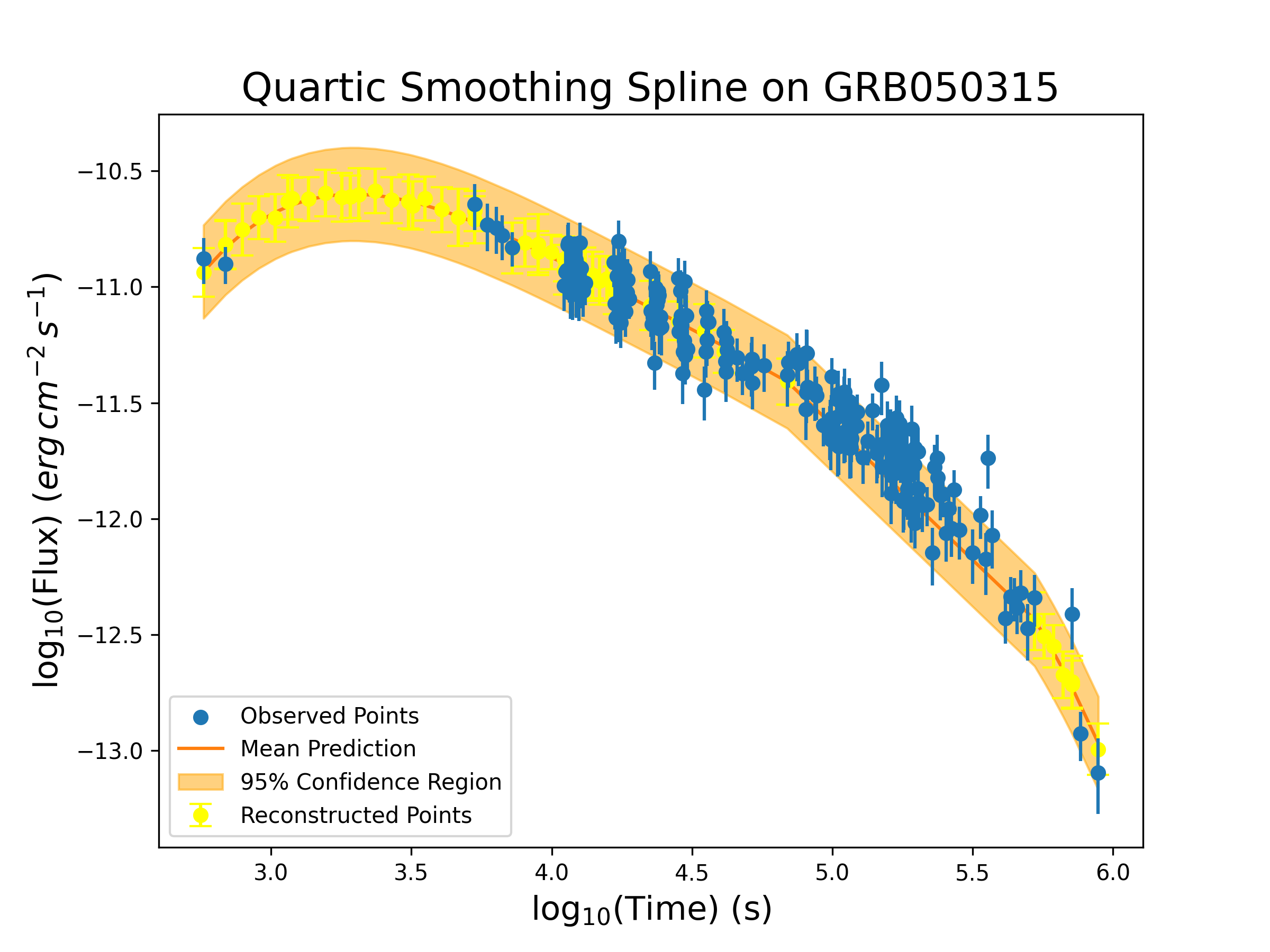}
    \includegraphics[width=.24\textwidth, height=.17\textwidth]{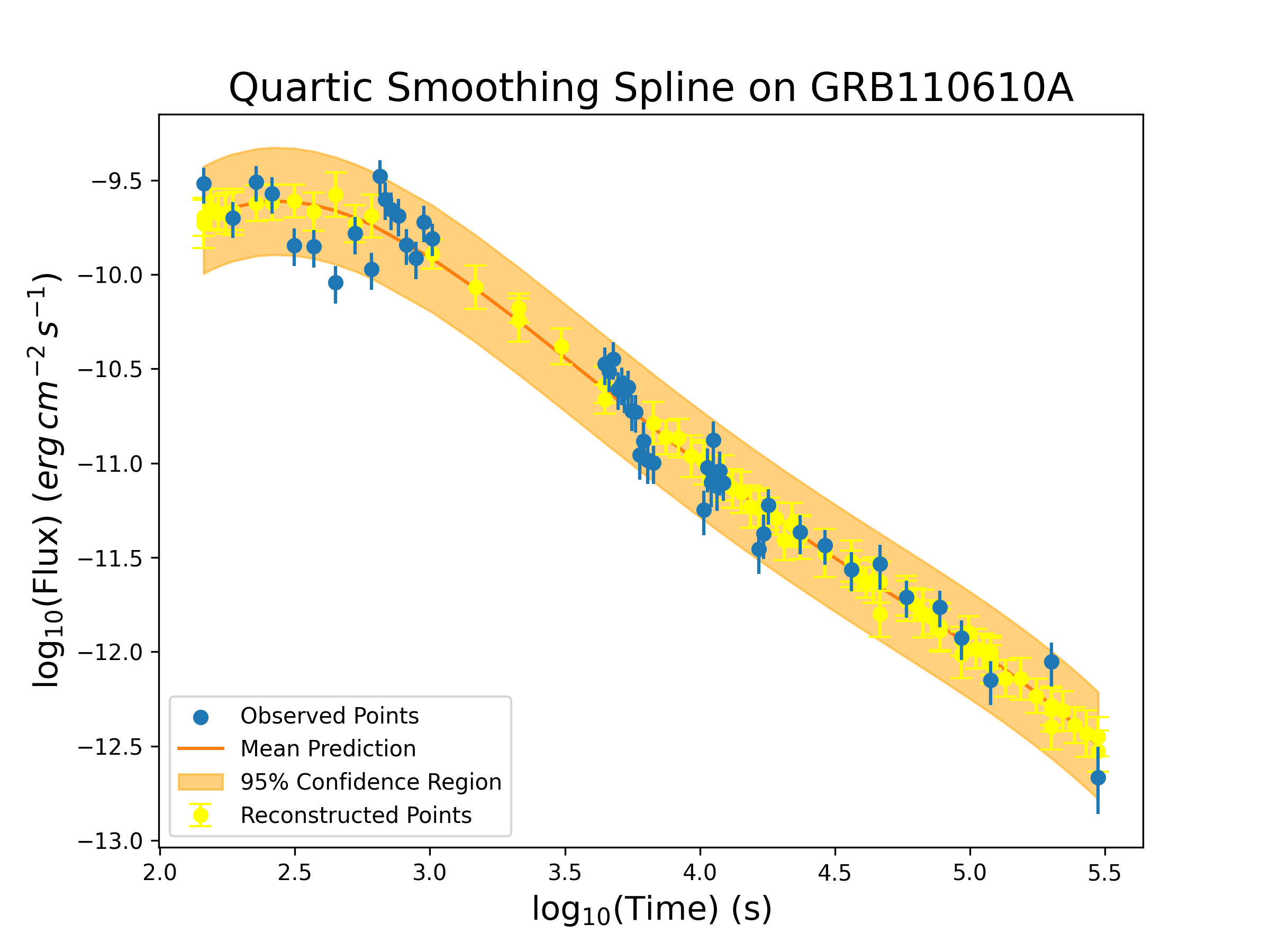}
    \includegraphics[width=.24\textwidth, height=.17\textwidth]{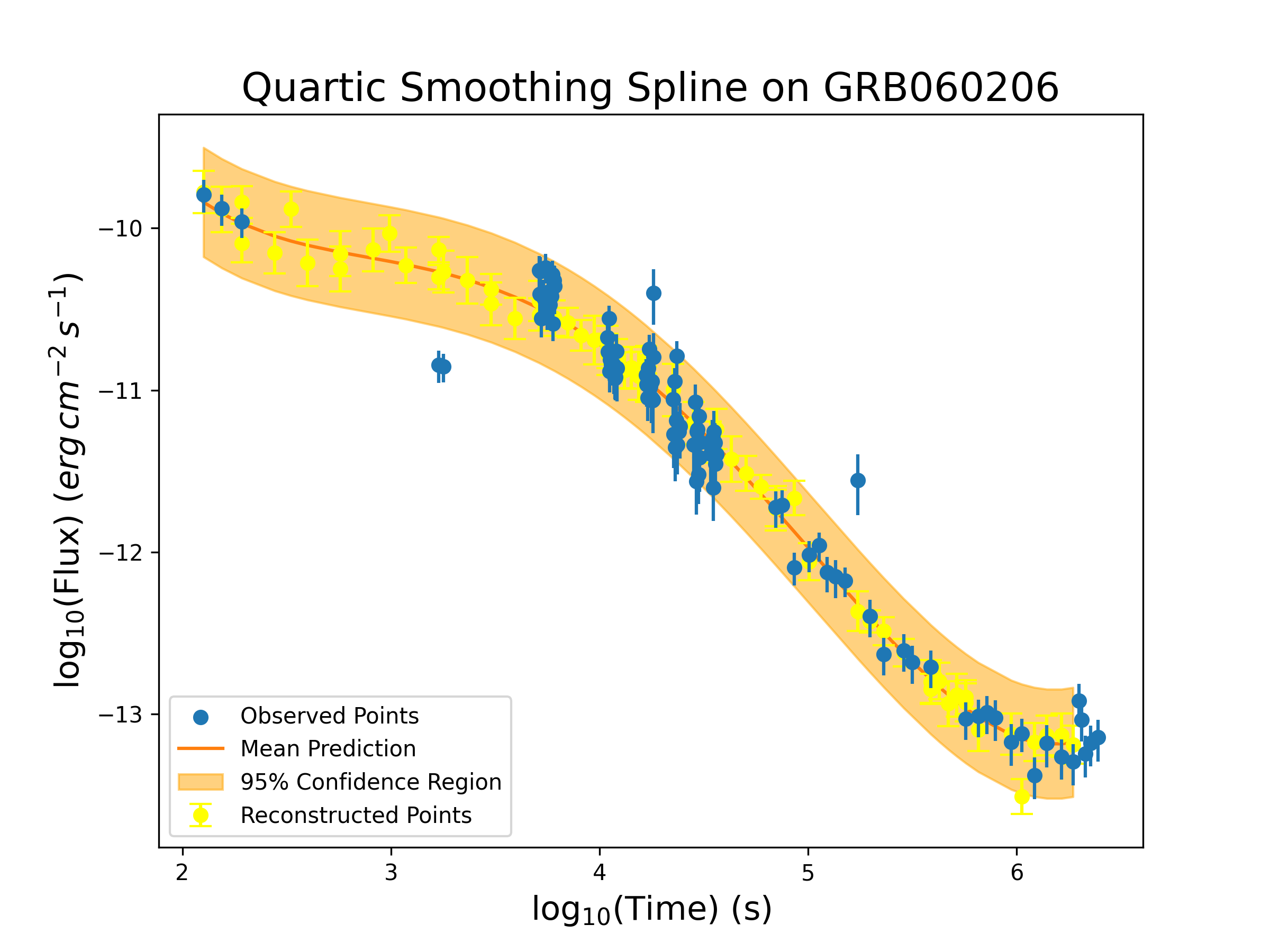}
\end{center}
    \caption{Reconstruction of LCs for all four varieties of GRBs are shown with four types of GRBs (left to right): i) Good GRBs (Column 1); ii) a GRB LC with a break towards the end (Column 2); iii) flares or bumps in the afterglow (Column 3); iv) flares or bumps with a double break towards the end of the LC (Column 4) and the models (top to bottom): i) Deep GP (Row 1); ii) Polynomial Fitting Model (Row 2); iii) TCN Model (Row 3); BNN Model (Row 4); CNN-BiLSTM Model (Row 5); Isotonic Regression Model (Row 6).}
    \label{fig: ALL-reconstruction}
    % Bi-MAMBA: i) Good GRBs (top left); ii) a GRB LC with a break towards the end (top right); iii) flares or bumps in the afterglow (bottom left); iv) flares or bumps with a double break towards the end of the LC (bottom right).
\end{figure*}

\begin{figure*}[htbp]
\begin{center}

%DGP-LCR
\includegraphics[width=.25\textwidth, height=.17\textwidth]{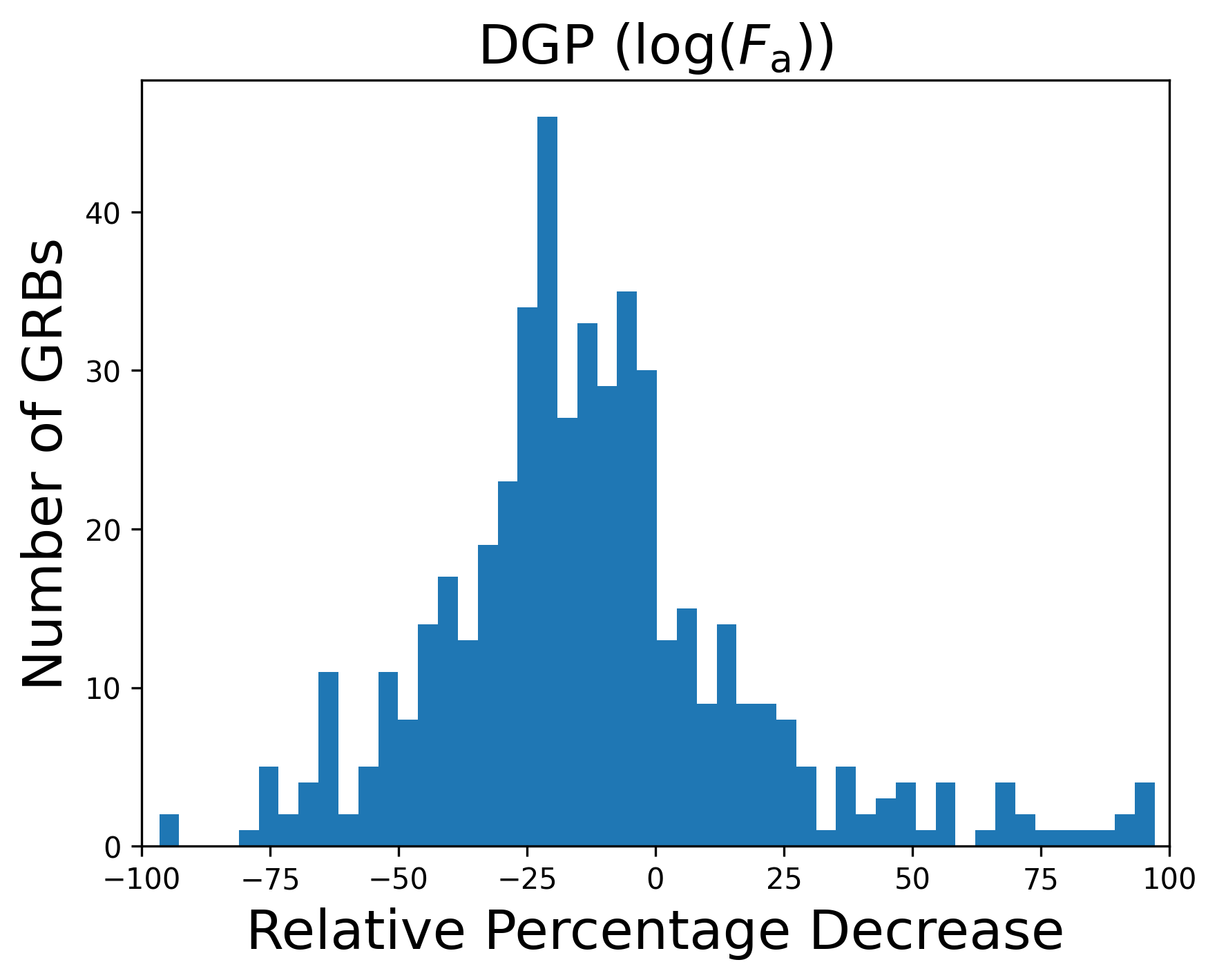}
\includegraphics[width=.25\textwidth, height=.17\textwidth]{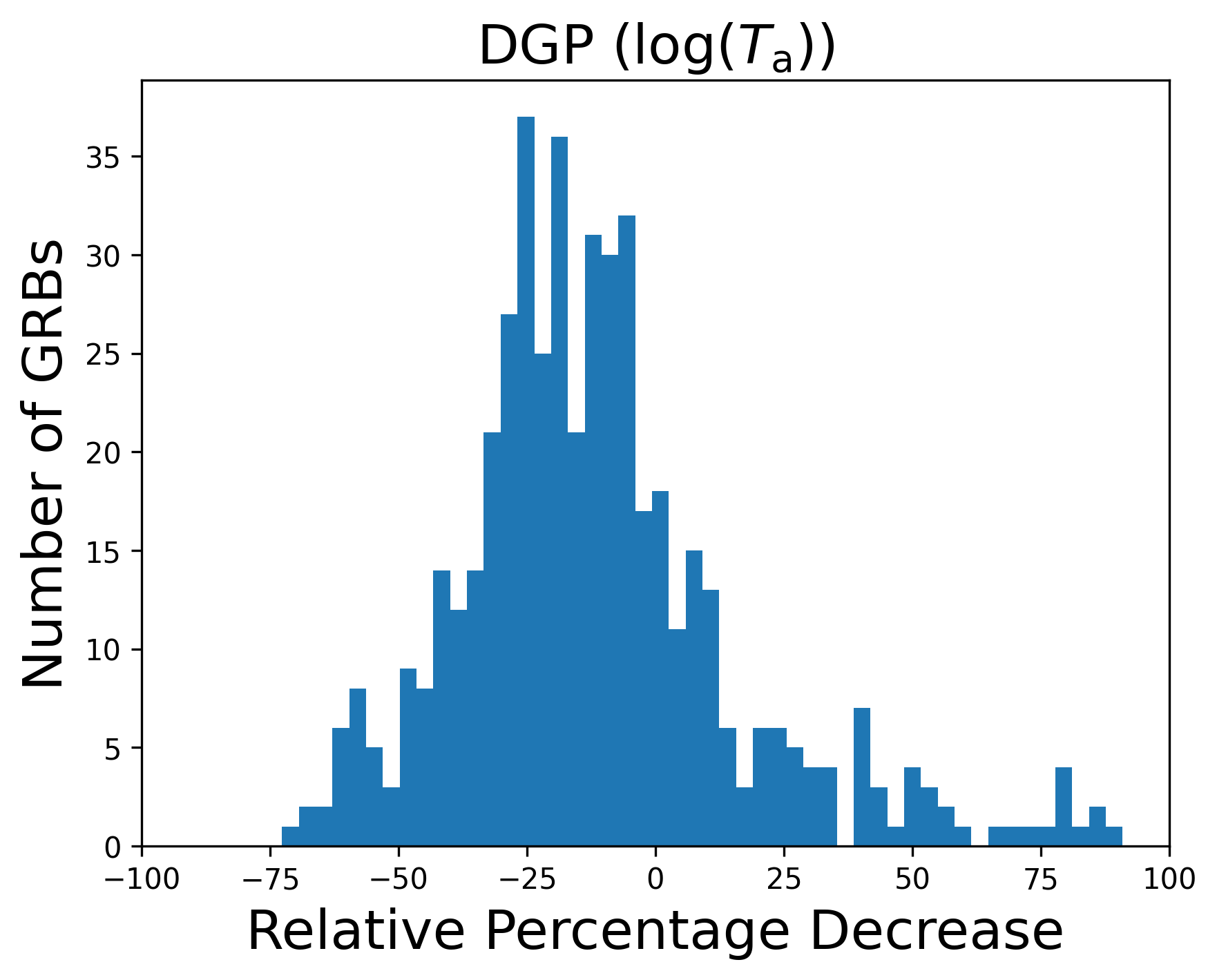}
\includegraphics[width=.25\textwidth, height=.17\textwidth]{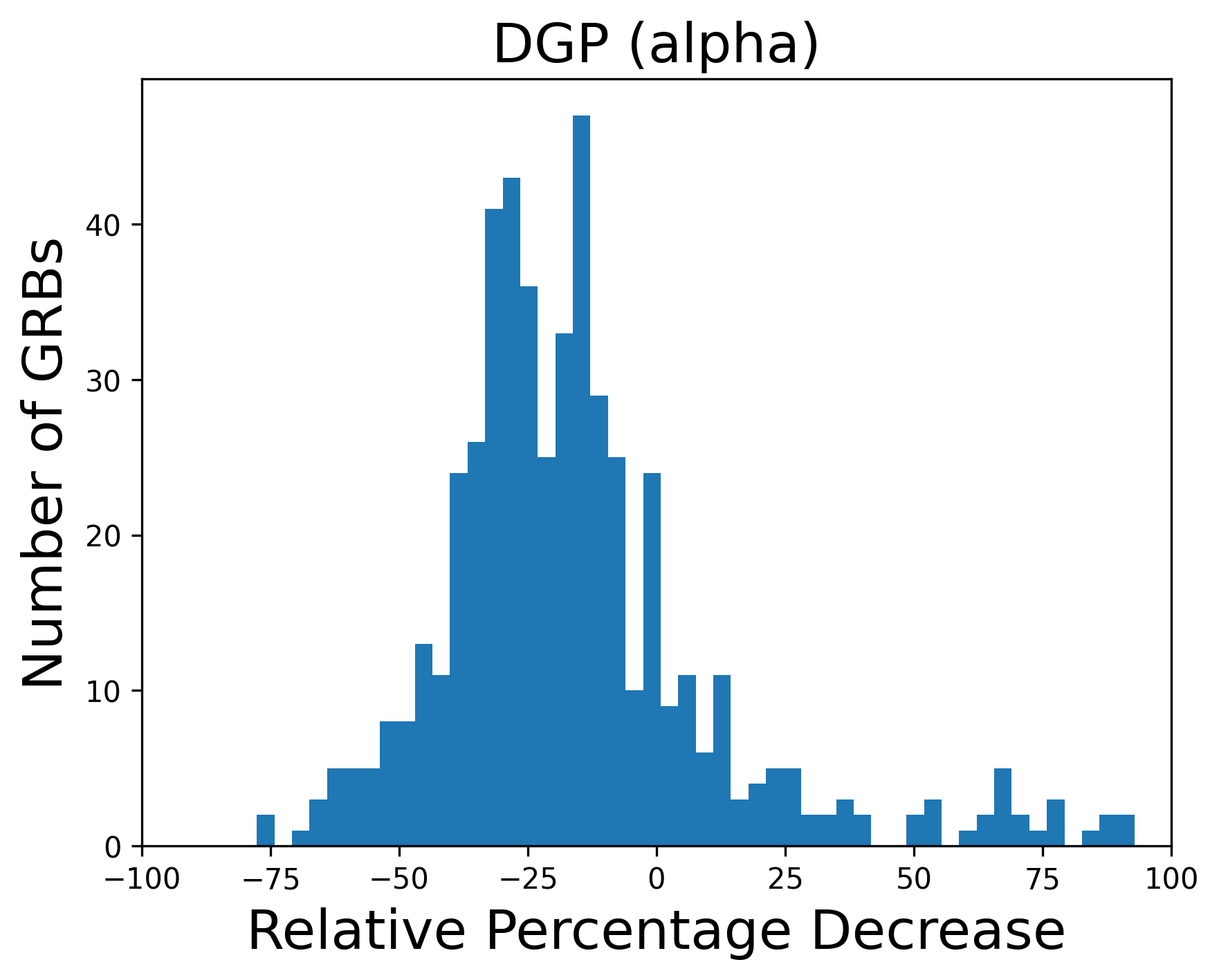}

%TCN-LCR
\includegraphics[width=.25\textwidth, height=.17\textwidth]{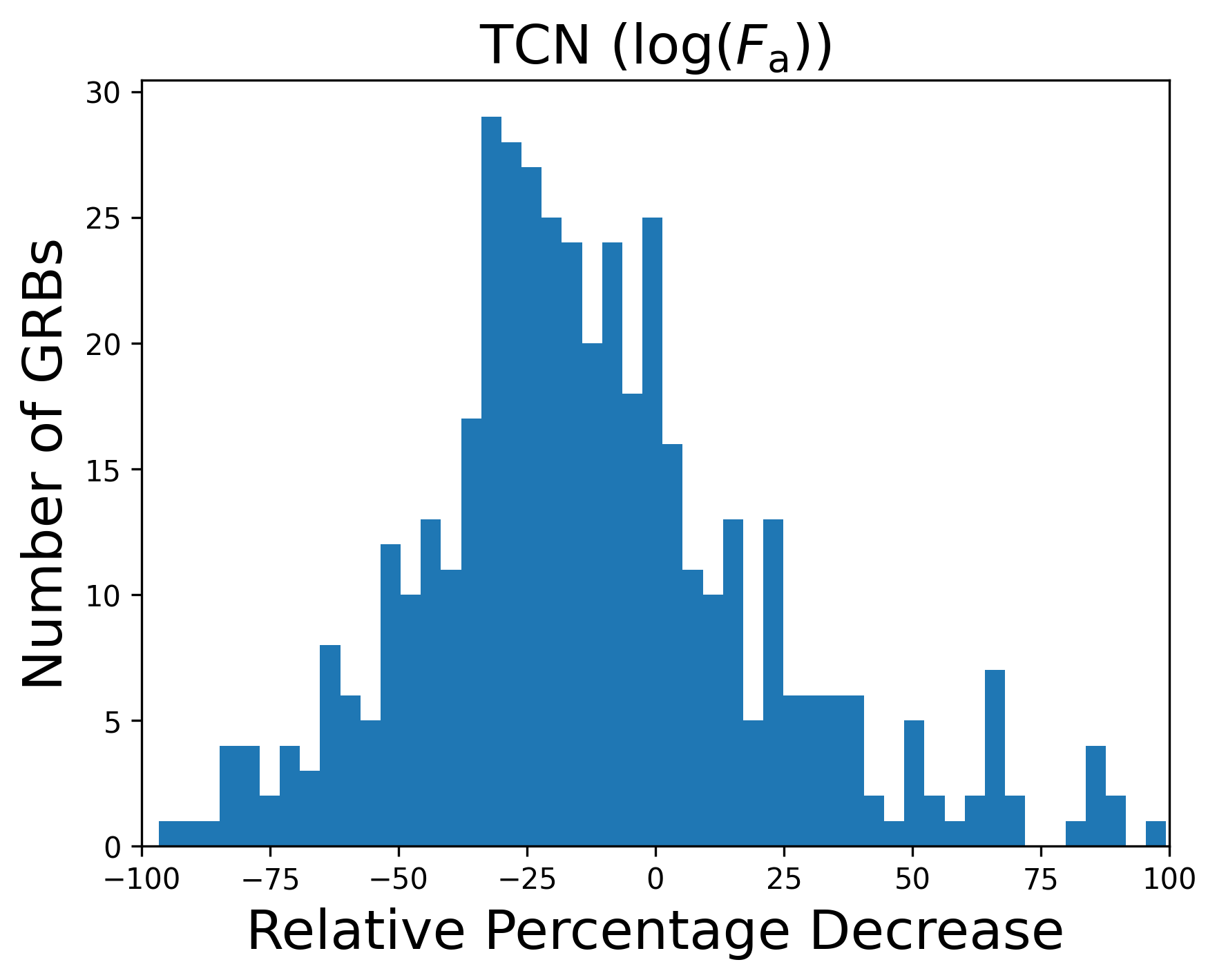}\includegraphics[width=.25\textwidth, height=.17\textwidth]{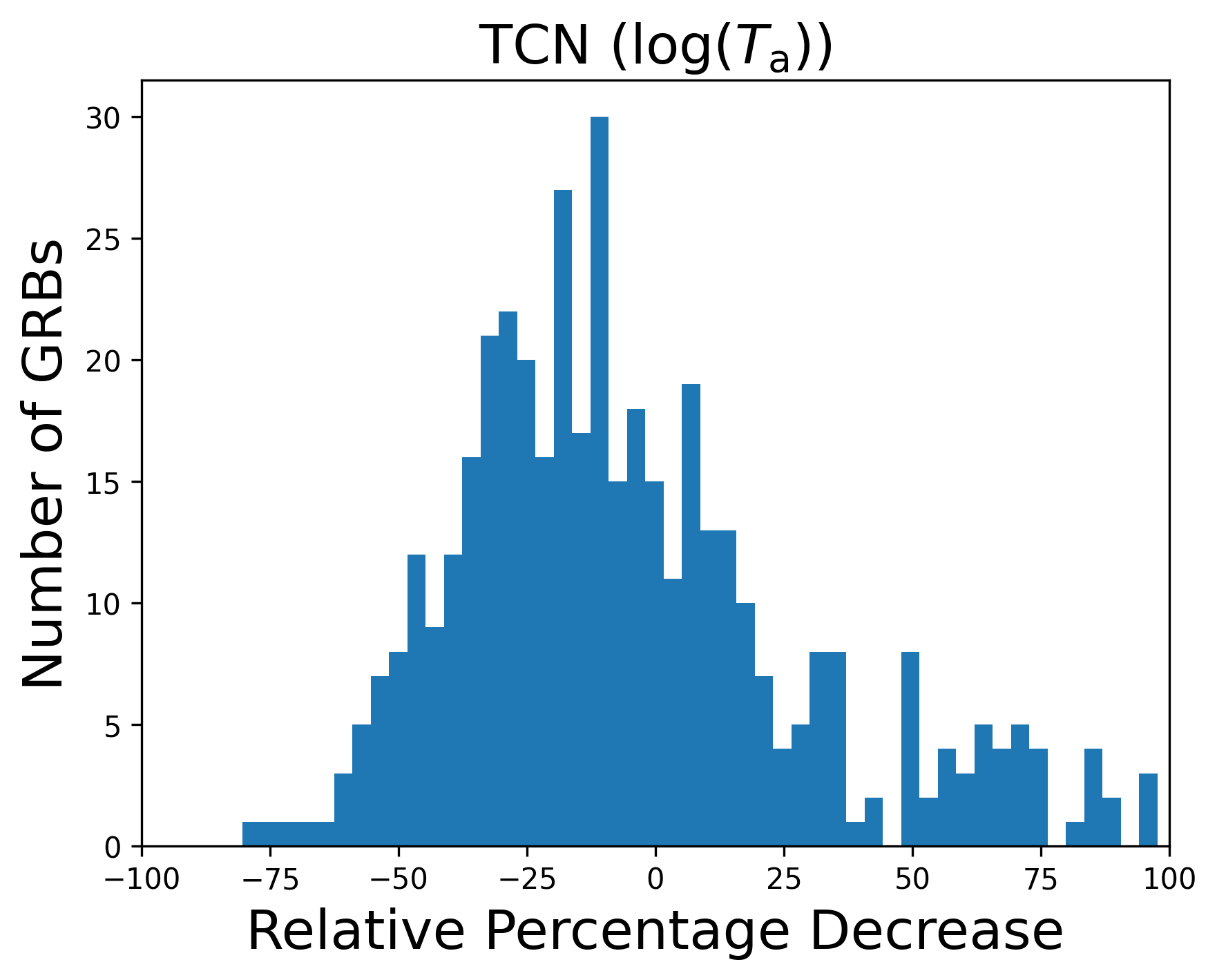}
\includegraphics[width=.25\textwidth, height=.17\textwidth]{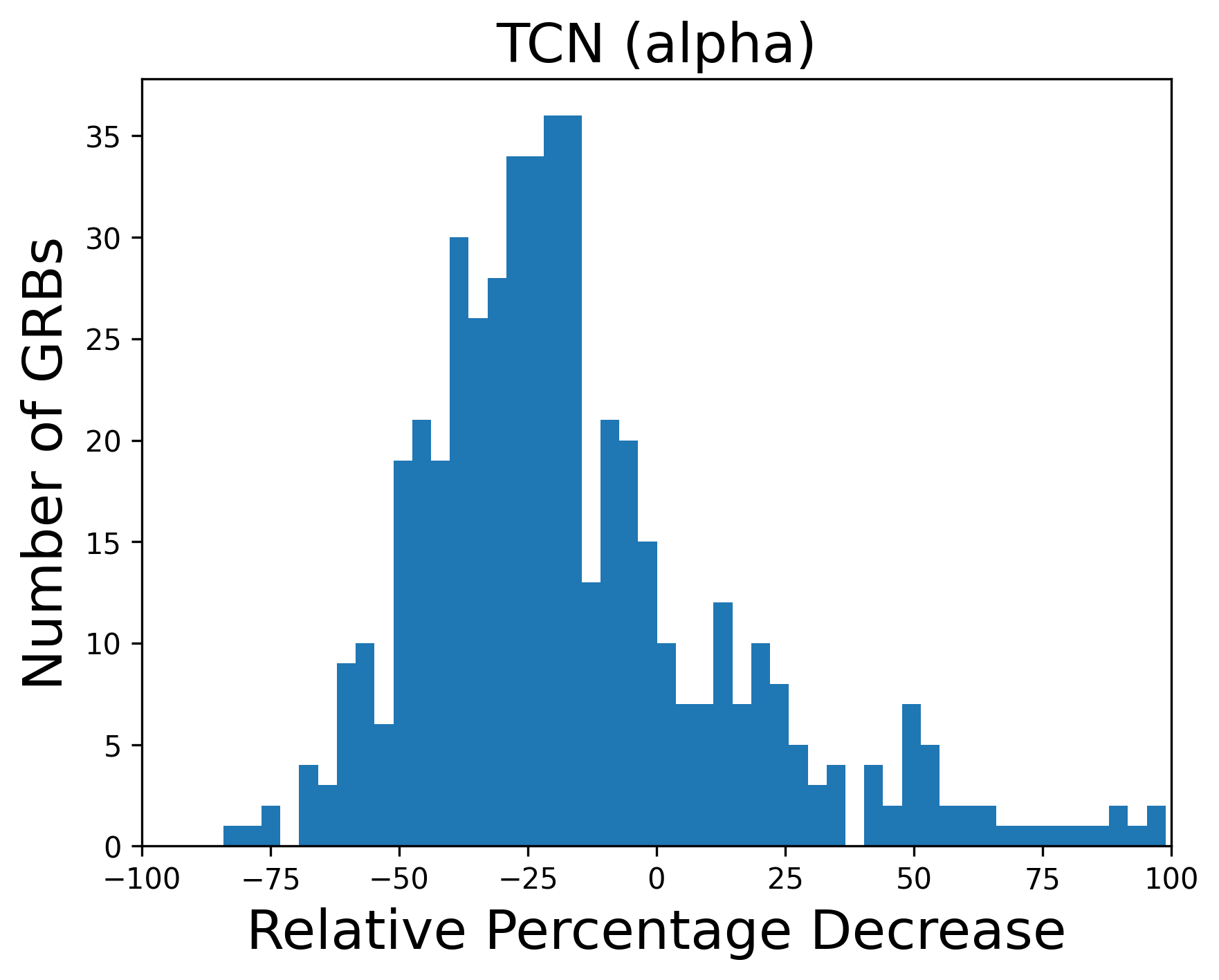}

%CNN-LSTM
\includegraphics[width=.25\textwidth, height=.17\textwidth]{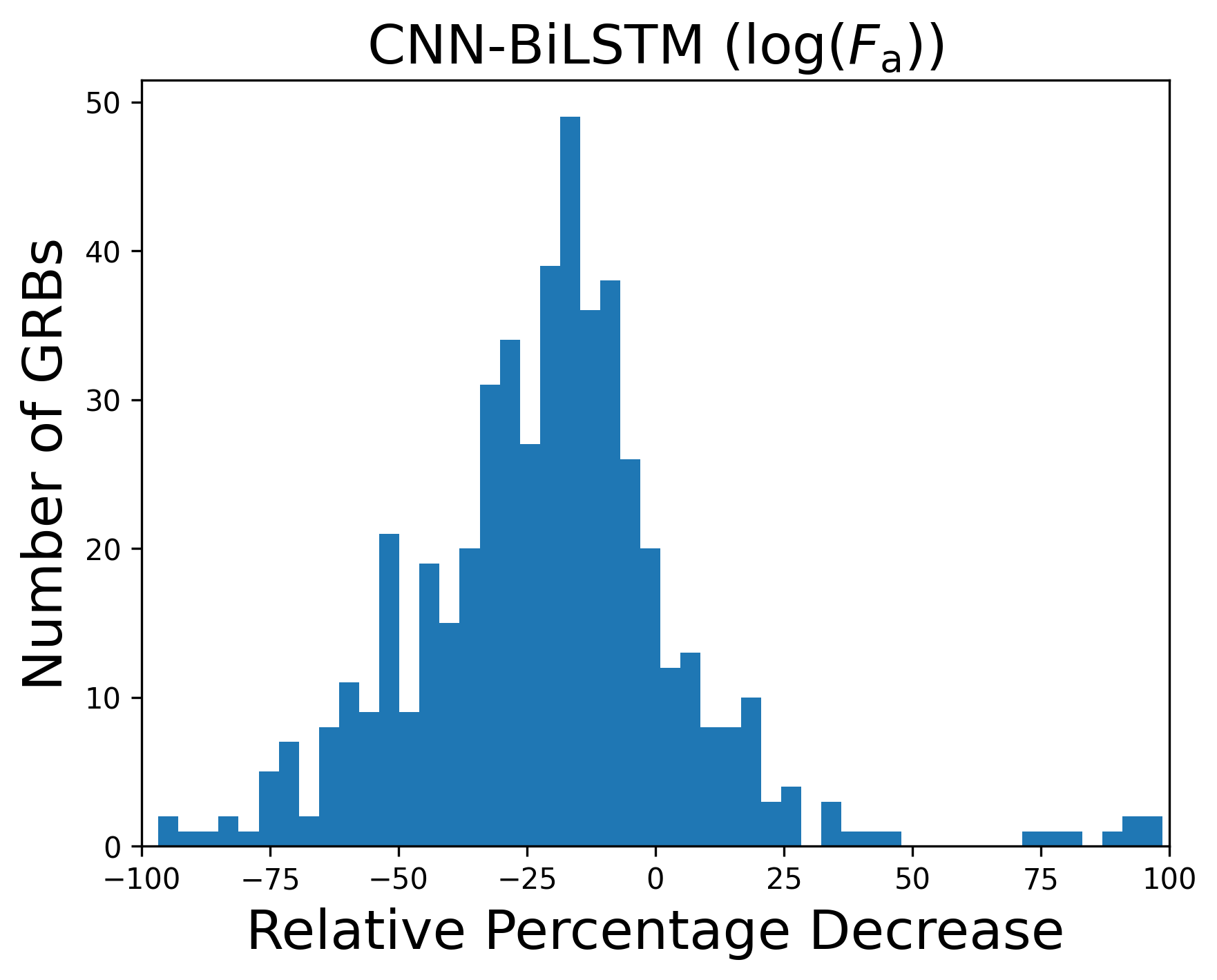}
\includegraphics[width=.25\textwidth, height=.17\textwidth]{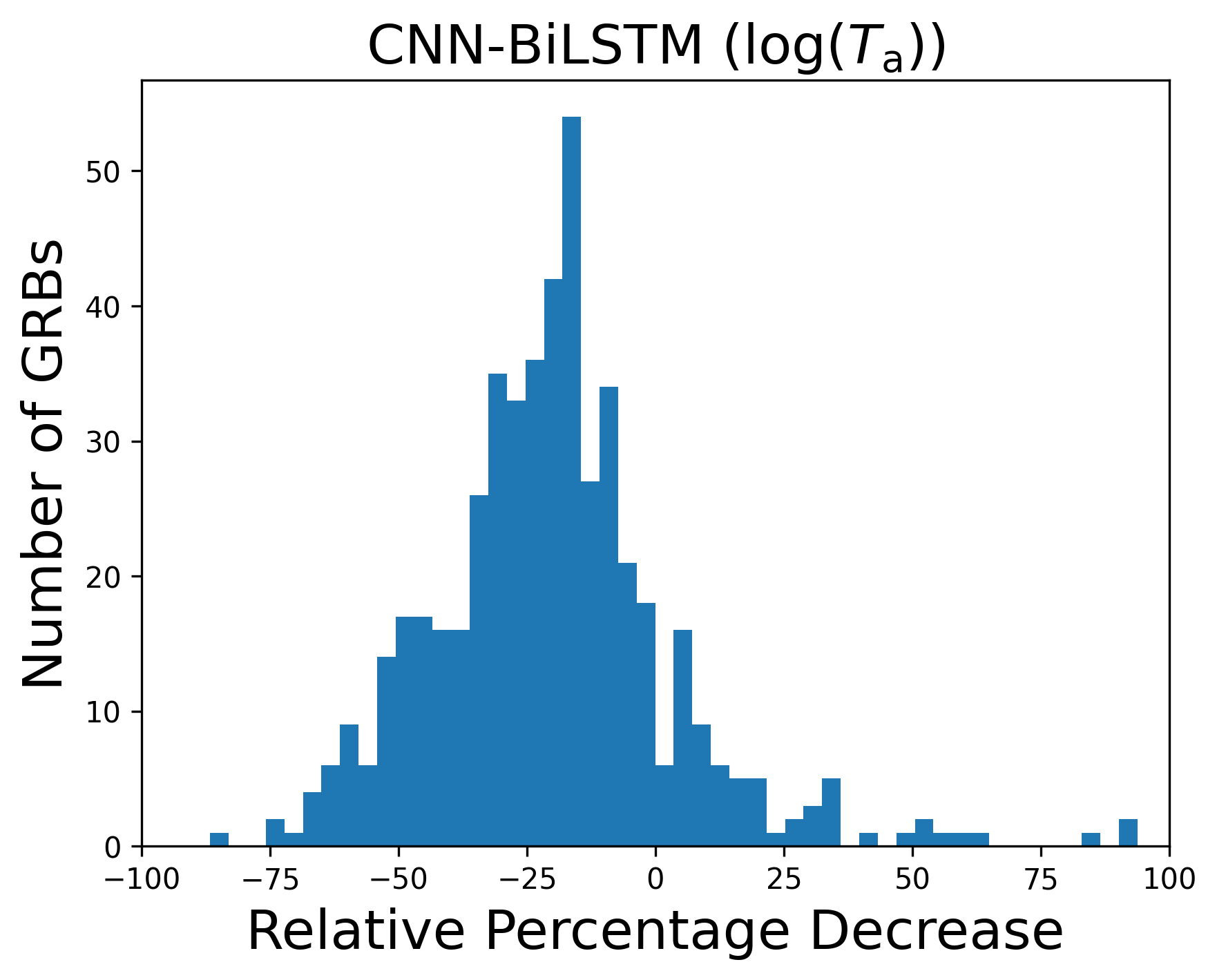}
\includegraphics[width=.25\textwidth, height=.17\textwidth]{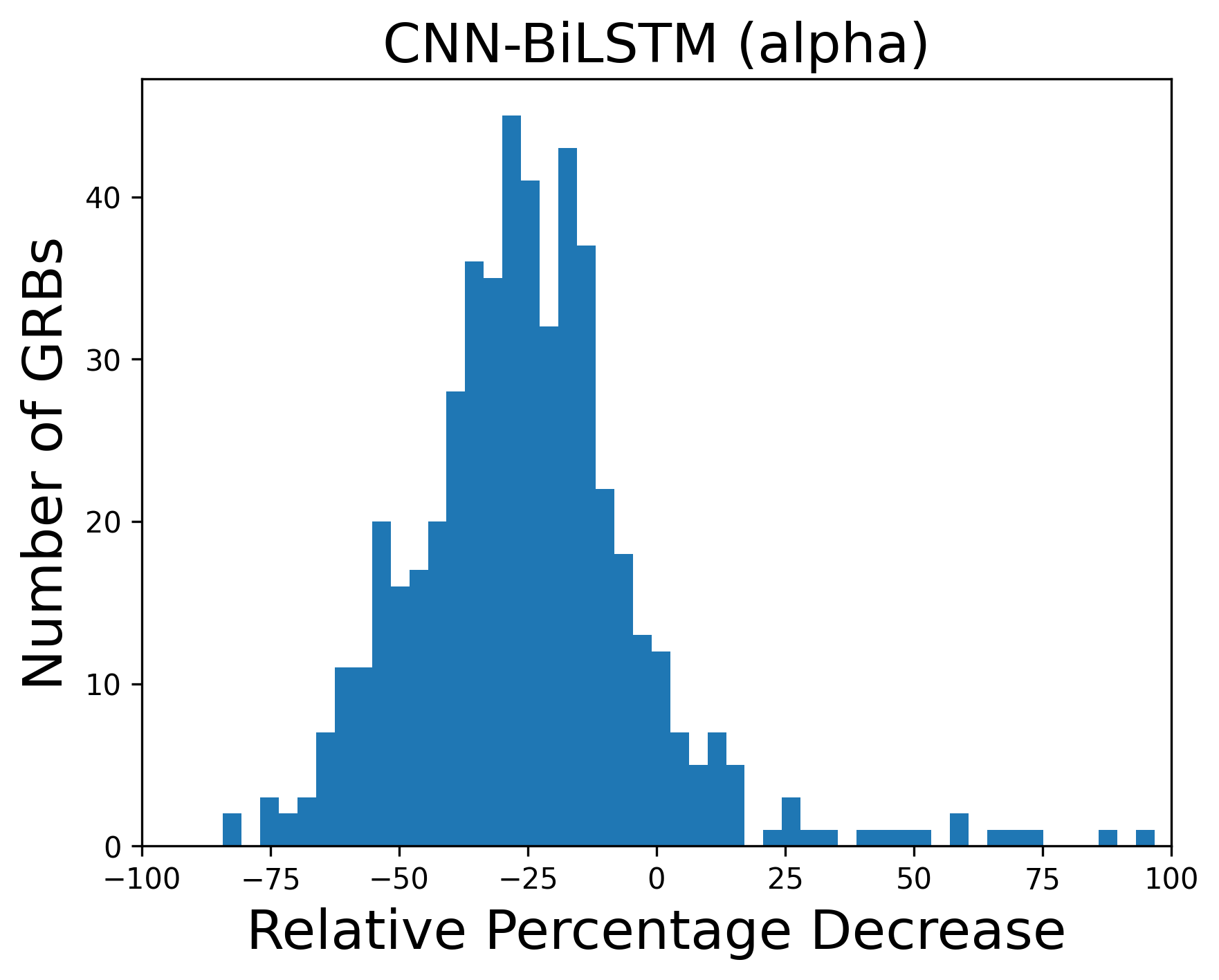}

%BNN-LCR
\includegraphics[width=.25\textwidth, height=.17\textwidth]{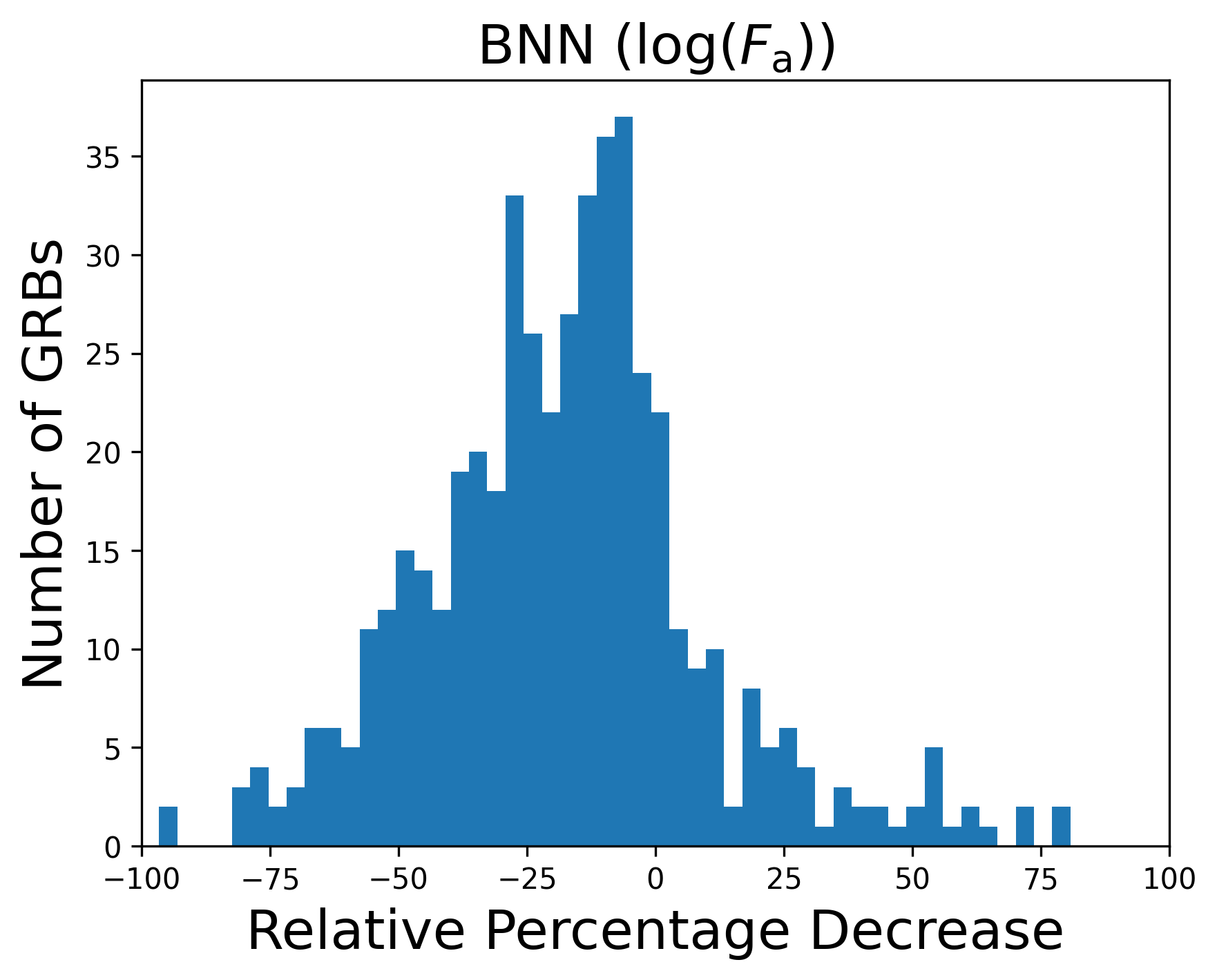}
\includegraphics[width=.25\textwidth, height=.17\textwidth]{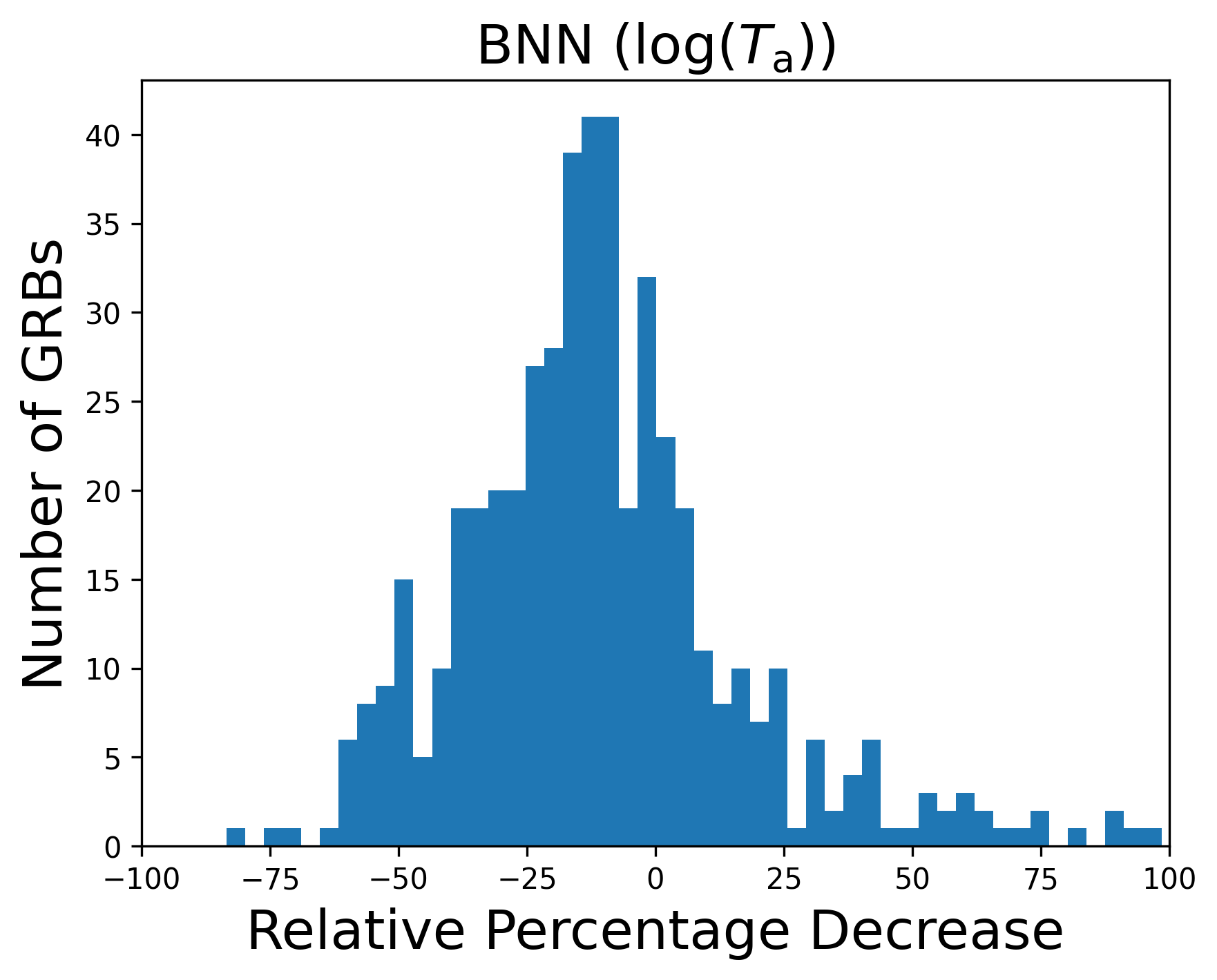}
\includegraphics[width=.25\textwidth, height=.17\textwidth]{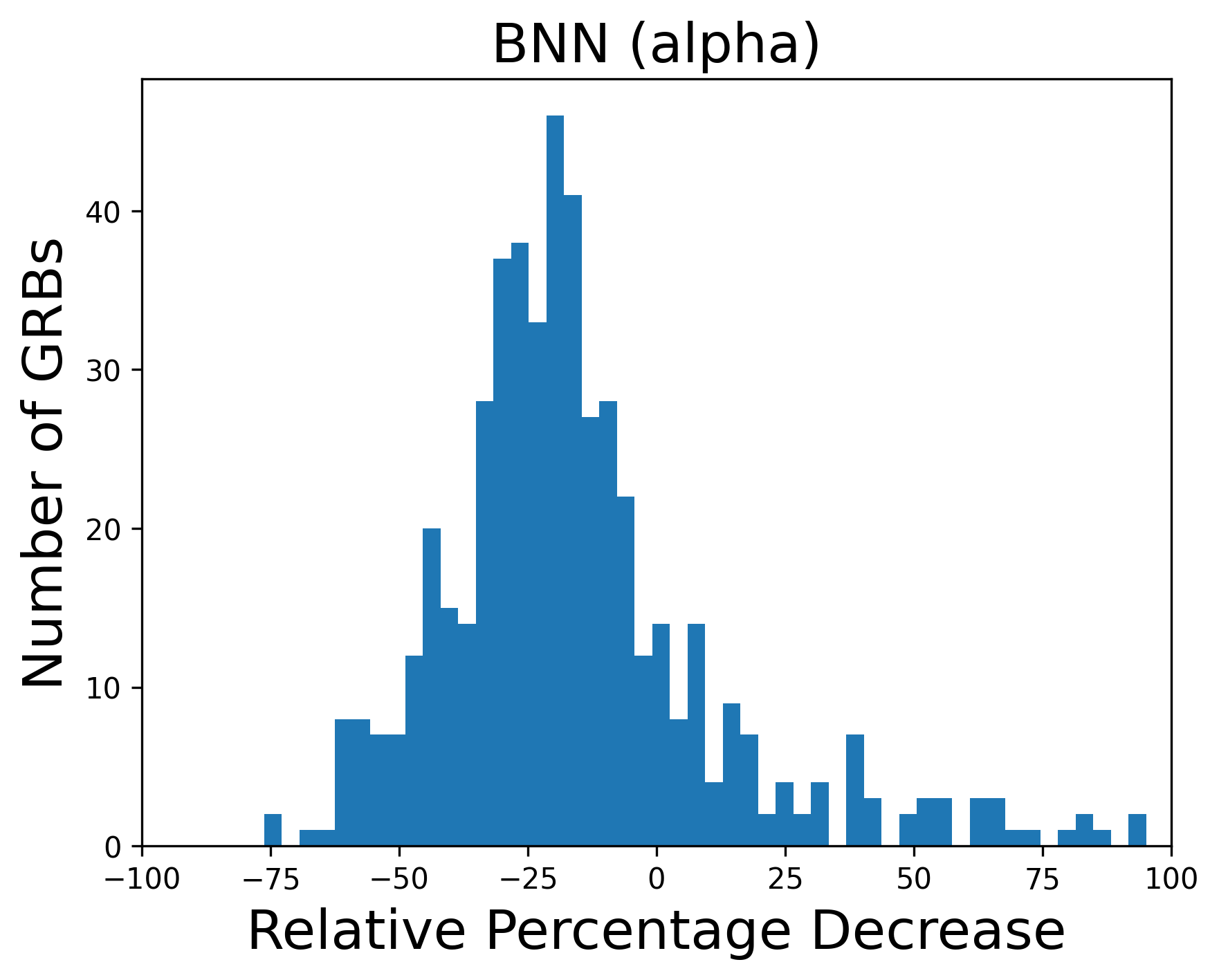}

%polynomial Curve Fitting
\includegraphics[width=.25\textwidth, height=.17\textwidth]{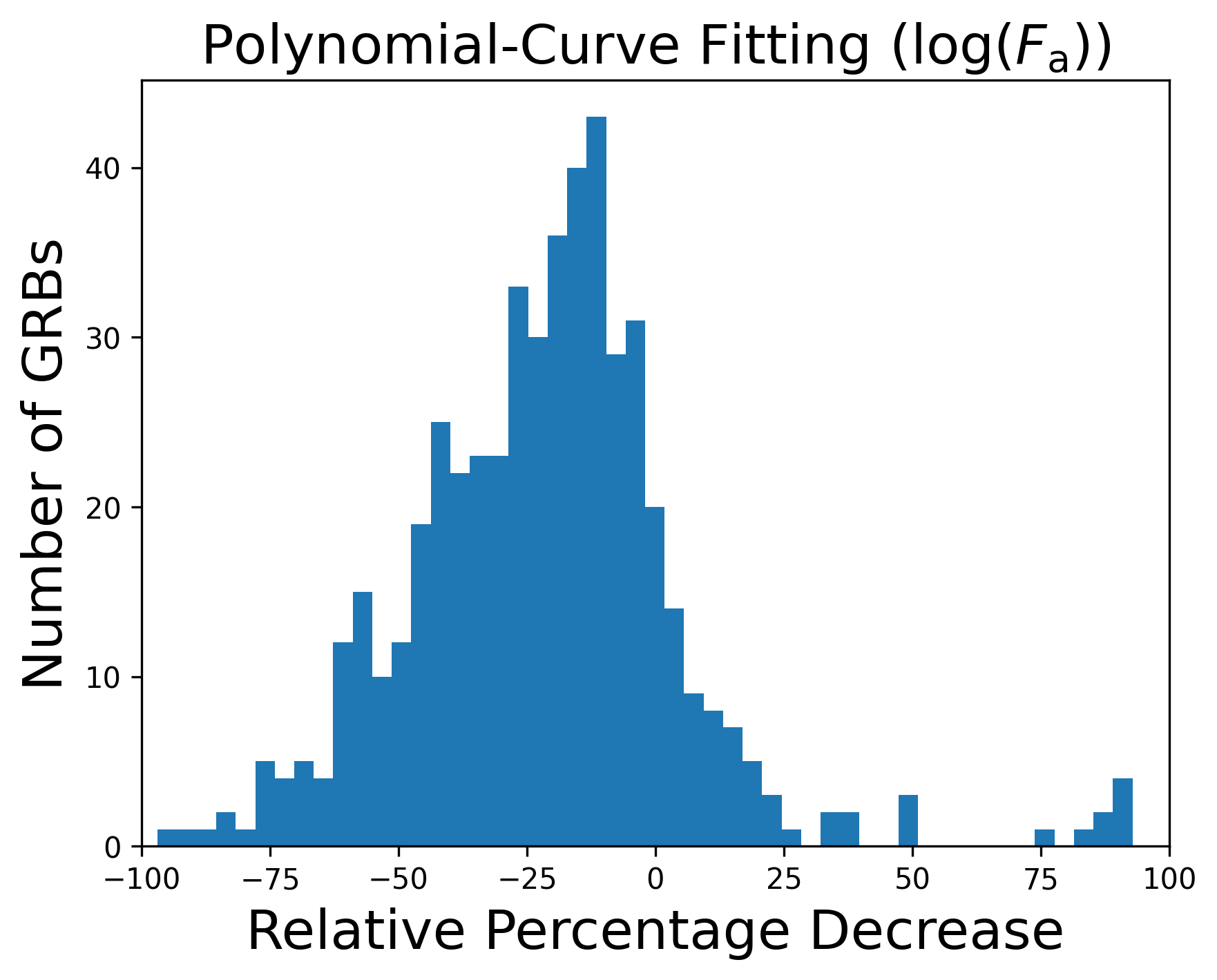}
\includegraphics[width=.25\textwidth, height=.17\textwidth]{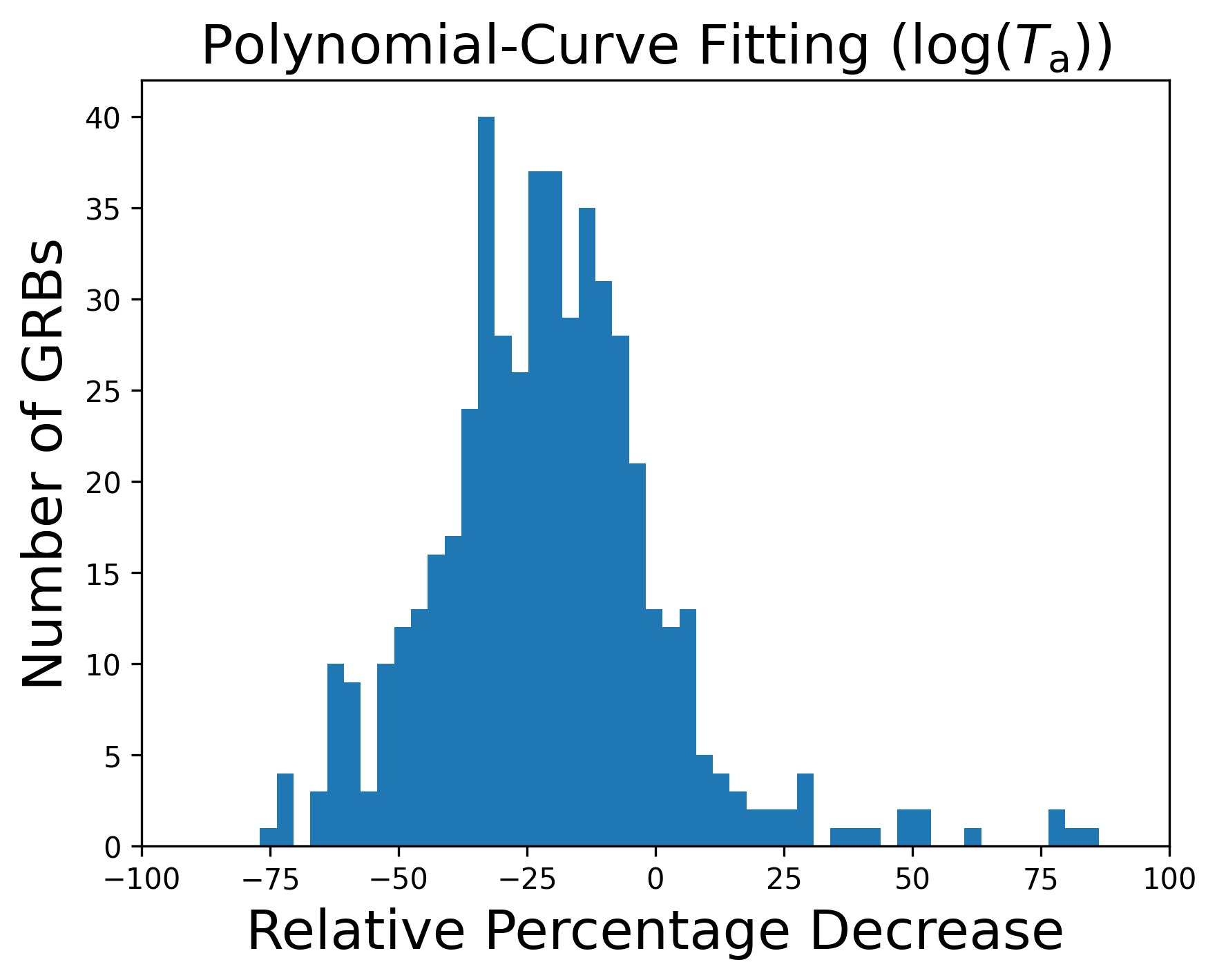}
\includegraphics[width=.25\textwidth, height=.17\textwidth]{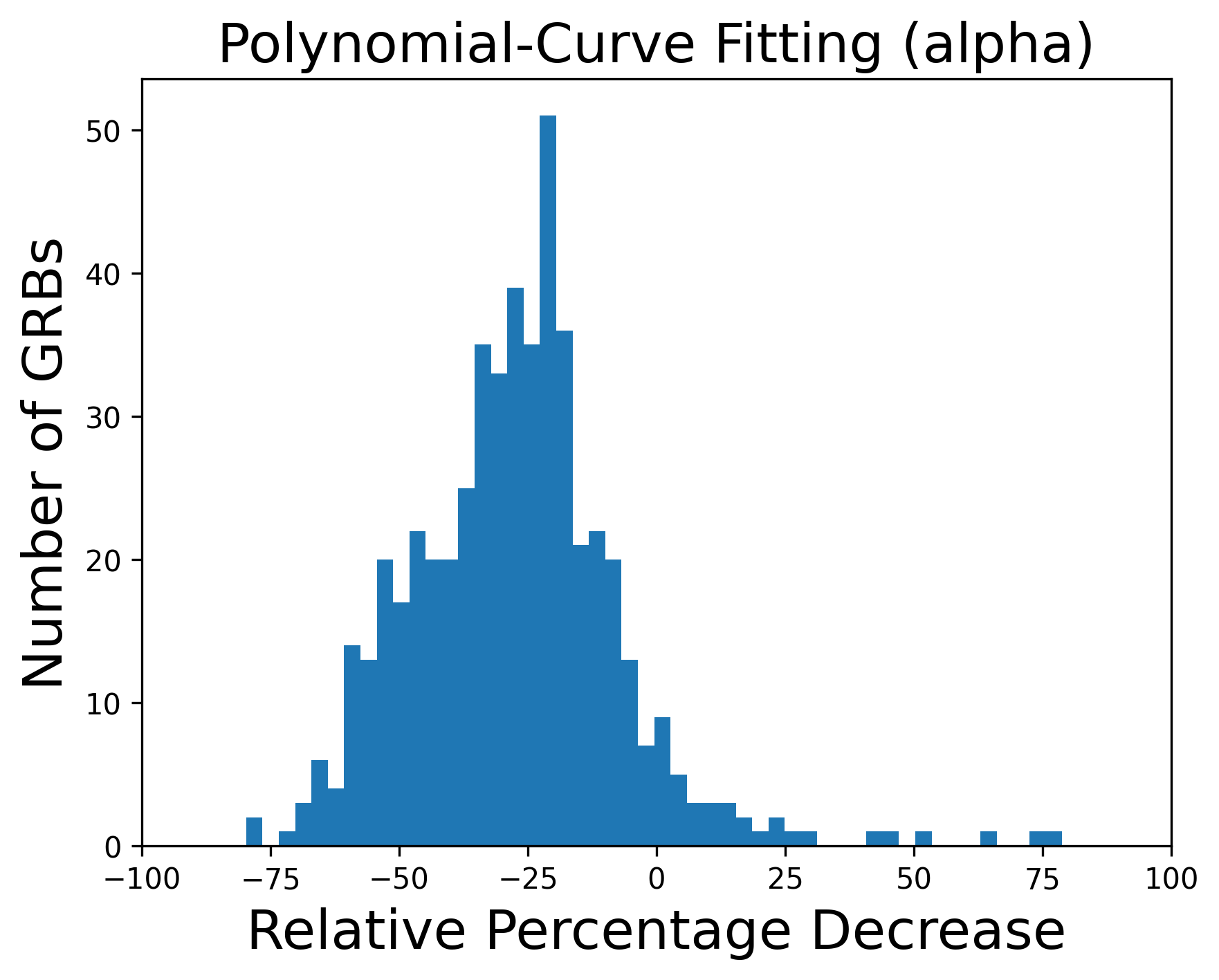}

%Isotonic
\includegraphics[width=.25\textwidth, height=.17\textwidth]{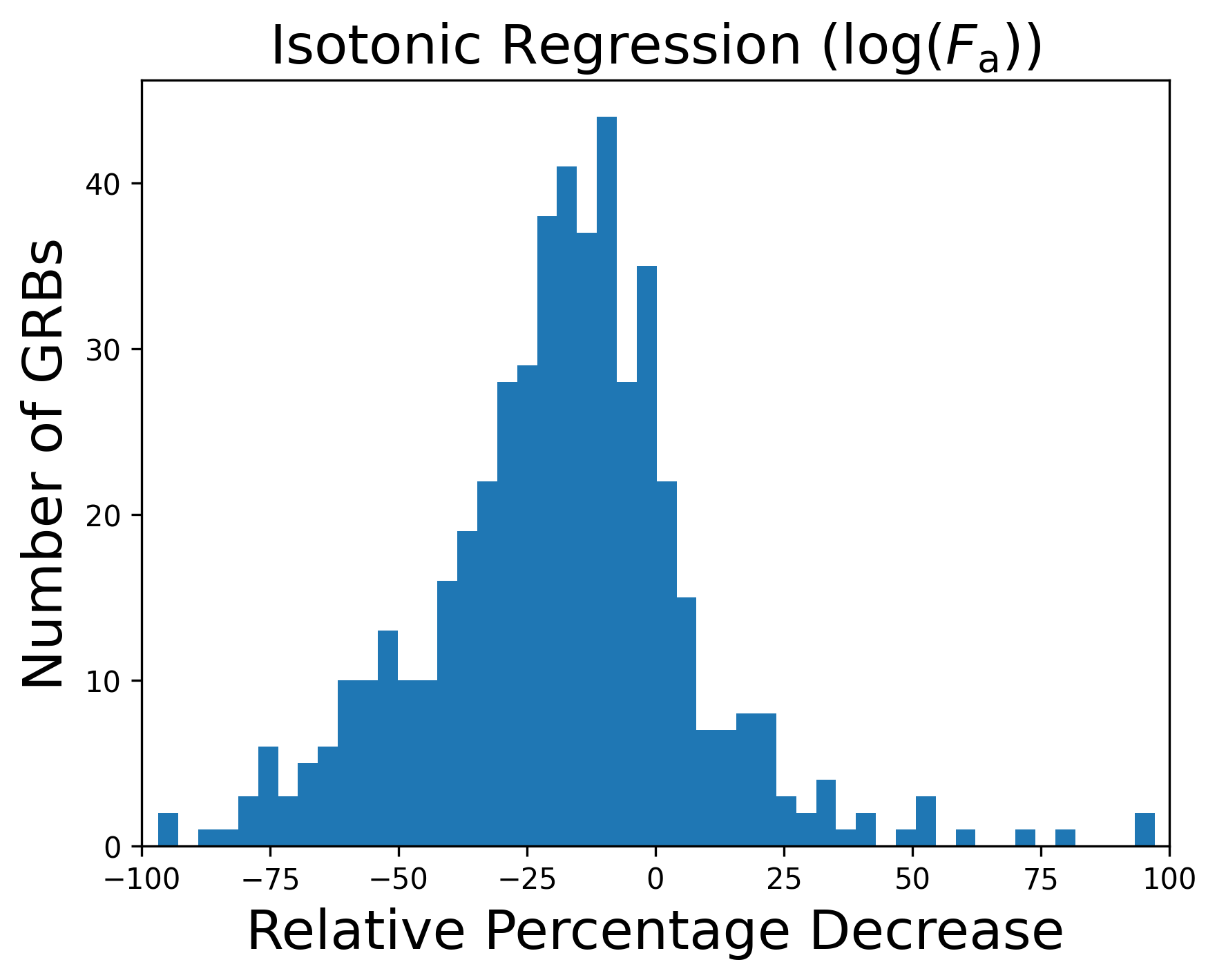}
\includegraphics[width=.25\textwidth, height=.17\textwidth]{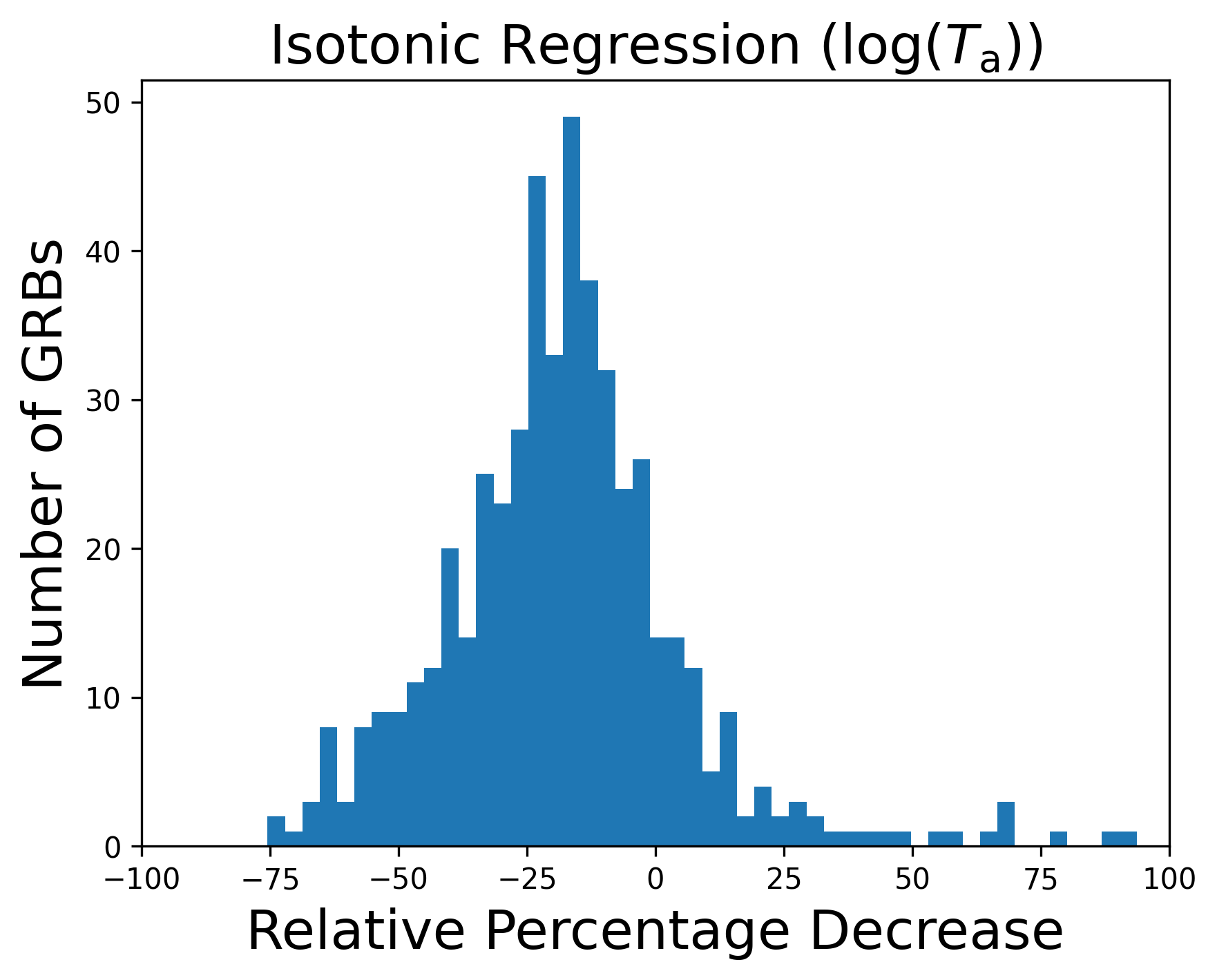}
\includegraphics[width=.25\textwidth, height=.17\textwidth]{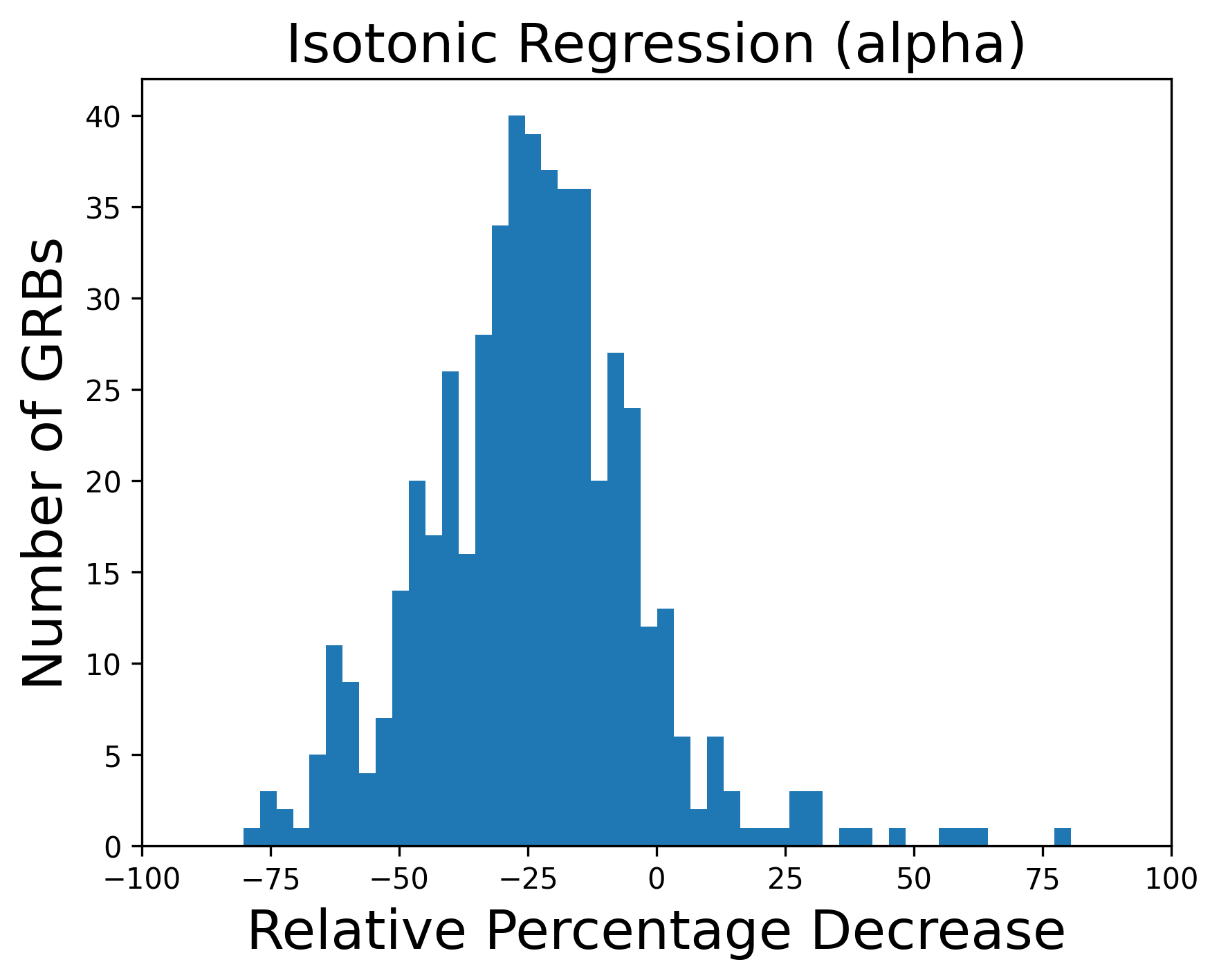}

%QSS
\includegraphics[width=.25\textwidth, height=.17\textwidth]{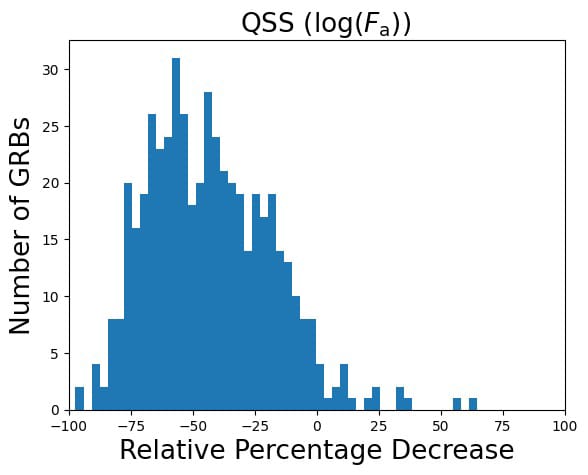}
\includegraphics[width=.25\textwidth, height=.17\textwidth]{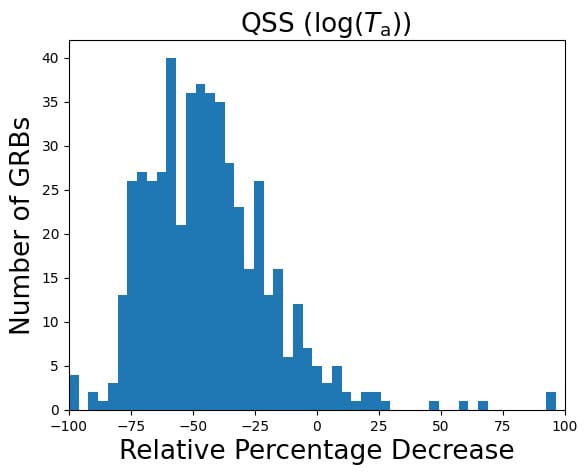}
\includegraphics[width=.25\textwidth, height=.17\textwidth]{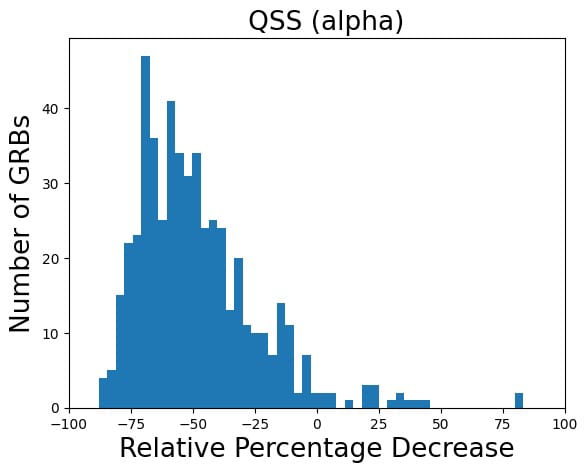}

\end{center}
    \caption{Distribution plot of all three W07 parameters in a grid with parameters (left to right): i) $\log F_a$ (Column 1) ii) $\log T_a$ (Column 2) iii) $\alpha$ (Column 3) and the models (top to bottom): i) Deep GP (Row 1); ii) TCN Model (Row 2); iii) CNN-BiLSTM Model (Row 3); iv) BNN Model (Row 4); v) Polynomial Curve Fitting Model (Row 5); vi) Isotonic Regression Model (Row 6); vii) QSS Model (Row 7)}
    \label{fig: ALL-results}
\end{figure*}

The reconstruction for each category of GRBs for all models is shown in Fig. \ref{fig: ALL-reconstruction}, and the histogram distribution of the relative percentage decrease for the three parameters is illustrated in Fig. \ref{fig: ALL-results} and tabulated in Table \ref{Table2}.

Table \ref{tab:reconstruction_results} shows how effectively the models' reconstructed data perform in comparison to the initial observed data for all GRBs. Although the models perform well for most GRB reconstructions, a few outliers are observed in the W07 parameters. We classify outliers as instances where the relative percentage increase exceeds 100\%. Table \ref{tab:model_attributes} shows the performance strengths and weaknesses of each model.

\begin{table*}[htbp]
\centering
\scalebox{1.0}
\scriptsize
\setlength{\tabcolsep}{4pt}
\renewcommand{\arraystretch}{1.1}
\begin{tabular}{|l|c|c|c|c|c|c|c|c|c|}
\hline
\textbf{GRB ID} & $EF_{\log_{10}(T_i)}$ & $EF_{\log_{10}(F_i)}$ & $EF_{\alpha_i}$ & $EF_{\log_{10}(T_i)}$ RC & $EF_{\log_{10}(F_i)}$ RC & $EF_{\alpha_i}$ RC & \%$_{\log_{10}(T_i)}$ & \%$_{\log_{10}(F_i)}$ & \%$_{\alpha_i}$ \\
\hline
\multicolumn{10}{|c|}{\textbf{TCN}} \\
\hline
070110 & 0.0126 & 0.00496 & 0.0326 & 0.00742 & 0.00299 & 0.0172 & -41.3 & -39.6 & -47.3 \\
070208 & 0.0410 & 0.0149 & 0.0524 & 0.0183 & 0.00395 & 0.0525 & -55.4 & -73.5 & 0.130 \\
070223 & 0.0481 & 0.0140 & 0.0800 & 0.119 & 0.0424 & 0.0752 & 147 & 203 & -5.96 \\
070306 & 0.00580 & 0.00297 & 0.0326 & 0.0129 & 0.00598 & 0.0606 & 123 & 101 & 86.0 \\
% 070517 & 0.0386 & 0.0113 & 0.140 & 0.0215 & 0.00555 & 0.0710 & -44.4 & -50.9 & -49.1 \\
070521 & 0.0136 & 0.00490 & 0.0389 & 0.0181 & 0.00763 & 0.0583 & 32.9 & 55.9 & 49.8 \\
070529 & 0.0220 & 0.00539 & 0.0288 & 0.0260 & 0.00565 & 0.0286 & 18.4 & 4.81 & -0.730 \\
\hline
\multicolumn{10}{|c|}{\textbf{CNN-BiLSTM}} \\
\hline
070110 & 0.0126 & 0.00496 & 0.0326 & 0.0119 & 0.00447 & 0.0255 & -5.78 & -9.71 & -21.8 \\
070208 & 0.0410 & 0.0149 & 0.0524 & 0.0126 & 0.00372 & 0.0444 & -69.3 & -75.1 & -15.3 \\
070223 & 0.0481 & 0.0140 & 0.0800 & 0.0263 & 0.00782 & 0.0495 & -45.3 & -44.1 & -38.1 \\
070306 & 0.00580 & 0.00297 & 0.0326 & 0.00576 & 0.00282 & 0.0282 & -0.700 & -5.20 & -13.4 \\
% 070517 & 0.0386 & 0.0113 & 0.140 & 0.0169 & 0.00484 & 0.0604 & -56.1 & -57.1 & -56.7 \\
070521 & 0.0136 & 0.00490 & 0.0389 & 0.00941 & 0.00375 & 0.0257 & -31.0 & -23.3 & -34.0 \\
070529 & 0.0220 & 0.00539 & 0.0288 & 0.0186 & 0.00437 & 0.0212 & -15.6 & -18.9 & -26.4 \\
\hline
\multicolumn{10}{|c|}{\textbf{BNN}} \\
\hline
070110 & 0.0126 & 0.00496 & 0.0326 & 0.0113 & 0.00420 & 0.0237 & -10.8 & -15.3 & -27.2 \\
070208 & 0.0410 & 0.0149 & 0.0524 & 0.0249 & 0.00482 & 0.0805 & -39.3 & -67.6 & 53.5 \\
070223 & 0.0481 & 0.0140 & 0.0800 & 0.0289 & 0.00802 & 0.0450 & -40.0 & -42.8 & -43.7 \\
070306 & 0.00580 & 0.00297 & 0.0326 & 0.00505 & 0.00237 & 0.0241 & -12.9 & -20.1 & -26.1 \\
% 070517 & 0.0386 & 0.0113 & 0.140 & 0.0239 & 0.00506 & 0.0729 & -38.0 & -55.1 & -47.8 \\
070521 & 0.0136 & 0.00490 & 0.0389 & 0.0124 & 0.00484 & 0.0386 & -9.24 & -1.22 & -0.660 \\
070529 & 0.0220 & 0.00539 & 0.0288 & 0.0240 & 0.00534 & 0.0321 & 9.03 & -0.870 & 11.2 \\
\hline
\multicolumn{10}{|c|}{\textbf{Polynomial Curve Fitting}} \\
\hline
070110 & 0.0126 & 0.00496 & 0.0326 & 0.0113 & 0.00432 & 0.0234 & -10.8 & -12.9 & -28.2 \\
070208 & 0.0410 & 0.0149 & 0.0524 & 0.0151 & 0.00461 & 0.0612 & -63.1 & -69.1 & 16.8 \\
070223 & 0.0481 & 0.0140 & 0.0800 & 0.0335 & 0.00877 & 0.0573 & -30.4 & -37.4 & -28.3 \\
070306 & 0.00580 & 0.00297 & 0.0326 & 0.00530 & 0.00245 & 0.0250 & -8.57 & -17.4 & -23.3 \\
% 070517 & 0.0386 & 0.0113 & 0.140 & 0.0185 & 0.00522 & 0.0659 & -52.0 & -53.7 & -52.8 \\
070521 & 0.0136 & 0.00490 & 0.0389 & 0.00938 & 0.00372 & 0.0255 & -31.2 & -24.1 & -34.5 \\
070529 & 0.0220 & 0.00539 & 0.0288 & 0.0187 & 0.00439 & 0.0218 & -14.8 & -18.5 & -24.3 \\
\hline
\multicolumn{10}{|c|}{\textbf{Isotonic Regression}} \\
\hline
070110 & 0.0126 & 0.00496 & 0.0326 & 0.0125 & 0.00482 & 0.0254 & -1.17 & -2.66 & -22.2 \\
070208 & 0.0410 & 0.0149 & 0.0524 & 0.0115 & 0.00357 & 0.0454 & -72.0 & -76.0 & -13.5 \\
070223 & 0.0481 & 0.0140 & 0.0800 & 0.0237 & 0.00670 & 0.0471 & -50.6 & -52.2 & -41.1 \\
070306 & 0.00580 & 0.00297 & 0.0326 & 0.00521 & 0.00247 & 0.0245 & -10.2 & -16.8 & -24.8 \\
% 070517 & 0.0386 & 0.0113 & 0.140 & 0.0171 & 0.00544 & 0.0644 & -55.7 & -51.9 & -53.8 \\
070521 & 0.0136 & 0.00490 & 0.0389 & 0.00976 & 0.00418 & 0.0286 & -28.4 & -14.5 & -26.4 \\
070529 & 0.0220 & 0.00539 & 0.0288 & 0.0183 & 0.00437 & 0.0193 & -16.7 & -18.9 & -32.9 \\
\hline
\multicolumn{10}{|c|}{\textbf{DGP}} \\
\hline
070110 & 0.0126 & 0.00496 & 0.0326 & 0.0126 & 0.00475 & 0.0255 & -0.770 & -4.18 & -21.9 \\
070208 & 0.0410 & 0.0149 & 0.0524 & 0.0160 & 0.00421 & 0.0518 & -60.9 & -71.8 & -1.25 \\
070223 & 0.0481 & 0.0140 & 0.0800 & 0.0323 & 0.00865 & 0.0515 & -32.8 & -38.2 & -35.6 \\
070306 & 0.00580 & 0.00297 & 0.0326 & 0.00500 & 0.00235 & 0.0236 & -13.8 & -20.9 & -27.4 \\
% 070517 & 0.0386 & 0.0113 & 0.140 & 0.0208 & 0.00634 & 0.0745 & -46.1 & -43.8 & -46.6 \\
070521 & 0.0136 & 0.00490 & 0.0389 & 0.00940 & 0.00360 & 0.0249 & -31.1 & -26.4 & -36.1 \\
070529 & 0.0220 & 0.00539 & 0.0288 & 0.0217 & 0.00509 & 0.0287 & -1.51 & -5.52 & -0.610 \\
\hline
\multicolumn{10}{|c|}{\textbf{QSS}} \\
\hline
070110 & 0.0126 & 0.00496 & 0.0326 & 0.00961 & 0.00364 & 0.0207 & -24.0 & -26.6 & -36.7 \\
070208 & 0.0410 & 0.0149 & 0.0527 & 0.0109 & 0.00322 & 0.0395 & -73.2 & -78.4 & -24.6 \\
070223 & 0.0481 & 0.0140 & 0.0800 & 0.0106 & 0.00312 & 0.0121 & -78.0 & -77.7 & -84.8 \\
070306 & 0.00580 & 0.00297 & 0.0326 & 0.00376 & 0.00170 & 0.0155 & -35.2 & -43.0 & -52.5 \\
070521 & 0.0136 & 0.00490 & 0.0389 & 0.00673 & 0.00278 & 0.0176 & -50.6 & -43.2 & -54.8 \\
070529 & 0.0220 & 0.00539 & 0.0288 & 0.00901 & 0.00229 & 0.00850 & -59.1 & -57.6 & -70.5 \\
\hline
\end{tabular}
\caption{Comparison of EF and RC values under different methods and noise levels.}\label{Table2}
\end{table*}

\begin{table*}[htbp] 
\begin{center}
\begin{tabular}{|l|c|c|c|c|c|c|}
\hline 
\textbf{Reconstruction Model} & \multicolumn{3}{|c|}{\textbf{Uncertainty Decrease}} & \multicolumn{3}{|c|}{\textbf{\% Outliers}} \\
\hline
 & \% log$_{10}$(T$_a$) & \% log$_{10}$(F$_a$) & \% $\alpha$ & \% logT$_a$ & \% logF$_a$ & \% $\alpha$ \\
\hline
\multicolumn{7}{|c|}{\textbf{521 GRBs}} \\
\hline
QSS & \textbf{-43.5} & \textbf{-43.2} & \textbf{-48.3} & 2.69 & 3.84 &\textbf{0.960} \\
Polynomial Curve Fitting & -20.8 & -21.6 & -27.4 & 2.88 & 2.30 & 1.15 \\
CNN-BiLSTM          & -20.3 & -20.9 & -25.1 & 3.07 & \textbf{2.69} & 0.768 \\
Isotonic Regression & -18.0 & -18.5 & -24.0 & 3.07 & 2.88 & \textbf{0.960} \\
BNN              & -10.9 & -17.6 & -15.9 & 5.95 & 5.76 & 2.69 \\
DGP             & -11.6 & -12.3 & -15.9 & 6.91 & 5.95 & 1.92 \\
TCN             & -5.31 & -12.7 & -16.2 & 18.5 & 14.6 & 4.62 \\

\hline
Attention U-Net  & -37.9 & -38.5 & -41.4 & \textbf{1.73} & 2.50 & 1.34 \\
MLP              & -37.2 & -38.0 & -41.2 & \textbf{1.73} & 2.30 & 1.34 \\
GP (W07)         & -16.9 & -18.6 & -24.3 & 3.07 & 3.45 & 1.54 \\
W07 model (10\%) & -18.0 & -19.1 & -25.2 & 2.30 & 2.30 & 2.11 \\
W07 model (20\%) & -15.8 & -17.8 & -23.8 & 2.30 & 2.69 & 2.30 \\

\hline
\multicolumn{7}{|c|}{\textbf{207 Good GRBs \citep{dainotti2023stochastic}}} \\
\hline
QSS & \textbf{-48.0} & \textbf{-48.8} & \textbf{-55.1} & \textbf{0} & 0.483 & \textbf{0} \\
CNN-BiLSTM & -23.1 & -25.2 & -29.1 & 1.93 & 1.93 & \textbf{0} \\
Polynomial Curve Fitting & -21.0 & -24.0 & -30.2 & \textbf{0} & \textbf{0} & \textbf{0} \\
Isotonic Regression & -19.4 & -21.6 & -28.0 & 0.966 & 0.966 & 0.483 \\
DGP & -14.8 & -16.5 & -20.7 & 4.35 & 3.86 & 0.966 \\
BNN & -11.2 & -19.7 & -19.1 & 2.42 & 2.90 & 0.483 \\
TCN & -3.56 & -16.4 & -16.9 & 21.7 & 18.8 & 4.83 \\

\hline
% Bi-Mamba         & -22.5 & -24.4 & -31.2 & 0.483 & 0.483 & 0 \\
Attention U-Net  & -38.8 & -40.3 & -44.0 & 0.483 & 0.483 & \textbf{0} \\
MLP              & -38.1 & -39.3 & -43.9 & 0.483 & 0.483 & \textbf{0} \\
GP (W07)         & -17.2 & -20.0 & -28.8 & 1.93 & 1.45 & \textbf{0} \\
W07 model (10\%) & -23.0 & -24.9 & -30.8 & 0.966 & 0.966 & \textbf{0} \\ 
W07 model (20\%) & -21.2 & -23.2 & -28.9 & \textbf{0} & \textbf{0} & \textbf{0} \\
\hline

\end{tabular}
\caption{Summary of average uncertainty decrease and outliers for different reconstruction models on 521 GRBs and the 207 good GRBs subset.}
\label{tab:reconstruction_results}
\end{center}
\end{table*}

%NEW TABLE
 % in preamble

\begin{table*}[htbp]
\centering
\resizebox{\textwidth}{!}{%
\begin{tabular}{|c|c|c|c|c|c|c|c|}
    \hline
    \textbf{Attributes} & \textbf{QSS} & \textbf{Polynomial Curve Fitting} & \textbf{CNN-BiLSTM} & \textbf{DGP} & \textbf{BNN} & \textbf{TCN} & \textbf{Isotonic Regression} \\ \hline
    High Uncertainty Reduction & \checkmark & \checkmark & & & & & \\ \hline
    Good Sparse Data Performance & \checkmark & \checkmark &  & \checkmark & \checkmark &  & \checkmark \\ \hline
    Almost Zero Outliers for Good GRBs & \checkmark & \checkmark & & & & & \\ \hline
    Low Computational Cost & \checkmark & \checkmark & & & & & \checkmark \\ \hline
    % Resilience to Noise Overfitting & \checkmark & \checkmark & \checkmark & \checkmark & \checkmark & \checkmark & \checkmark \\ \hline
    Captures Complex Patterns & \checkmark &  & \checkmark & \checkmark & \checkmark & \checkmark &  \\ \hline
    % Models Data Distribution & \checkmark & \checkmark & \checkmark & \checkmark & \checkmark & \checkmark & \checkmark \\ \hline
\end{tabular}
}
\caption{Comparison of the models presented here on various attributes.}
\label{tab:model_attributes}
\end{table*}

 %As shown in Table \ref{tab:reconstruction_results}, the CNN-LSTM-LCR model yields the least number of outliers for both the $\log(T_a)$ and $\log(F_a)$ parameters, at 2.57\% and 2.94\%, respectively. For the $\alpha$ parameter, the CNN-LSTM-LCR model again performs best, with only 0.55\% of the values classified as outliers.

\section{Summary and Conclusion}\label{section:conclusion}

We showcase the efficacy of the proposed methods in modeling LCs and bridging the temporal gaps in the GRB LCs. The main conclusions of our studies are outlined below:

\begin{itemize}

\item The QSS model achieves outstanding reductions in uncertainty: 43.5\% for $\log T_a$, 43.2\% for $\log F_a$, and 48.3\% for $\alpha$. QSS's reduction is almost double the reductions obtained by the other five models presented in this article. These reductions are the lowest we have achieved, particularly for $\alpha$, where it outperformed several deep learning models. For Good GRBs, it has strong reductions of 48.0\%, 48.8\%, and 55.1\% for $\log(T_a)$, $\log(F_a)$, and $\alpha$ respectively. It maintains zero outliers across $\log(T_a)$ and $\alpha$ parameters for Good GRBs. QSS remains a satisfactory baseline, particularly strong for the $\alpha$ parameter.

\item The Polynomial Curve Fitting model achieves good reductions in uncertainty - $20.8\%$ for $\log T_a$, $21.6\%$ for $\log F_a$, and $27.4\%$ for $\alpha$. These reductions are the highest among our newly introduced models, particularly for $\alpha$, where it outperformed several deep learning models. It maintains zero outliers across all three parameters for Good GRBs. Polynomial Curve Fitting model is the fourth top-performing baseline, among all the models we have tried so far. This model's performance is comparable to the Bi-Mamba model.

\item The CNN-BiLSTM model achieves moderate performance across all parameters, 20.3\%, 20.9\%, and 25.1\% for $\log(T_a)$, $\log(F_a)$, and $\alpha$ respectively. It maintains zero outliers for the $\alpha$ parameter for Good GRBs. This model shows good reduction across all GRB parameters with the lowest outlier rate for $\alpha$ (0.768\%). Similarly, the DGP model demonstrates uncertainty reduction of $11.6\%$ for $\log T_a$, $12.6\%$ for $\log F_a$, and $15.9\%$ for $\alpha$ parameters.

\item The Isotonic Regression models lowers uncertainty by $18.0\%$ for $\log T_a$, $18.5\%$ for $\log F_a$, and $24.0\%$ for $\alpha$. The BNN model showcases moderate reductions of 10.9\%, 17.6\%, and 15.9\% for $\log(T_a)$, $\log(F_a)$, and $\alpha$ respectively. While the TCN model achieves reductions of $5.31\%$ for $\log T_a$, $12.7\%$ for $\log F_a$, and $16.2\%$ for $\alpha$.

\item Most of the outliers found in the study are GRBs which have at least two of the variations–flares, bumps, breaks, and double breaks. Very few Good GRBs (0-4\% across the three parameters) are present in the outliers, as demonstrated in the table featuring Good GRB outlier rates in the paper. This concludes that the models perform better on Good GRBs as compared to the other three types.

\item The platinum sample, composed of 50 GRBs from \cite{dainotti2020a} is benefitting the most from this analysis and it will be then used for theoretical model discrimination and as a cosmological tool with the advantage of having a decreased uncertainties on the cosmological parameters as it was discussed in \cite{dainotti2022g} where a reduction of 47.5\% on the uncertainties parameters allow to reach the precision of SNe Ia in \citep{Conley2011} already now and the one reached by \citep{Betoule} in 17 years from now. So this reduced percentages allow to reach these goals in a few years less.
% and had an average test MSE value of 0.215.    
% \item  It worsens upon the single-layer GP baseline, as well as the GP-RF model. The comparison is shown in Table \ref{tab:dgp_gp}. This highlights that additional layers do not improve upon the basic GP architecture.

% \item The TCN model reduces the W07 parameters by the lowest percentage for the full dataset as well as the 207 Good GRBs, achieving $5.31\%$ for $\log T_a$, $12.7\%$ for $\log F_a$, and $16.2\%$ for $\alpha$.

We consider these $logT_a$, $logF_a$, and $\alpha$ as the most important outputs of our reconstruction. The parameter $\alpha$ describes the post-plateau decay behavior, relevant for the testing of the standard fireball model \citep{piran1999gamma}, and the reduction of the $\alpha$ parameter is crucial to obtain reduced uncertainties on the closure relationship \citep{dainotti2021closure, dainotti2024analysis}. The parameters $\log T_a$ and $\log F_a$ are key for identifying empirical correlations, such as the Dainotti relations in two and three dimensions \citep{Dainotti2008, dainotti2010a, dainotti2011a, dainotti2013determination, dainotti2016fundamental, dainotti2020a, dainotti2020b, dainotti2022b, levine2022}. Among all the models tested, QSS demonstrates the most significant reduction in error for all three parameters. Comparing it with our previous models, QSS persisted as the most effective reconstruction strategy. Thus, we recommend QSS as the preferred model when accurate estimation of these physically meaningful quantities is the main objective.

\end{itemize}

These reconstructions achieve a significant reduction in uncertainty, which improves the reliability of the GRB plateau parameters for both model estimator and for their use as standard candles. By achieving reduced uncertainty, the relationships associated with plateau emissions can be considered as standard candles for cosmological research \citep{dainotti2022g, dainotti2022c, dainotti2023b}.  Additionally, the precision in determining cosmological parameters has also been improved. The results of this work can approach the estimation accuracy of dark matter from Type Ia supernovae \citep{dainotti2022g, Betoule}. 

These methods can also be explored in future studies, including cutting-edge deep learning approaches such as neural ordinary differential equations (neural ODEs), Physics Informed Neural Network (PINN), and evolutionary algorithms. We analysed a time-series model; for details, refer to Appendix \ref{sec:appendixA}. Additionally, the potential of transformer architectures and large language models to capture complex temporal patterns warrants investigation. When combined with new observational data, it is anticipated that these methods will greatly improve LCR's accuracy and interpretability.

While significant uncertainty reduction was achieved on the parameters, the challenge of providing global influence in relation to GRBs as distributions still remains. Our current models exhibit limited global influence during parameter tuning
and do not account when GRBs are considered altogether, because we analyze GRBs singularly. 
The models used in this analysis
primarily function as curve-fitters with deep local influence, inferring/extracting local representations to fill LC gaps. The shared hyperparameters across GRBs contribute to a shared influence between GRBs; however, this influence is limited and warrants further exploration. Optimizing the hyperparameters for 16 GRBs, taken from a set of 4 GRBs belonging to  each class (Good, Flares/Bumps, Breaks, Flares with bumps + double breaks) and freezing these hyperparameters for all the 521 GRBs provides a first step towards a global framework for reconstructing the samples that is not GRB specific. Future work could build on this methodology to address sparsity and obtain consistent time-length samples. These reconstructed samples could then be fed into temporal aware ML/ DL methods to extract parameters across several GRBs. This will allow us to include global GRB parameters in the latent features, rather than fine tuning singular GRBs.
% Future work could extend our methodology to obtain consistent time-length samples with removed sparsity to incorporate this global influence and leverage multiple GRBs at the same time with their latent representations which would then improve reconstruction results.

Though we have only used Swift LCs to develop this reconstruction framework thus far, we plan to apply it to other ongoing missions like SVOM \citep{atteia2022svom} and Einstein Probe \citep{yuan2022einstein}, as well as upcoming missions like THESEUS \citep{amati2018theseus} and HiZ-GUNDAM \citep{yonetoku2024high}. Additionally, future analysis will provide data from other wavelengths. Given the availability of the most comprehensive optical catalogue to date, it is particularly appealing to extend this work to optical wavelengths \citep{dainotti2020optical, dainotti2022b, dainotti2024analysis, dainotti2024optical}.

\section{Appendix A. TN-ODE}
\label{sec:appendixA}
This section discusses an additional model we analyzed with, namely the Time-aware Neural Ordinary Differential Equation (ODE). Although this model did not yield satisfactory outcomes for our particular application, we believe it may prove more beneficial in other hybrid configurations or in future research endeavors. 

% \subsection{Time-aware Neural ODE (TN-ODE)}

We developed the \textbf{TN-ODE} model to handle time-series data in which the measurements are irregular, as is frequently the case with GRB LCs. The model consists of three main components:

\begin{enumerate}
    \item \textbf{Encoder:} After reading the supplied data, this module compresses it into a representation with fewer dimensions. It uses a time-aware LSTM network \citep{baytas2017patient}, a variant of RNNs that incorporates the time gaps between observations to control how much past information is retained or forgotten.
    
    \item \textbf{Neural ODE Block:} The compressed latent representation from the encoder is then passed into a Neural ODE block \citep{chen2018neural}. This component models the continuous-time dynamics of the latent state, simulating how it evolves over time. The neural ODE enables interpolation and extrapolation at arbitrary time points, even between known data samples \citep{rubanova2019latent}.
    
    \item \textbf{Decoder:} Finally, the evolved latent representation is decoded into predictions of the original signal (e.g., GRB light intensity) at specified time points.
\end{enumerate}

To incorporate uncertainty and improve generalization, we employ \textbf{variational sampling} \citep{kingma2013auto} during encoding. Stochasticity is introduced into the latent vectors via sampling from a training-learned distribution. This step allows the model to generate a distribution over possible outputs, enabling it to produce not only point forecasts but also confidence intervals, which quantify the model's uncertainty.

During training and inference, the full process is repeated multiple times with different random samples. This Monte Carlo sampling technique provides probabilistic forecasts and confidence bounds.

In our analysis, TN-ODE demonstrated strong performance in modeling and forecasting GRB LCs with irregular time intervals. However, due to the model's complexity and large number of parameters, careful hyperparameter tuning was required to prevent overfitting and mitigate sensitivity to noise. While the uncertainty estimates were generally informative, they were sometimes unstable because of the stochastic sampling process.

\section{Appendix B. Shift Problem Resolution}
\label{sec:appendixB}

\begin{figure}[h]
    \centering
    \includegraphics[width=0.8\linewidth]{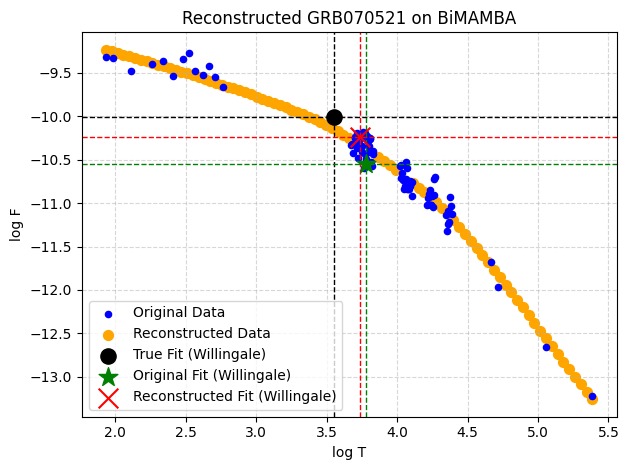}
    \caption{Systematic shift observed in the platinum sample post Bi-MAMBA Model reconstruction. Units: $\log F$ (erg cm $^{-2}$s$^{-1}$) and $\log T$ (s).}
    \label{fig:ann_shift}
\end{figure}

\begin{figure}[h]
    \centering
    \includegraphics[width=0.8\linewidth]{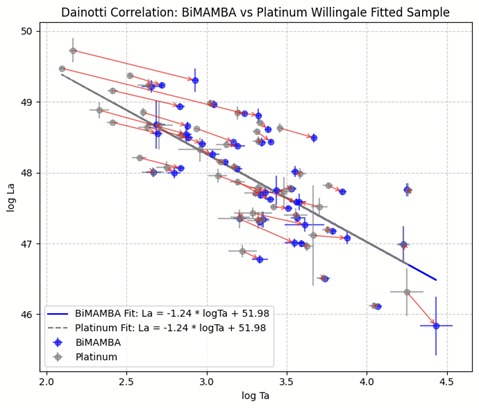}
    \caption{Systematic shift observed in the platinum sample post BiMamba Model reconstruction for $\log L_a$ (erg s$^{-1}$) and $\log T_a$ (s).}
    \label{fig:bimamba_shift}
\end{figure}

In this section, we detail an investigation into systematic shifts at larger times in the reconstructed $\log T_a$ and $\log F_a$ parameters, relative to the original GRBs (see Figure~\ref{fig:ann_shift}). The systematic shift in the platinum sample is shown in Figure \ref{fig:bimamba_shift}. This issue was persistent across all models utilized in this study. This deviation consequently produced an elevated scatter value in the reconstructed platinum sample, and a different slope of the 2-D Dainotti relation between the rest-frame end time $T_a$ and its luminosity, $L_a$. a result that is contradictory to the research's main goal.

To resolve this anomaly, the following investigative methods were employed:
\begin{itemize}
    \item \textbf{Noise Recalibration:} The reconstruction's noise generation protocol was recalibrated. Instead of assuming uniform or Gaussian distributions, samples were drawn from the best-fit flux error distribution. This change was implemented to mitigate artificial bias in the post-$T_a$ regions. However, it was observed that the noise level was not the cause of the systematic shift.
    
    \item \textbf{Hybrid Model Testing:} We explored multiple hybrid models, such as interfacing an ML model with a power-law decay curve. A separate strategy applied a double-power law to the ML model output following a GRB-specific threshold. None of these hybrid attempts, however, produced a significant reduction in the observed shift or the overall scatter.
    
    \item \textbf{$\sigma$-Cut Filtering:} A filtering mechanism using a $\pm2\sigma$ cut was applied to the $(\log T_a, \log F_a)$ distribution. The purpose was to eliminate extreme outliers that were presumed to be artifacts of the reconstruction process. Also, this cut did not resolve the issue.
    
    \item \textbf{Simulation Testing:} We simulated synthetic GRBs with predefined $\log T_a$, $\log F_a$, and $\alpha$ parameters. By introducing controlled data gaps of varying magnitudes, we analyzed the conditions that produced the shift. The analysis revealed that the shift was predominantly caused by large flux gaps located immediately after the $T_a$ point.
    
    \item \textbf{Data-Augmentation and gap-aware reconstruction:} To address this core issue, we developed a `gap-aware' reconstruction algorithm. This method first identifies significant data gaps (defined as $\Delta \log t > 0.05$). It then adaptively inserts new data points exclusively into these identified sparse regions. The quantity of new points is determined adaptively as a percentage of the original LC's length: 5\% (for $>500$ points), 10\% (for 250--500 points), 30\% (for 100--250 points), and 40\% (for $<100$ points, minimum 20). This ensures that larger gaps are filled with proportionally more data, enhancing continuity without oversampling the denser segments.
\end{itemize}

\begin{figure}[h]
    \centering
    \includegraphics[width=0.8\linewidth]{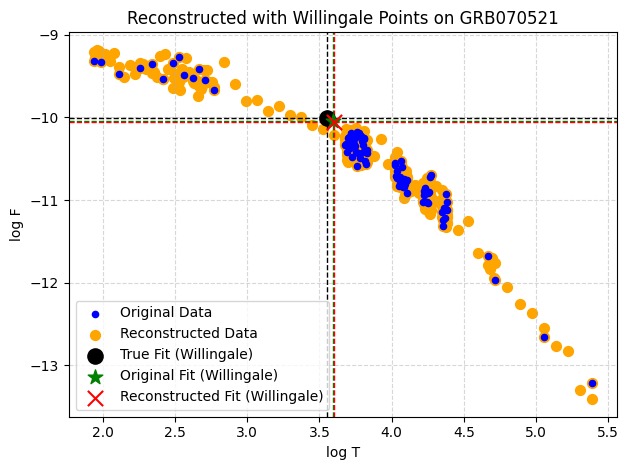}
    \caption{Resolved shift post Bi-MAMBA Model reconstruction. Units: $\log F$ (erg cm $^{-2}$s$^{-1}$) and $\log T$ (s).}
    \label{fig:ann_fixed}
\end{figure}

\begin{figure}[h]
    \centering
    \includegraphics[width=0.8\linewidth]{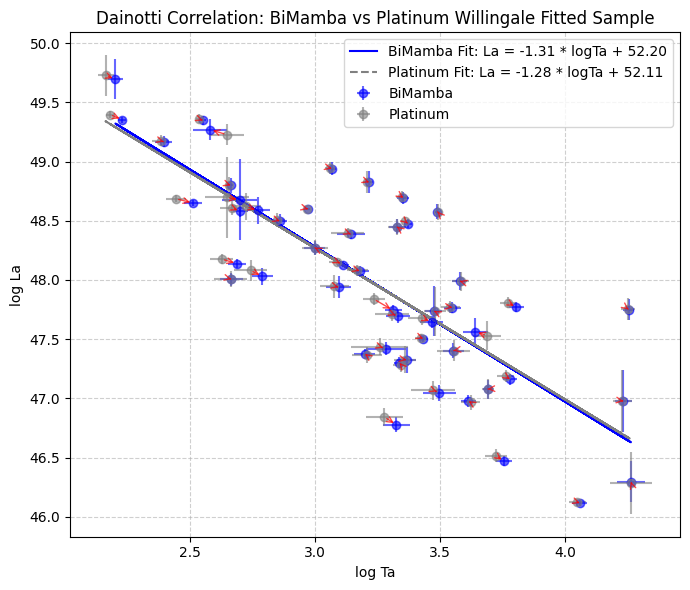}
    \caption{Resolved shift post gap-aware BiMamba Model reconstruction for $\log L_a$ (erg s$^{-1}$) and $\log T_a$ (s).}
    \label{fig:bimamba_fixed}
\end{figure}

The gap-aware reconstruction method succeeded in resolving the systematic shift (Figure \ref{fig:ann_fixed}) and was then applied across the complete 521 GRB sample. After implementing these corrections, the systematic shift was gone (Figure~\ref{fig:bimamba_fixed}), and the reconstructed GRBs achieved a scatter ($\sigma_{\text{BiMamba}} = 0.5134$) statistically comparable to the original platinum sample ($\sigma_{\text{Platinum}} = 0.5131$), remaining consistent within $1\sigma$. 



\section*{CRediT authorship contribution statement}
 \textbf{A. Kaushal:} Conceptualization, Methodology, Supervision, Project Administration, Investigation, Software, Writing Original draft - Review \& Editing.  
 \textbf{A. Manchanda:} Conceptualization, Methodology, Supervision, Project Administration, Investigation, Software, Writing Original draft - Review \& Editing.
 \textbf{M. G. Dainotti:} Conceptualization, Methodology, Supervision, Project Administration, Software, Resources, Writing Original draft - Review \& Editing.
 \textbf{Krishnanjan Sil:} Conceptualization, Methodology, Software, Validation, Formal Analysis, Investigation, Data Curation, Writing - Original Draft, Writing - Review \& Editing, Visualization. 
 \textbf{K. Gupta:} Conceptualization, Methodology, Software, Validation, Formal Analysis, Investigation, Writing - Original Draft, Writing - Review \& Editing.
 \textbf{Z. Nogala:} Methodology, Software, Validation.  
 \textbf{ A. Madhan:} Methodology, Software, Validation. 
 \textbf{S. Naqi:} Methodology, Software, Validation. 
 \textbf{R. Kumar:} Methodology, Software, Validation, Formal analysis, Optimization, writing -OriginalDraft.
 \textbf{V. Oad:} Methodology, Software, Validation. 
 \textbf{N. Indoriya:} Methodology, Software, Validation.
 \textbf{D.H. Hartmann:} Writing - Review \& Editing. 
 \textbf{M. Bogdan:}  Methodology. 
 \textbf{A. Pollo:} Methodology. 
 \textbf{N. Fraija:} Writing - Review \& Editing.  
\textbf{J.X. Prochaska:} Methodology. 
\textbf{N. Fraija:} Methodology.
% \textbf{K. Gupta:} . \textbf{Krishnanjan Sil:} Writing - Original Draft, Writing - Review \& Editing, Methodology, Software, Conceptualization. \textbf{M. G. Dainotti:} . \textbf{D. H. Hartmann:} . \textbf{M. Bogdan:} . \textbf{A. Pollo:} . \textbf{N. Fraija:} . \textbf{E. Gangler:} . \textbf{Shigehiro Nagataki:} . \textbf{Jurgen Mifsud:} . \textbf{Jackson Levi Said:} . \textbf{JX. Prochaska:} .

\section*{Declaration of competing interest}

The authors declare that they have no known competing financial
interests or personal relationships that could have appeared to influence the work reported in this paper.

\section*{Data availability}

Data will be made available upon reasonable request to the corresponding author. 
% \textcolor{yellow}{The data is publicly available/The data supporting the findings of this study are based on simulations that are proprietary and specifically for ...}

%

\section*{Acknowledgements}
The authors thank Aditya Narendra, Nikita Khatiya, and Dhruv Bal for their important inputs regarding the examination of the models. We thank Spencer James Gibson and Federico Da Rold for their helpful recommendations for exploring new kinds of algorithms for the reconstruction process. We also extend our gratitude to Shaivi Malik for her co-implementation of the TCN model, and to Himanshu Gupta and Sudhipta Roy for their assistance with running the model.

We are grateful to Dr. Jurgen Mifsud, Dr. Purba Mukherjee, Dr. Konstantinos F. Dialektopoulos, Prof. Emmanuel Gangler, and Prof. Jackson Said for their insightful feedback and discussion of our analysis. M.G. D. is supported by JSPS Grant-in-Aid Scientific Research (KAKENHI) (A), Grant Number JP25H00675.

\bibliographystyle{elsarticle-harv} 
\bibliography{bibliography}

\end{document}